\documentclass[fleqn,usenatbib]{mnras}

\usepackage{newtxtext,newtxmath}
\usepackage{subcaption}
\usepackage[version=4]{mhchem}
\usepackage{tabularray}
\usepackage{graphicx} 
\usepackage{xspace}
\usepackage[T1]{fontenc}

\DeclareRobustCommand{\VAN}[3]{#2}
\let\VANthebibliography\thebibliography
\def\thebibliography{\DeclareRobustCommand{\VAN}[3]{##3}\VANthebibliography}

\usepackage{graphicx}
\usepackage{amsmath}

\newcommand{\one}{3DOneExpl\xspace}
\newcommand{\two}{3DTwoExpl\xspace}
\newcommand{\oneno}{3DOneExplNoHe\xspace}
\newcommand{\twono}{3DTwoExplNoHe\xspace}
\newcommand{\oneO}{1DOneExpl\xspace}
\newcommand{\twoO}{1DTwoExpl\xspace}
\newcommand{\onenoO}{1DOneExplNoHe\xspace}
\newcommand{\twonoO}{1DTwoExplNoHe\xspace}
\newcommand{\mass}{\mbox{M$_{\odot}$}}

\newcommand{\fe}{SN 2011fe\xspace}
\newcommand{\bt}{SN 2006bt\xspace}

\newcommand{\submch}{sub-$M_\mathrm{Ch}\xspace$}
\newcommand{\dmm}{$\Delta m_{15}\xspace$}
\newcommand{\dmb}{$\Delta m_{15}(B)\xspace$}
\newcommand{\dmv}{$\Delta m_{15}(V)\xspace$}

\title[Will the secondary white dwarf endure?]{%
  On the fate of the secondary white dwarf in
  double-degenerate \\double-detonation Type~Ia supernovae - II. 3D synthetic observables}

\author[J. M. Pollin et al.]{J. M. Pollin,$^{1}$\thanks{E-mail: jpollin02@qub.ac.uk}
S. A. Sim,$^{1}$
R. Pakmor,$^{2}$
F. P. Callan,$^{1}$
C. E. Collins,$^{3}$
L. J. Shingles,$^{3}$
F. K Röpke $^{4,5}$
\newauthor
and S. Srivastav$^{6}$
\\
$^{1}$Astrophysics Research Center, School of Mathematics and Physics, Queen’s University Belfast, Belfast BT7 1NN, Northern Ireland, UK\\
$^{2}$Max-Planck-Institut für Astrophysik, Karl-Schwarzschild-Str. 1, D-85748, Garching, Germany\\
$^{3}$GSI Helmholtzzentrum für Schwerionenforschung, Planckstraße 1, 64291 Darmstadt, Germany\\
$^{4}$Heidelberger Institut für Theoretische Studien, Schloss-Wolfsbrunnenweg 35, 69118 Heidelberg, Germany\\
$^{5}$Zentrum für Astronomie der Universität Heidelberg, Institut für Theoretische Astrophysik, Philosophenweg 12, 69120 Heidelberg, Germany\\
$^{6}$Department of Physics, University of Oxford, Denys Wilkinson Building, Keble Road, Oxford OX1 3RH, UK\\ }

\date{Accepted XXX. Received YYY; in original form ZZZ}

\pubyear{2024}

\begin{document}
\label{firstpage}
\pagerange{\pageref{firstpage}--\pageref{lastpage}}
\sloppy
\maketitle

\begin{abstract}
A leading model for Type~Ia supernovae involves the double-detonation of a sub-Chandrasekhar mass white dwarf. Double-detonations arise when a surface helium shell detonation generates shockwaves that trigger a core detonation; this mechanism may be triggered via accretion or during the merger of binaries. Most previous double-detonation simulations only included the primary white dwarf; however, the fate of the secondary has significant observational consequences. Recently, hydrodynamic simulations accounted for the companion in double-degenerate double-detonation mergers. In the merger of a 1.05\mass\space primary white dwarf and 0.7\mass\space secondary white dwarf, the primary consistently detonates while the fate of the secondary remains uncertain. We consider two versions of this scenario, one in which the secondary survives and another in which it detonates. We present the first 3D radiative transfer calculations for these models and show that the synthetic observables for both models are similar and match properties of the peculiar 02es-like subclass of Type~Ia supernovae. Our calculations show angle dependencies sensitive to the companion's fate, and we can obtain a closer spectroscopic match to normal Type~Ia supernovae when the secondary detonates and the effects of helium detonation ash are minimised. The asymmetry in the width-luminosity relationship is comparable to previous double-detonation models, but the overall spread is increased with a secondary detonation. The secondary detonation has a meaningful impact on all synthetic observables; however, multidimensional nebular phase calculations are needed to support or rule out either model as a likely explanation for Type~Ia supernovae.

\end{abstract}

\section{Introduction}
\label{sec:Introduction}

There is general agreement that the explosion mechanism responsible for a Type Ia supernovae (SNe~Ia) involves a carbon-oxygen (CO) white dwarf (WD) \citep{Hoyle1960} in a close binary. However, it remains an open question if the WD undergoes a thermonuclear explosion near the Chandrasekhar mass ($M_\mathrm{Ch}$) or whether sub-Chandrasekhar mass models (sub-$M_\mathrm{Ch}$) produce SNe~Ia (see e.g.\,\citealt{Liu2023} for a review). Similarly, an extensive debate persists regarding whether the single-degenerate \citep{Whelan1973} or double-degenerate \citep{Iben1984} scenario is responsible for producing SNe~Ia. Several observational motivations exist for favouring WD binaries for SNe~Ia; the double-degenerate scenario would occur in rates consistent with SNe~Ia \citep{Ruiter2009}, the lack of hydrogen in spectra is easily explained \citep{Leonard2007}, and they can account for the delay time distribution of SNe~Ia \citep{Maoz2012}.

Previous investigations into detonations of CO \submch~WDs have been shown to produce good agreement with observations of SNe~Ia \citep{sim2010,Blondin2017,Shen2018,Shen2021a} and have some ability to reproduce the width-luminosity relation \citep{philips}. The general success of these simple models has motivated further exploration of scenarios that can produce \submch~CO detonations. Due to the unknown nature of the companion, there are multiple potential detonation mechanisms, with the double-detonation mechanism being of particular interest. In the double-detonation scenario, a detonation first occurs in a layer of helium on the surface of the \submch WD, and this triggers the detonation of the underlying CO core  \citep{Taam1980,Livne1990,Nomoto1980,Hoeflich1996,Nugent1997}. 

Early investigations necessitated slow helium accretion rates \citep{Taam1980b,Taam1980,Nomoto1982,Woosley1986} to ensure a carbon detonation occurs, hence early double-detonation models possessed a $\sim$0.4\mass \ thick helium shell \citep{Livne1990,Livne_Glasner1990,Livne1995}. These large helium shell detonations produce a considerable amount of intermediate-mass elements (IMEs) and iron-group elements (IGEs) \citep{kromer2010,Polin2019}, which yield spectroscopic features that more closely resemble that of peculiar SNe~Ia subclasses \citep{Inserra2016,Jiang2017,De2019,Dong2022}. There has been increased scrutiny on the amount of helium required to produce a detonation and investigations have suggested that a detonation is possible even for significantly less massive shells \citep{Bildsten2007,Shen2009,fink2010,Shen2010}. \cite{kromer2010} highlighted the substantial impact of helium shell composition on photospheric phase features and noted that the effect of the lines produced by the optically thick IGEs in the shell detonation could be minimised by reducing the helium shell mass, which has been shown to produce more favourable synthetic observables \citep{Townsley2019,Shen2021b,Boos2021}; as such, the mechanism remains widely investigated.

The double-detonation mechanism can be applied to a dynamically driven double-degenerate merger scenario where the primary WD encounters unstable helium accretion from the secondary WD just before the merger, which heats the helium shell on the primary. The presence of dynamical instabilities in the helium accretion stream leads to a thermonuclear runaway in the helium shell, leading to the double-detonation of the primary WD \citep{Guillochon2010,Pakmor2013,Kashyap2015,Tanikawa2018,Boos2021,Gronow2020,Shen2021b,pakmor_2021}, for which there is an increasingly growing observational \citep{Dong2022,Liu2023_he}, and theoretical motivation \citep{Collins2023}. A notable difference between the double-degenerate double-detonation scenario and a violent merger scenario \citep{Pakmor2012b} lies in the detonation timing. In the violent merger scenario, the secondary WD is disrupted during the merging of the CO cores, which destroys the whole system and has been suggested as the progenitor system of the peculiar 02es-like SNe~Ia \citep{Maguire2011,Ganeshalingam2012,Pakmor2013} due to the subclass possessing peak magnitudes and decline rates, comparable to normal SNe~Ia while having spectrocopic features similar to 91bg-like SNe~Ia \citep{stefan_sn_review}. Additionally if a CO detonation was to occur later after the dynamical merger in a "helium-ignited violent merger" it would produce a highly asymmetric supernova remnant \citep{Tanikawa2015}. Unlike the violent merger scenarios the detonation in the double-degenerate double-detonation scenario occurs earlier during the dynamical inspiral of the helium shell material. As a result of the earlier detonation of the primary, uncertainty exists on the companion's fate as, at the time of the first thermonuclear runaway, it remains intact. There is also theoretical argument that a secondary WD detonation would produce peculiar SNe~Ia due to the large amount of $^{56}$Ni synthesized \citep{Tanikawa2019}. As a consequence of the uncertainty surrounding the secondary WD and how the total $^{56}$Ni produced would effect synthetic observables, \cite{pakmor_2021} investigated two models in which the only difference is the fate of the secondary WD. The investigation showed that regardless of the fate of the secondary WD, the synthetic observables are remarkably similar. Moreover, \citet{Boos2024} has also shown that a double-degenerate double-detonation explosion scenario can produce synthetic observables in 2D, which appear similar to thin-shelled helium detonations and hence appear similar to normal SNe~Ia. 

In this work, we expand on the promising work of \cite{pakmor_2021} by investigating the explosion models using 3D radiative transfer. By exploring the models in 3D, it will be possible to determine if line-of-sight variation in the synthetic observables produces distinctive characteristics that can be used to favour or disfavour either model for SNe~Ia. Many double-detonation models appear too red when compared to observations, which can be attributed to the elements synthesized in the outer ejecta \citep{kromer2010, Polin2019, Gronow2020, Shen2021b, collins_double_detonation}. The models investigated here can be classified as having low-mass helium shells compared to previous double-detonation models \citep{Nugent1997} as each WD possesses a total helium shell of 0.03\mass. To further explore the effect of the helium shell detonation, the models with their helium detonation ash removed developed by \citet{pakmor_2021} are also investigated here and are motivated by other investigations finding that minimal helium shell masses produce observables that more closely resemble that of observations \citep{Townsley2019,Shen2021b}. In Section \ref{sec:Method}, we describe the simulation codes and explosion models employed. In Section \ref{sec:Results}, we show light curves and spectra and compare these synthetic observables to observational data. Finally, we discuss the results in a broader context and additional future work we aim to conduct on the dynamically driven double-degenerate double-detonation scenario in Section \ref{sec:Discussion and Conclusions}.

\section{Methods}
\label{sec:Method}
 
\subsection{Radiative Transfer}
\label{sec:Radiative Transfer}

We perform our radiative transfer calculations using the 3D Monte Carlo radiative transfer code \textsc{artis} \citep{sim_2007,Kromer2009}. The methods used by \textsc{artis} are based on \cite{Lucy2002,lucy2003,Lucy2005}. The work performed here uses the approximate NLTE (non-local thermodynamic equilibrium) treatment described by \cite{Kromer2009}. In this approach the ionisation state is estimated using the photoionisation rates produced by the radiation field model and excitation states are controlled via a Boltzmann distribution. The explosion model densities and abundances were mapped to a $50^{3}$ Cartesian grid for the radiative transfer calculations. The models are assumed to be in homologous expansion, and each simulation used 100 logarithmically placed time steps from 0.04 days to 60 days and $6.38$x$10^7$ Monte Carlo packets. Line-of-sight synthetic observables are constructed by binning escaping packets. These packets are binned into 100 uniformly spaced solid angle bins in both $\cos{\theta}$ and $\phi$, where $\theta$ is the angle from the positive z-axis, and $\phi$ is the degree of rotation of the projection in the xy-plane (i.e $\phi$ rotates anti-clockwise from $\phi=0$ which is indicated by the directional arrow in Figure~\ref{Fig: Model abundances plot}).

The radiative transfer calculations in this work used the atomic data set cd23\_gf-20 compiled from the data of \cite{Kurucz1995} as described by \cite{Kromer2009}.
This atomic data set produces optical spectra similar to more extensive data sets with larger line lists; however, we note that the near-infrared region is sensitive to the choice of atomic data sets \citep{Kromer2009}. Previous investigations using \textsc{artis} \citep{Kromer2009} found that a larger atomic data set can result in the emergence of a secondary maximum in the I-band. Additionally we note that the feature is sensitive to the detailed micro-physics \citep{Kasen2006,Kromer2009,Jack2015} of the simulation.
In particular, the timing and strength of the secondary maximum have been shown to change with a complete NLTE solution \citep{Blondin2022}. Hence, discrepancies between observations and synthetic spectra may be partially reduced by using larger atomic data sets with more extensive line lists and implementing full NLTE modeling.

\subsection{Models}
\label{sec:Models}

\begin{figure*}
\centering
\begin{subfigure}[b]{1\textwidth}
   \includegraphics[width=1\linewidth]{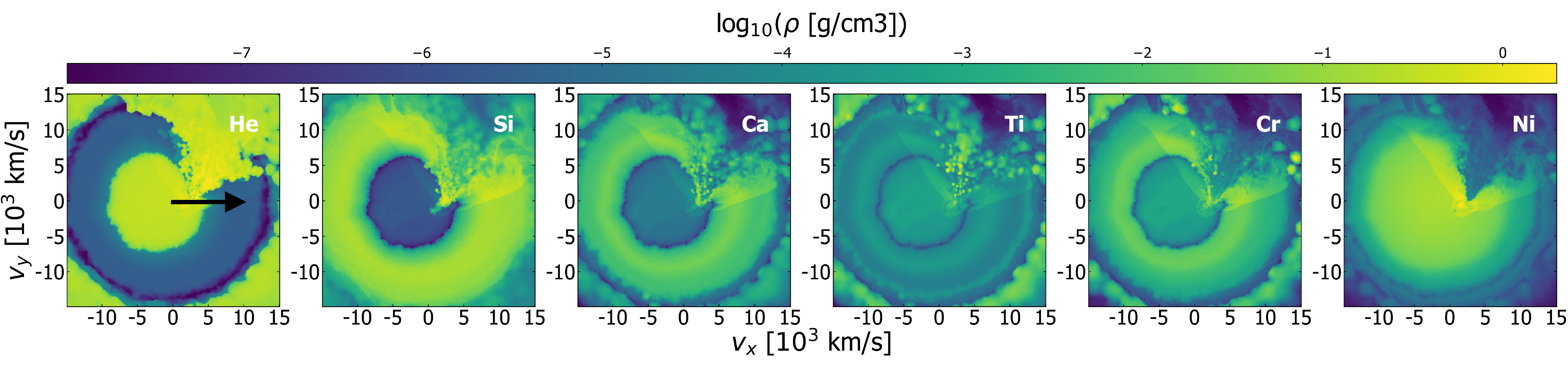}
\end{subfigure}

\begin{subfigure}[b]{1\textwidth}
   \includegraphics[width=1\linewidth]{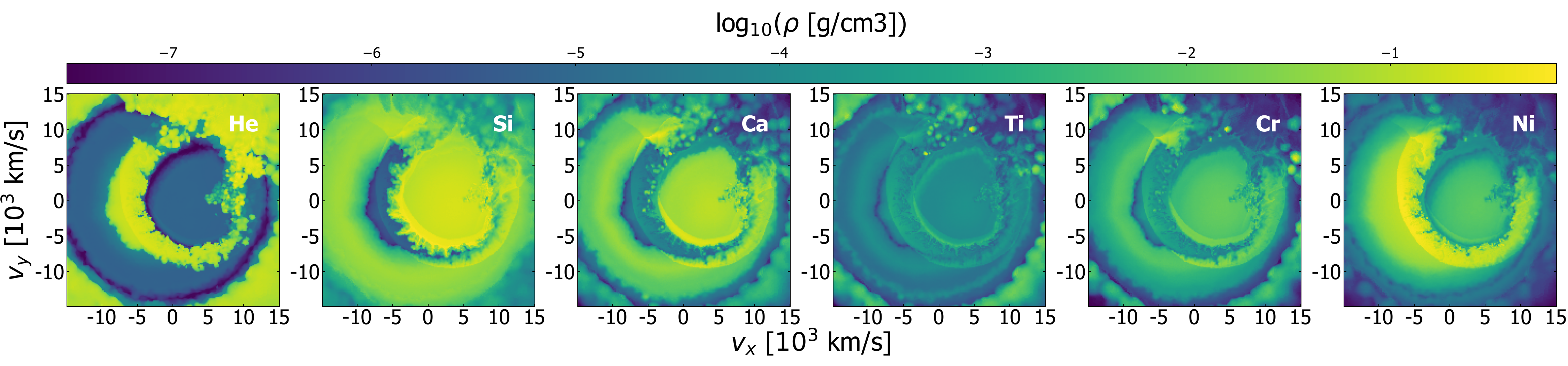}
\end{subfigure}
\caption{Density of key species in the \one (top) and \two (bottom) models, sliced in the X-Y plane at Z=0 where the arrow in the top left panel represents the direction of $\phi=0$. The colour scale indicates the logarithmic density of each species.}
\label{Fig: Model abundances plot}
\end{figure*}

The four 3D models investigated in this work are the \one, \two, \oneno, and \twono simulations of \cite{pakmor_2021}. The models originate from a primary CO WD of mass 1.05\mass\space and a secondary CO WD of 0.7\mass, with each possessing a helium shell of 0.03\mass. The \one model assumes that the secondary survives and the ejecta mass involved in this model is 1.09\mass. The \two assumes that both primary and secondary detonate, such that the total mass of the ejecta is 1.75\mass. The abundances of key species for the \one and \two models are shown in Figure \ref{Fig: Model abundances plot}, and while the outer ejecta show only moderate differences, the inner ejecta display significantly different structures. We also examined two 3D artificial models, which we refer to as \oneno and \twono. These models differ from the others as the helium shell ash has been removed from the ejecta in postprocessing by omitting tracer particles containing helium in their initial composition \citep{pakmor_2021} when mapping to the grid used for the radiative transfer calculations. Within this work, the models that have had their helium detonation ash removed are collectively referred to as the NoHe models, whereas the models which still contain the helium detonation ash are collectively referred to as the full models. We have also replicated the 1D simulations originally produced by \cite{pakmor_2021} to illustrate the effects of multidimensional radiative transfer on the hydrodynamical models. To construct a 1D model the full 3D hydrodynamic simulations are averaged in 100 spherical shells to impose spherical symmetry. The 1D explosion models investigated are referred to as the \oneO, \twoO, \onenoO, and \twonoO.

\section{Results}
\label{sec:Results}

We present the computed light curves and spectra for each explosion model in Sections~\ref{sec:Light Curves} and \ref{sec:Spectra} respectively. We compare the synthetic observables to observational data in Section \ref{Sec: Observational Comparisons}.

\subsection{Light Curves} 
\label{sec:Light Curves}
\subsubsection{Angle-averaged light curves}
\label{sec:Angle Averaged Light Curves}

\begin{figure*}
    \centering
        \includegraphics[width=1\linewidth,height=10cm]{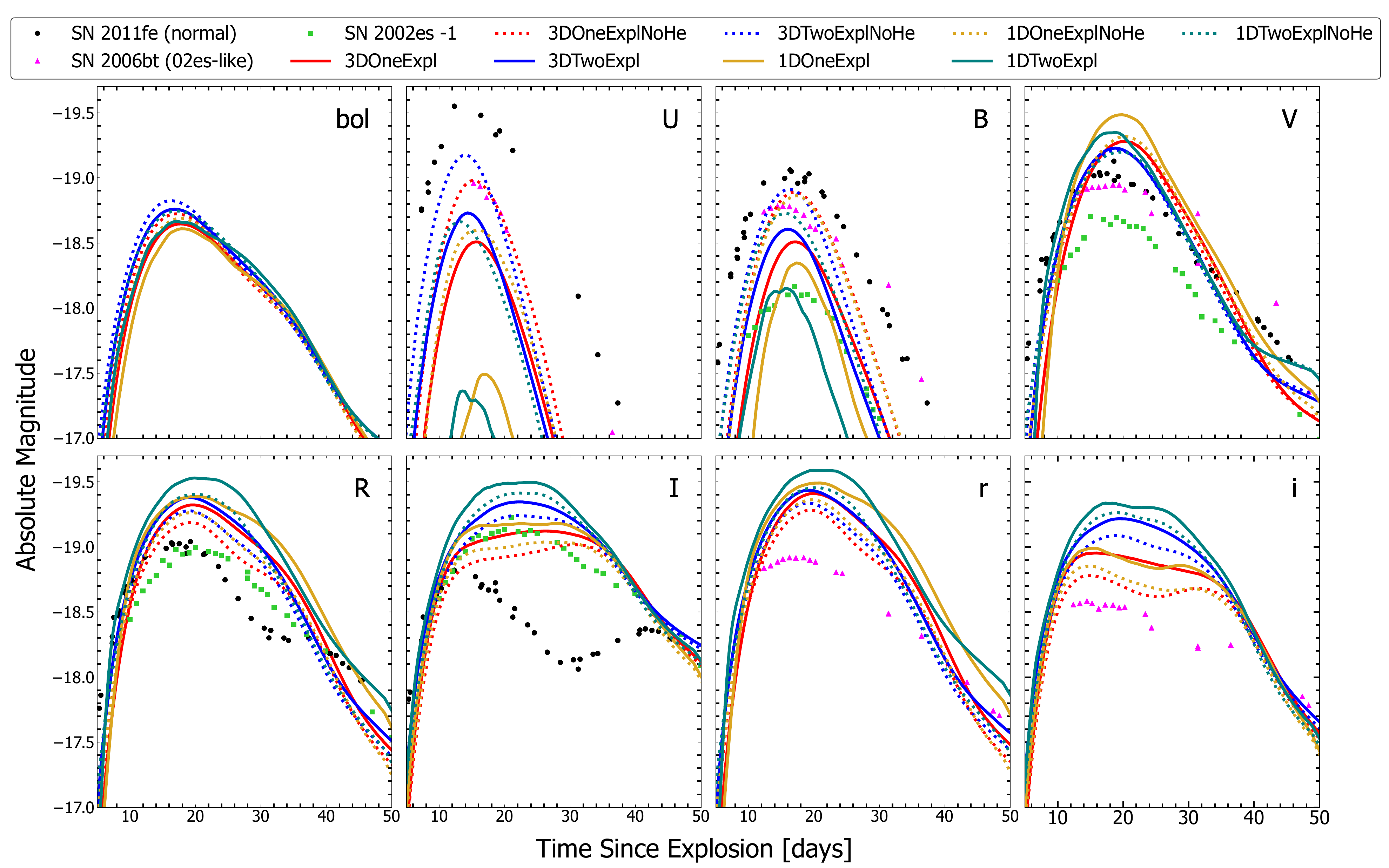}
    \caption{Bolometric and UBVRri band light curves for each model are displayed over 50 days and compared to three observed SNe~Ia. Solid lines represent the full models, while dashed lines represent the NoHe models. Observations of SN 2002es have been shifted by 1 mag for ease of comparison.}
    \label{Fig: spherically averaged light curves}
\end{figure*}

Before examining the line-of-sight variations, we first review the average light curves. Figure \ref{Fig: spherically averaged light curves} shows the light curves for the 1D and 3D angle-averaged models. These synthetic observables are also compared to normal and peculiar SNe~Ia observations which we discuss in Section \ref{Sec: Observational Comparisons}. 

The full 3D simulations produce bolometric light curves that are within $\sim$0.3 mag of the 1D full models at all epochs, which demonstrates the ability of the 1D simulations to capture the overall energy released throughout the simulation. A key difference between the simulations is that the 1D approximation produces bolometric rise times longer than their 3D equivalent ($\sim$0.7 days) and marginally fainter ($\sim$0.06 mag) at peak magnitude. In contrast to the bolometric light curves, the band-limited light curves show large differences, particularly in bluer bands. It can be seen that both the \one and \two models are significantly brighter ($\sim$1 mag) in the U-band than their 1D equivalent. Conversely, the \onenoO and \twonoO models exhibit a similar peak brightness to the 3D simulations, however, they still remain 0.36 mag and 0.45 mag fainter than their respective 3D simulation at peak brightness. This difference can be primarily attributed to the 1D simulation using a spherically-averaged composition, resulting in a model which has a radially varying distribution of ash produced by the helium shell detonation. However, the helium shell detonation is not symmetric and has radial and angular variation, which the spherically-averaged composition cannot display. The overall brighter mean of the 3D simulations is a result of the line-of-sight variation produced by accounting for both the radial and angular nature of the explosion models. The effect of line-of-sight variation is discussed further when orientation effects are presented below.

\subsubsection{Angle-averaged colour evolution}
\label{sec:colour evolution}

\begin{figure*}
    \centering
    \includegraphics[scale=0.25, width=1\linewidth]{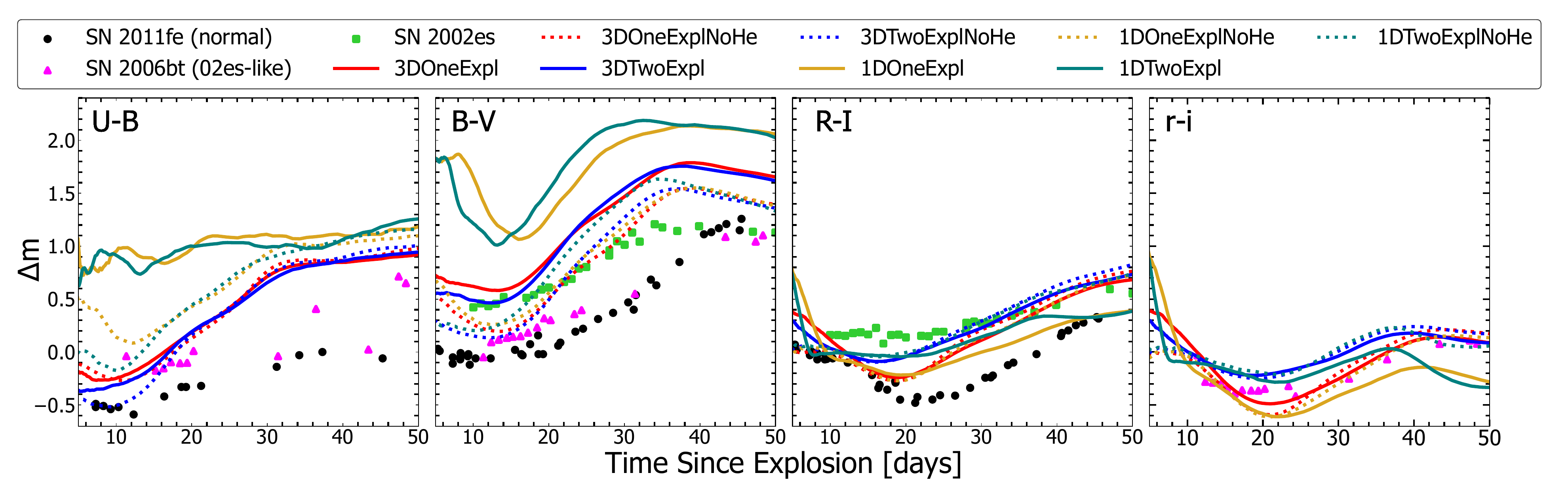}
        \caption{Colour evolution for U-B, B-V, R-I and r-i bands over 50 days compared to three different SNe~Ia. Solid lines represent the full models, while dashed lines represent the NoHe models.}
    \label{Fig: colour evolution spherically averaged}
\end{figure*}

Figure \ref{Fig: colour evolution spherically averaged} displays the 3D angle-averaged and 1D colour evolution for all four models alongside observed supernovae. Initially examining the difference between the colour evolution in 1D and 3D, we find that the 1D colour evolution for \oneO and \twoO models in U-B and B-V appear significantly redder than their 3D equivalent. However, in R-I and r-i, the evolution is marginally bluer compared to the 3D counterpart. The absence of this trend in \onenoO and \twonoO suggests this is linked to line blanketing produced by the helium shell detonation ash. We find that the \two model also exhibits a bluer colour evolution when compared to that of the \one model in U-B and B-V at early times (before $\sim$20 days), while appearing redder in R-I and r-i. We find that the color evolution of the 3D NoHe models in U-B and B-V mainly differs before $\sim$ 15 days, with the \twono model being marginally bluer than the \oneno model. However, in R-I and r-i, the \oneno model is bluer than the \twono model by $\sim$0.2 mag and $\sim$0.5 mag, respectively. In general we find that the effect of the secondary detonation has a more significant effect on the colour evolution than the helium shell detonation ash, with the largest deviation occurring at early times ($\sim$20 days).

\subsubsection{Line-of-sight light curves}
\label{Sec:Line of sight light curves}

\begin{figure*}
\centering
\begin{subfigure}[b]{\textwidth}
    \centering
   \includegraphics[width=0.99\linewidth,trim={0cm 2.8cm 0cm 0cm},clip]{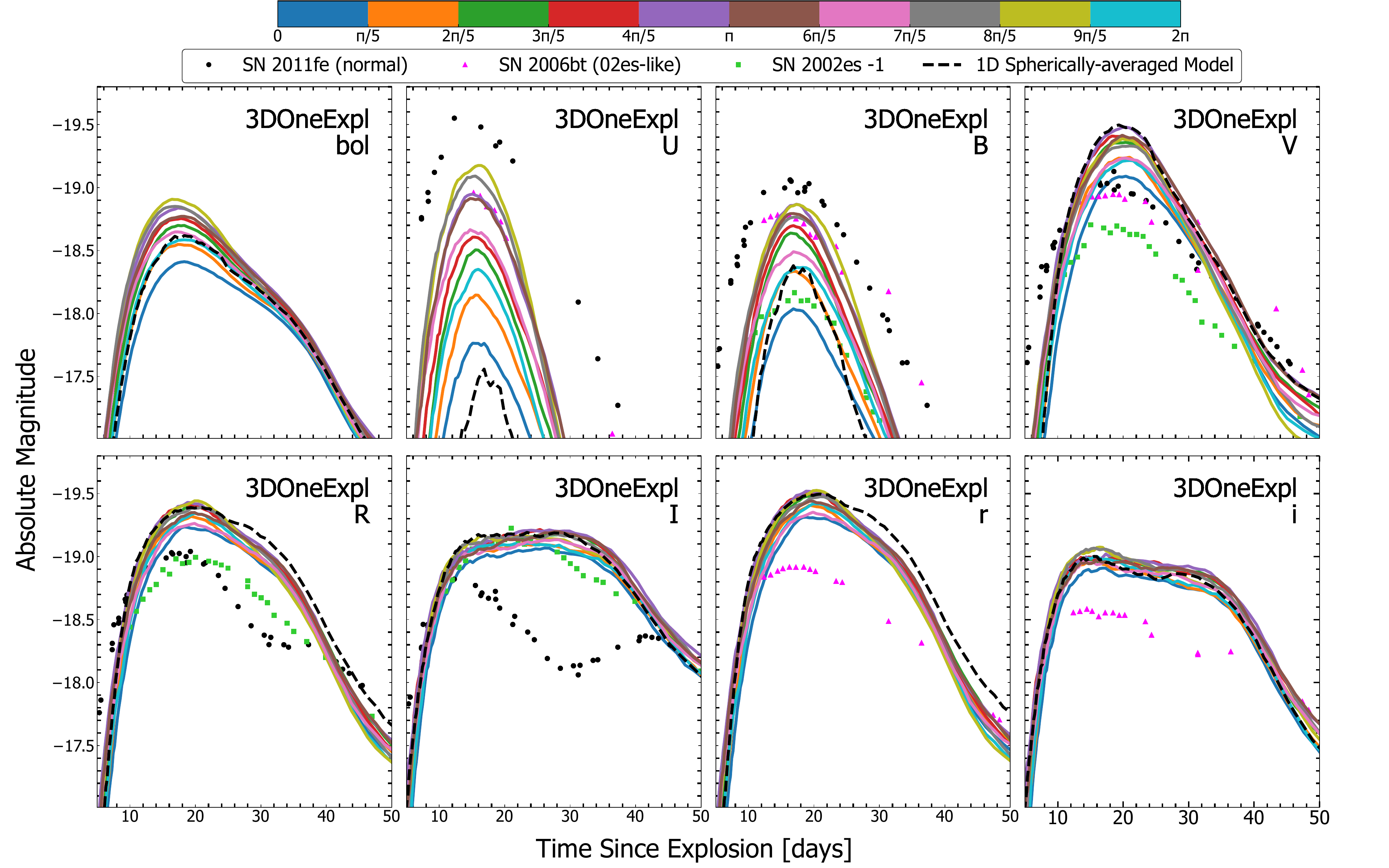}
   \label{fig:Line of sight One Explosion} 
\end{subfigure}
\begin{subfigure}[b]{\textwidth}
    \centering
   \includegraphics[width=0.99\linewidth,trim={0cm 0cm 0cm 4.1cm},clip]{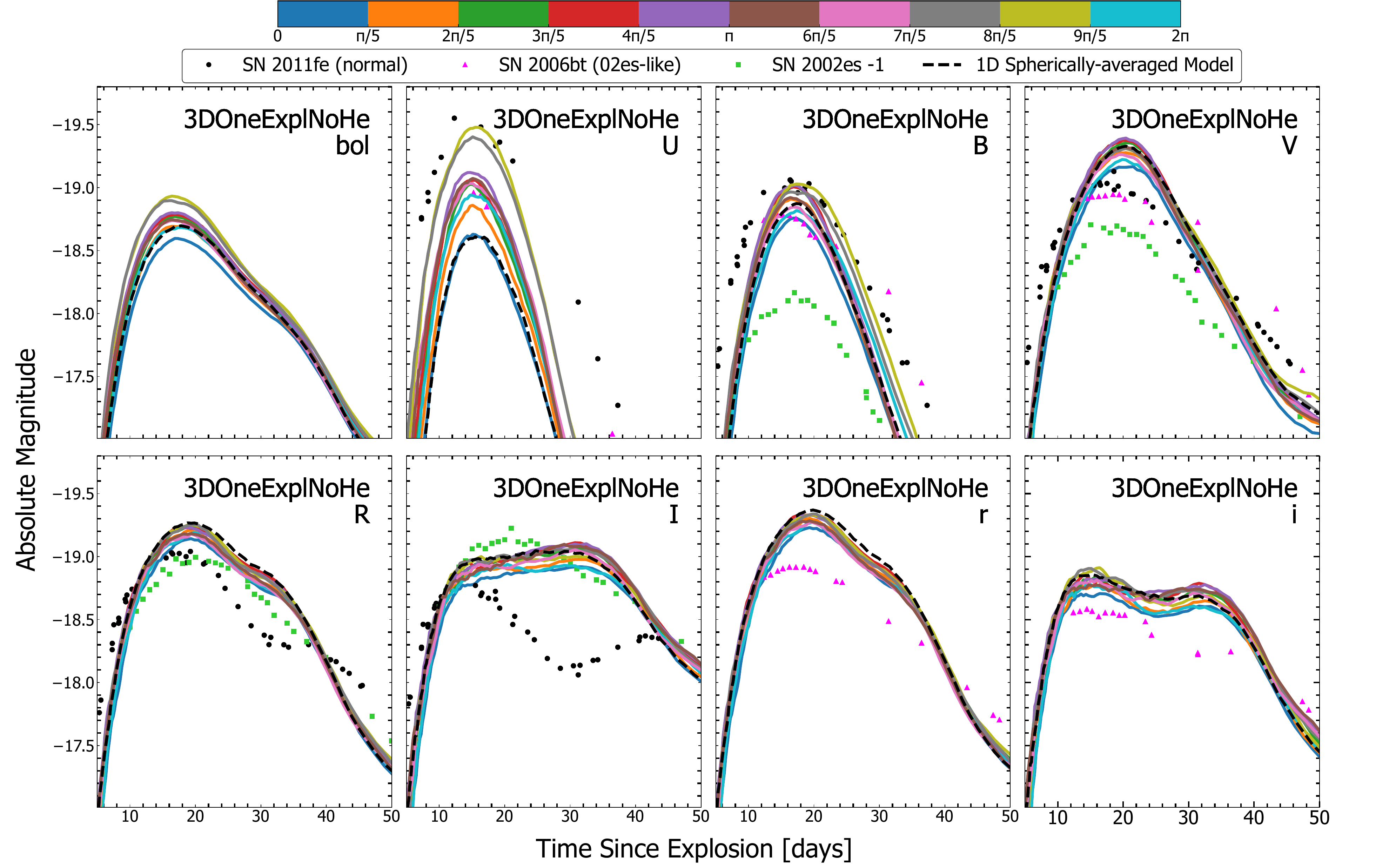}
   \label{fig:Line of sight Two Explosion}
\end{subfigure}

\caption{Bolometric and UBVRIri band light curves displayed over 50 days for all 10 lines-of-sight in the $0.0\le\cos{\theta}<0.2$ plane for the \one, and \oneno models. The line-of-sight light curves are compared to three observed SNe~Ia and the corresponding 1D simulation has been over-plotted for ease of comparison.}
\label{fig:Line of sight Viewing Angles}
\end{figure*}

\begin{figure*}
\begin{subfigure}[b]{\textwidth}
    \centering
   \includegraphics[width=0.99\linewidth,trim={0cm 2.8cm 0cm 0cm},clip]{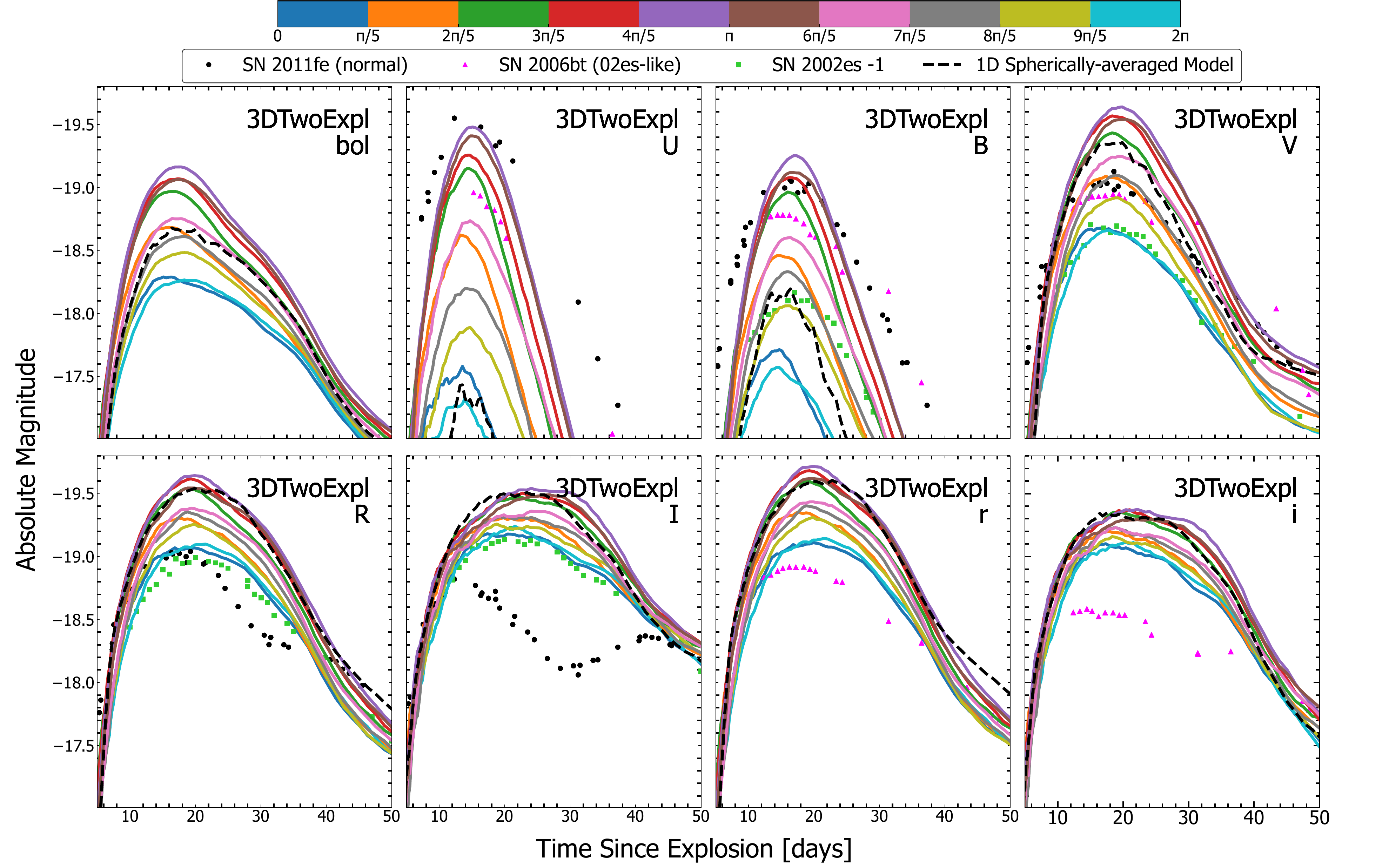}
   \label{fig:Line of sight One Explosion no he} 
\end{subfigure}
\begin{subfigure}[b]{\textwidth}
    \centering
   \includegraphics[width=0.99\linewidth,trim={0cm 0cm 0cm 4.1cm},clip]{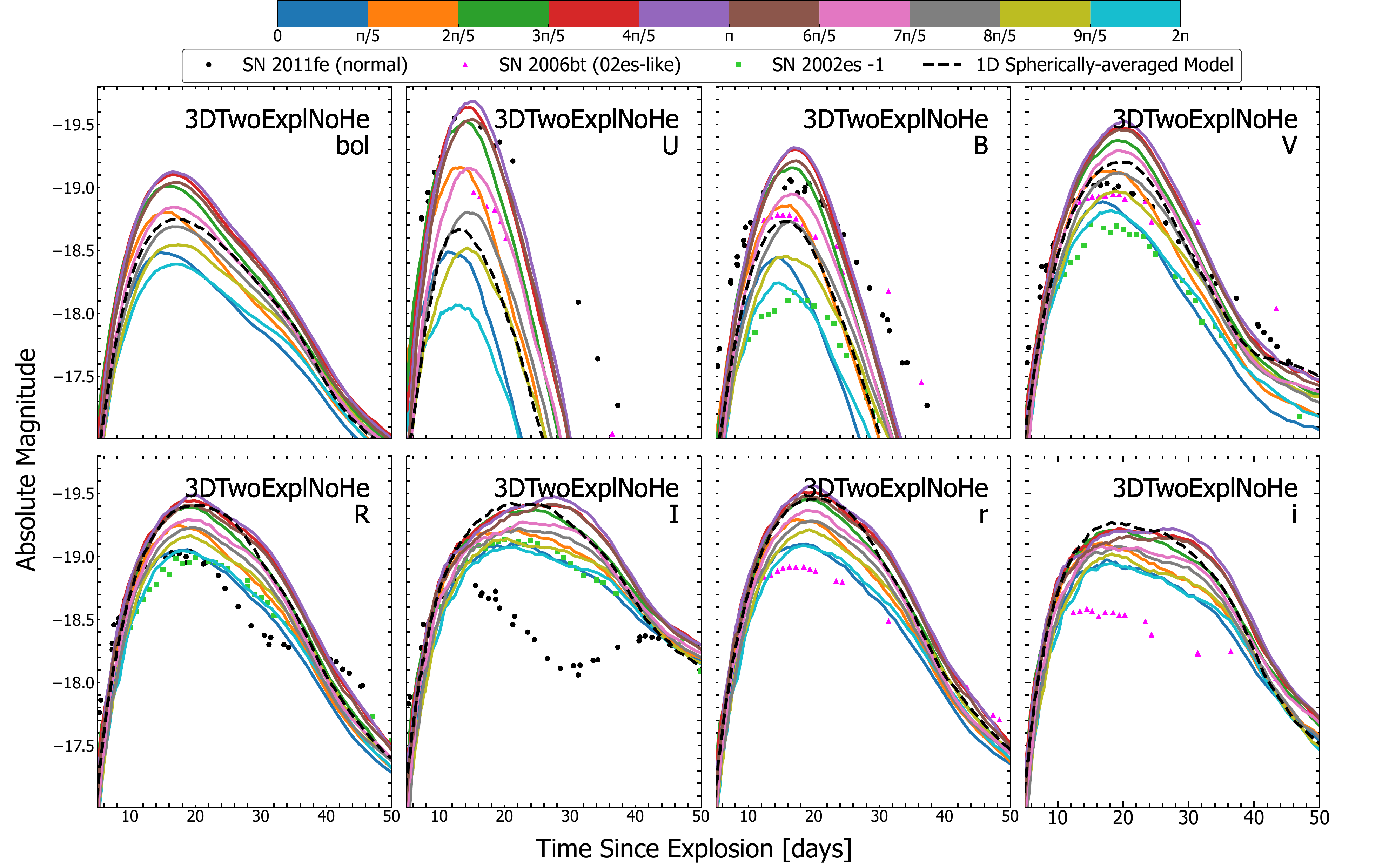}
   \label{fig:Line of sight Two Explosion no he}
\end{subfigure}

\caption{Same as Figure \ref{fig:Line of sight Viewing Angles} for \two and \twono.}
\label{fig:Line of sight Viewing Angles no he}
\end{figure*}

In Figure \ref{fig:Line of sight Viewing Angles} and \ref{fig:Line of sight Viewing Angles no he}, line-of-sight light curves are displayed for the \one, \two, \oneno and \twono models and compared to the same classes of supernovae shown in Section \ref{sec:Angle Averaged Light Curves}. To highlight the degree of variation that can be seen in the synthetic observables due to the different line-of-sight properties, we show synthetic observables in the $0.0 \le \cos{\theta} < 0.2$ plane between $0 \le \phi <2\pi$ which is split into 10 uniformly spaced $\phi$ bins; we note that this interval of $\cos{\theta}$ is close to the merger plane. Although the explosion models display variations for different $\cos{\theta}$ values, the most significant effect is in $\phi$ around the plane close to the merger.

The \one model has a spread of peak bolometric magnitude with observer orientation of $\sim$0.6 mag, while the \two model has a wider spread of $\sim$1 mag. For both the \one and \two models, the U and B band light curves show the most viewing-angle dependency while the redder bands show less variation. This is broadly similar to findings from previous double-detonation models \citep{kromer2010,Shen2021a,collins_double_detonation} that did not include the companion. The variation produced by the line-of-sight light curves in this work are more extensive than the 2D light curves produced by \citet{Boos2024}, which also displays a significant degree of angle dependency that is most extensive in the bluest bands. We find that the \one model displays less variation than the \two model and when the helium ash is removed the line-of-sight variation is reduced significantly in both scenarios. This reduction is most prominent in the U-band where the spread in peak luminosity has been reduced by $\sim$0.6 mag in both scenarios. This confirms that the ash plays a critical role in causing the extreme line-of-sight variations.

The exploration of line-of-sight light curves has highlighted a crucial difference between 1D and 3D shell detonation simulations; very few lines-of-sight resemble their 1D counterpart in the bluest bands. Simultaneously in the reddest bands this trend is reversed and only the extremely bright lines-of-sight resemble their 1D counterpart. Hence, for these models, the 1D calculations do not produce observables that capture the average observer orientation but instead more closely correspond to extreme cases predicted in the 3D calculation. Upon further examination, it can be seen that the \onenoO and \twonoO models produce lines-of-sight closer to matching those of the full 3D simulations.

\subsubsection{Line-of-sight colour evolution}
\label{sec:Line of sight Colour Evolution}

\begin{figure*}
\centering
\begin{subfigure}[b]{\textwidth}
   \includegraphics[scale=0.2,width=1\linewidth,height=5cm,trim={0cm 2.6cm 0cm 0cm},clip]{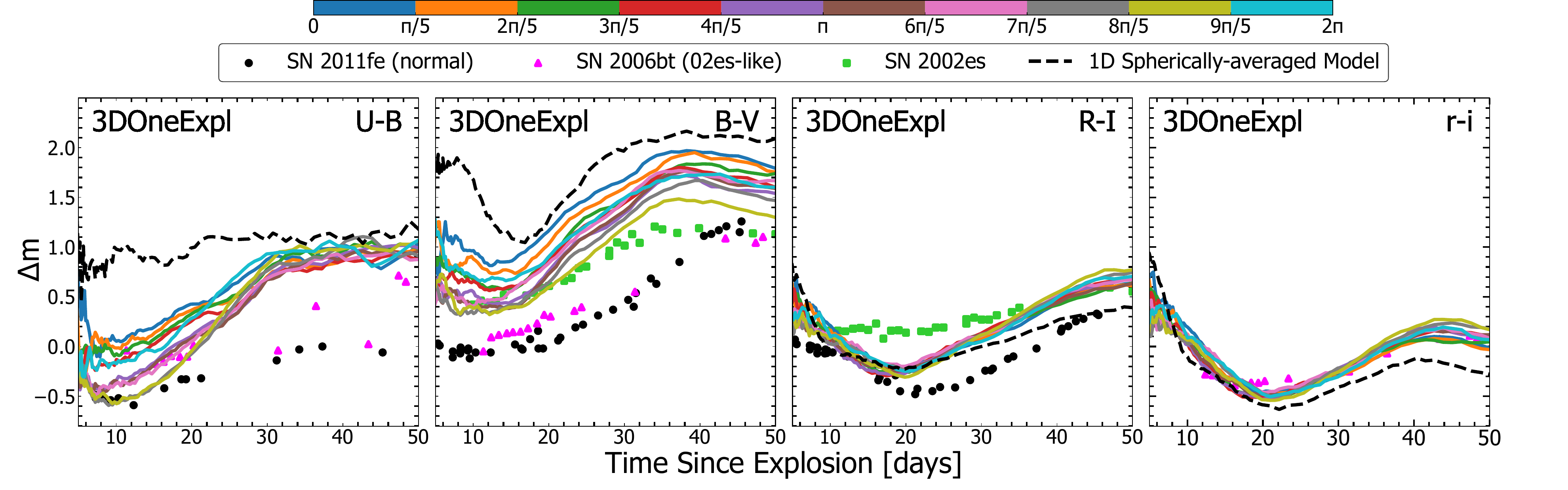}
   \label{fig:line of sight colour evolution One Explosion} 
\end{subfigure}
\begin{subfigure}[b]{\textwidth}
   \includegraphics[scale=0.2,width=1\linewidth,height=3.8cm,trim={0cm 2.6cm 0cm 4.0cm},clip]{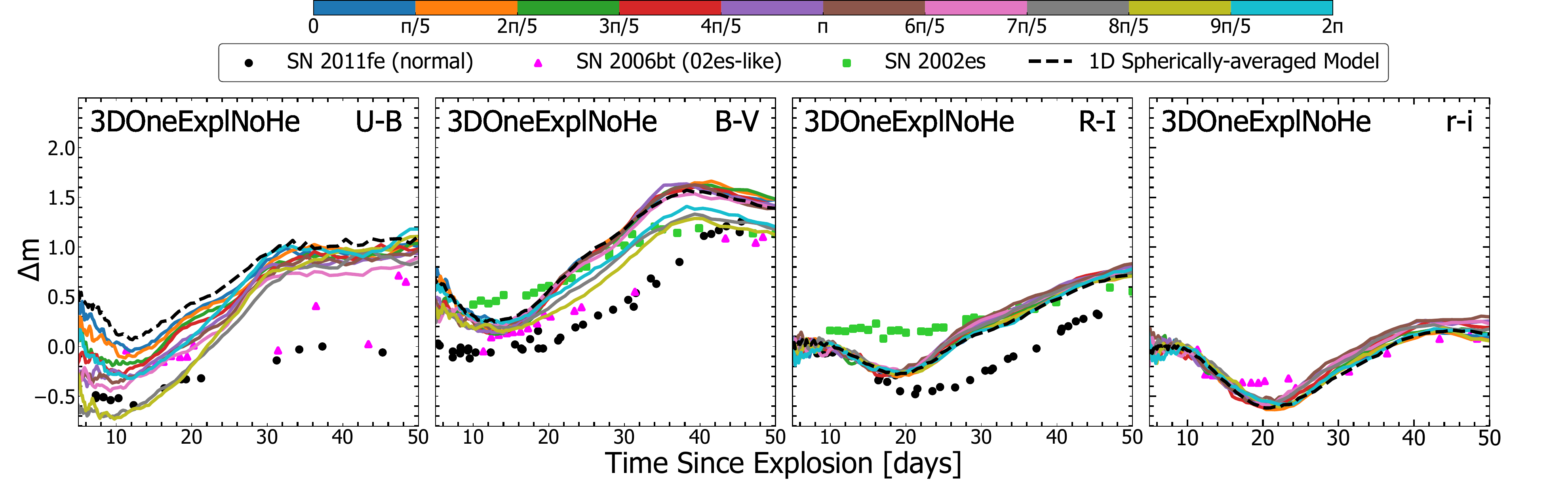}
   \label{fig:line of sight colour evolution Two Explosion}
\end{subfigure}
\begin{subfigure}[b]{\textwidth}
   \includegraphics[scale=0.2,width=1\linewidth,height=3.8cm,trim={0cm 2.6cm 0cm 4.0cm},clip]{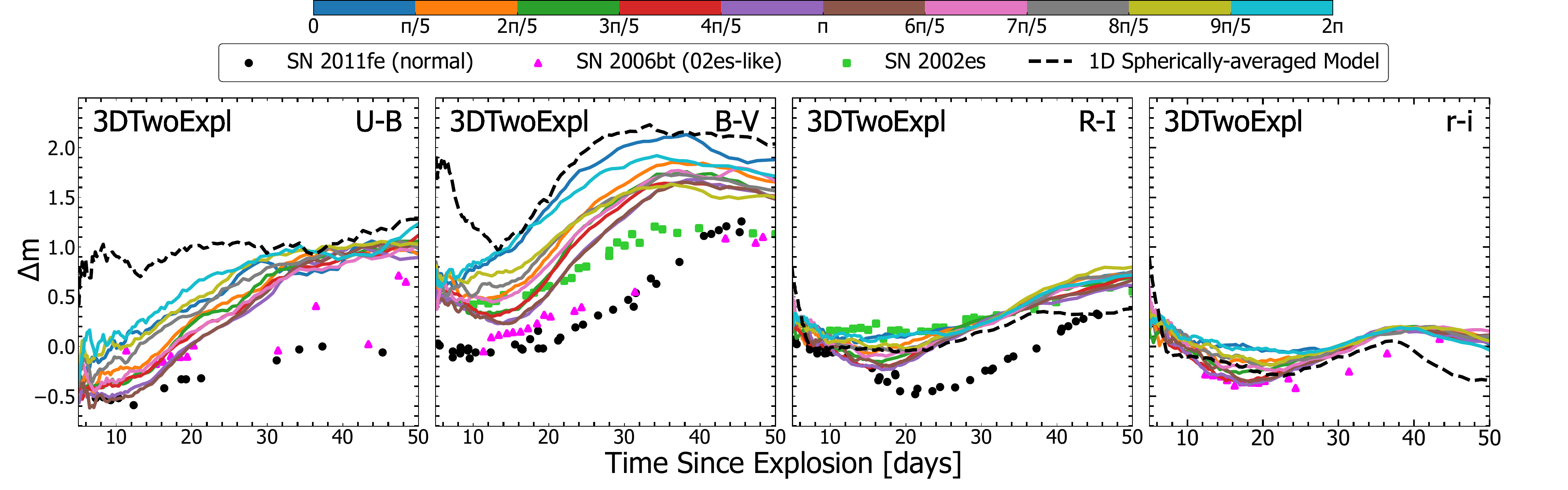}
   \label{fig:line of sight colour evolutionOne Explosion No He} 
\end{subfigure}
\begin{subfigure}[b]{\textwidth}
   \includegraphics[scale=0.2,width=1\linewidth,height=4.4cm,trim={0cm 0.5cm 0cm 4.0cm},clip]{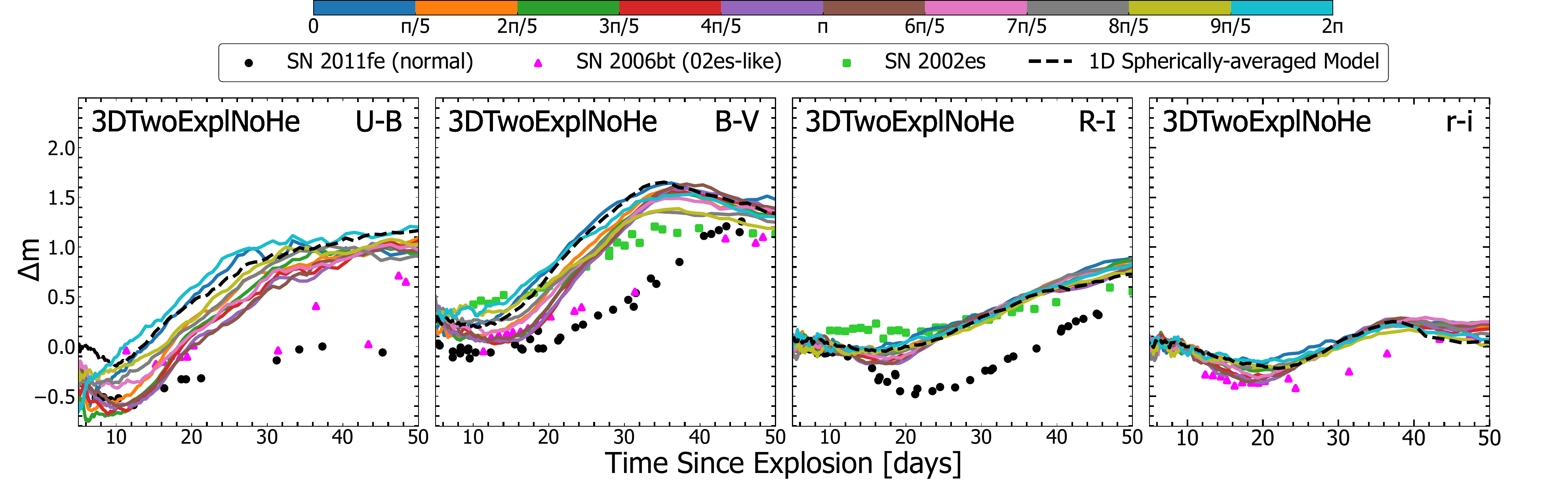}
   \label{fig:line of sight colour evolution Two Explosion No He}
\end{subfigure}

\caption{Line-of-sight colour evolution of \one, \two, \oneno and \two models for the U-B, B-V, R-I and r-i bands displayed over 50 days. The lines-of-sight are compared to three observed supernovae and the corresponding 1D simulation has also been over-plotted for ease of comparison.}
\label{fig:Line of sight colour evolution viewing angles}
\end{figure*}

In Figure \ref{fig:Line of sight colour evolution viewing angles}, line-of-sight colour evolution is shown for the \one, \two, \oneno and \twono models and compared to the same three supernovae as in Section \ref{sec:Angle Averaged Light Curves}. The colour evolution is shown for the same 10 lines-of-sight displayed in the previous Section, which encompass the most substantial viewing-angle variation, thus showing the most distinct colour evolution the model can possess. As can be anticipated from Section \ref{Sec:Line of sight light curves}, colour evolution derived from redder bands show little orientation dependence while the largest line-of-sight variations appear in the U-B and B-V colours. The \two model shows more extensive orientation dependence than the \one model which is apparent in the B-V evolution where the \two model displays a consistently broader spread of $\sim$0.1 mag.
We find that the NoHe models exhibit a reduced overall color spread; however, the variation remains largest in U-B and B-V. The investigation of line-of-sight colour evolution again reveals a notable difference between 1D and 3D simulations: we find that at early times in U-B and B-V, and at later times in R-I and r-i, the 1D simulations of the full models cannot adequately capture the average line-of-sight colour evolution, instead appearing closer to the extreme lines-of-sight. Similarly to the line-of-sight light curves, the 1D NoHe models produce an overall colour evolution closer to matching the full 3D simulations. This difference highlights that 1D radiative transfer calculations of shell detonations do not accurately represent the typical colours from 3D models. Our calculations predict that the 1D colour evolution is only consistent with the most extreme lines-of-sight in 3D.

\subsubsection{Width-luminosity relation}
\label{sec: Width-luminosity relation}

\begin{figure}
    \centering
        \includegraphics[width=1\linewidth]{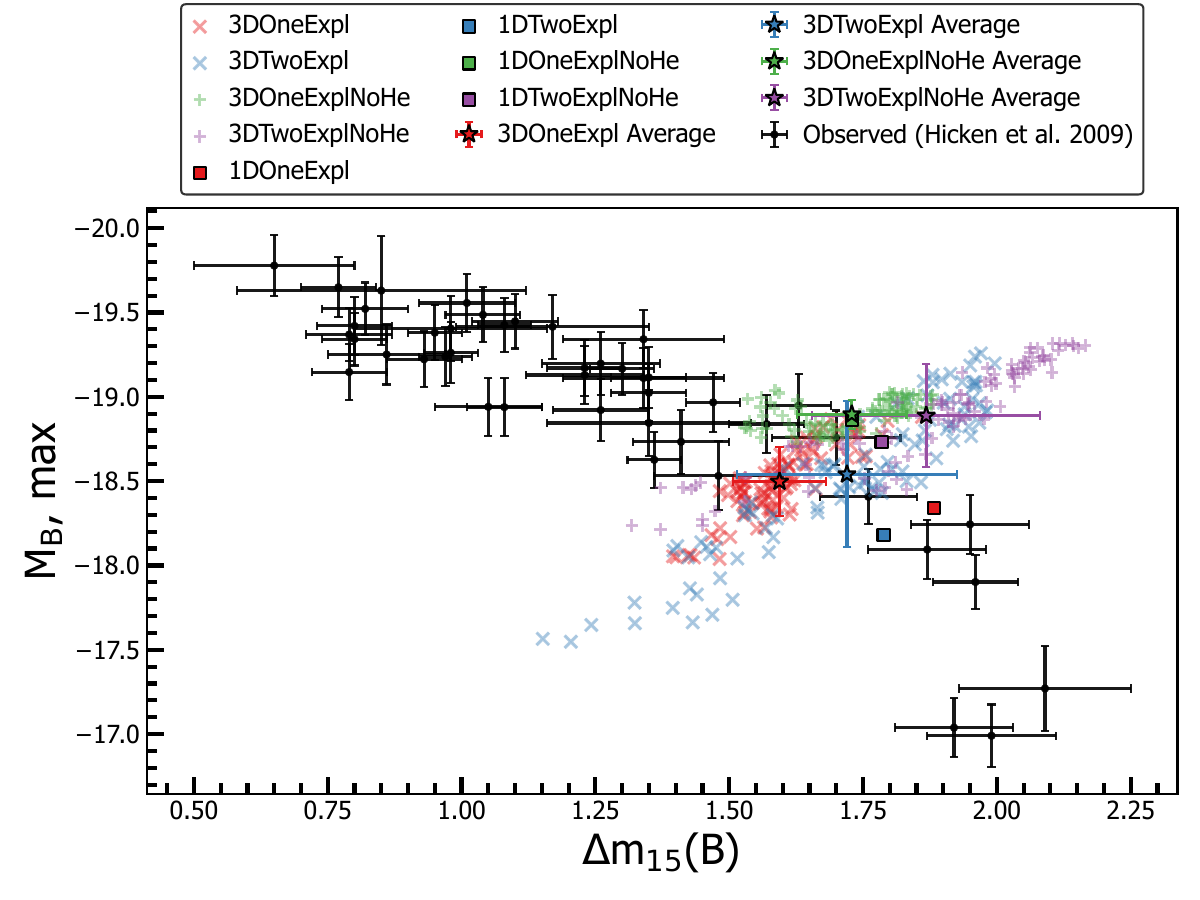}
        \caption{Peak B-band magnitude plotted against \dmb. For each 3D model, the angle-averaged values are represented by $\star$. The 100 different viewing-angles are represented by $\times$ for the full 3D models and $+$ for the 3D NoHe models; the corresponding error bars represent the standard deviation of the 100 viewing-angles. We also compare the 3D simulations to the 1D simulations ($\square$). These simulations have been compared to an observational sample taken from \protect\cite{Hickensample}. Supernovae with a distance modulus with $\mu < 33$ have been excluded.}
    \label{Fig:delta_m15_B}
\end{figure}

\begin{figure}
    \centering
        \includegraphics[width=1\linewidth]{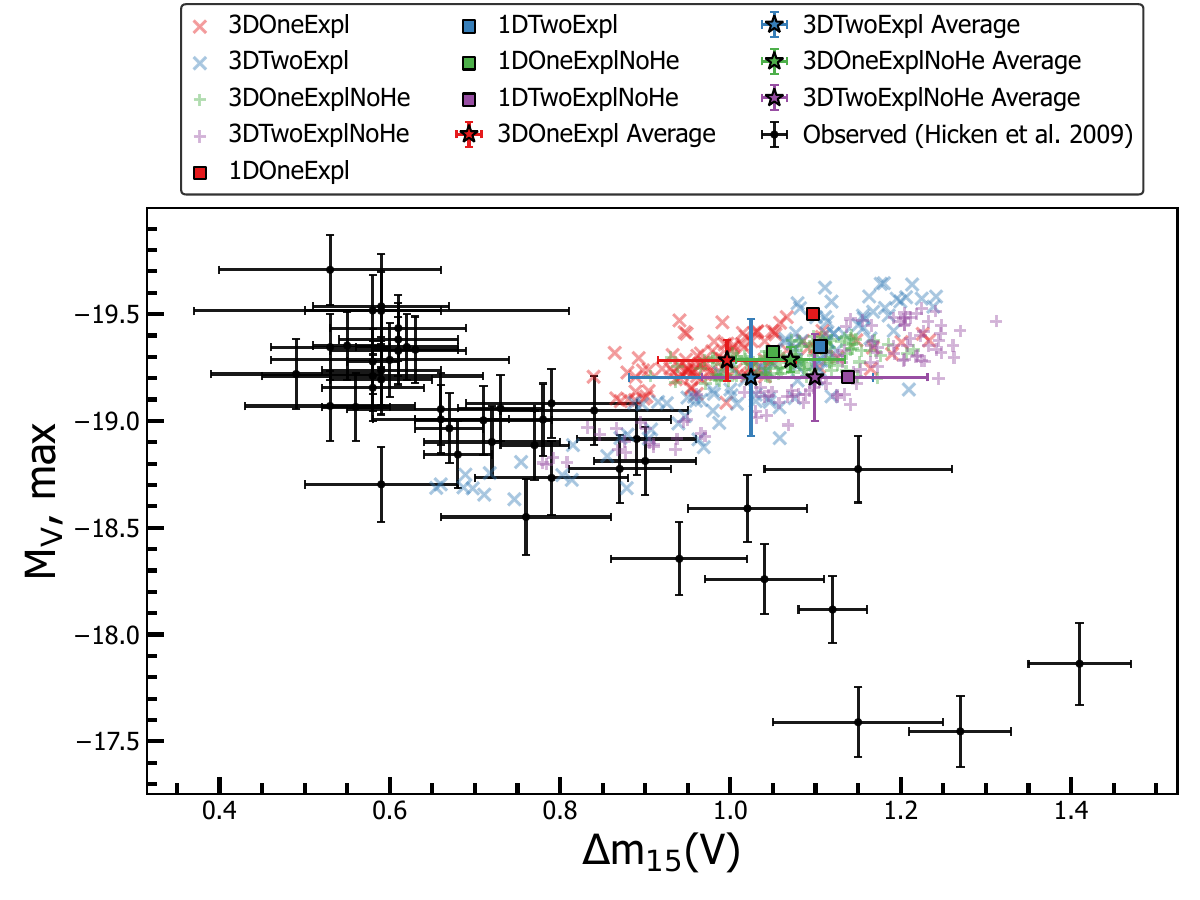}
        \caption{Same as Figure \ref{Fig:delta_m15_B} but for Peak V-band magnitude and $\Delta M_{15}(V)$.}
    \label{Fig:delta_m15_V}
\end{figure}

In this Section, we investigate the width-luminosity relationship produced by the models in both 1D and 3D. We compare the synthetic observables to a homogeneous sample of nearby SNe~Ia \citep{Hickensample} for both \dmb~and $\Delta m_{15}(V)$ which can be seen in Figure~\ref{Fig:delta_m15_B} and \ref{Fig:delta_m15_V} for the B and V bands, respectively.

For the models presented here, we find that the fate of the secondary has a significant impact on the line-of-sight scatter in the width-luminosity relationship. We also find that the helium shell detonation ash produces a notable effect in the overall shape of the scatter which is strongest in \dmb. It can also be seen that the 1D simulations of the full models produce substantially faster decline rates than the 3D lines-of-sight at a comparable luminosity. In contrast, the 1D NoHe models produce decline rates which are consistent with their full 3D calculation. We also find that the full 1D models do not produce a \dmv\xspace that represents a typical line-of-sight from the full 3D models; however, the decline rate of the 1D models is closer to matching that of the angle-averaged 3D models.

Recently, the effect of ignition mechanisms on \dmm~has been investigated; \cite{collins_double_detonation} found that double-detonation models could not reproduce the entire distribution of the width-luminosity relationship. We find that the models investigated in this work also cannot reproduce the full scope of the width-luminosity relationship and have a similar anti-correlated distribution to \cite{collins_double_detonation}.

\subsubsection{B-V at B-band maximum}
\label{sec: B-V at B band Maximum}

\begin{figure}
    \centering
        \includegraphics[width=1\linewidth,height=7cm]{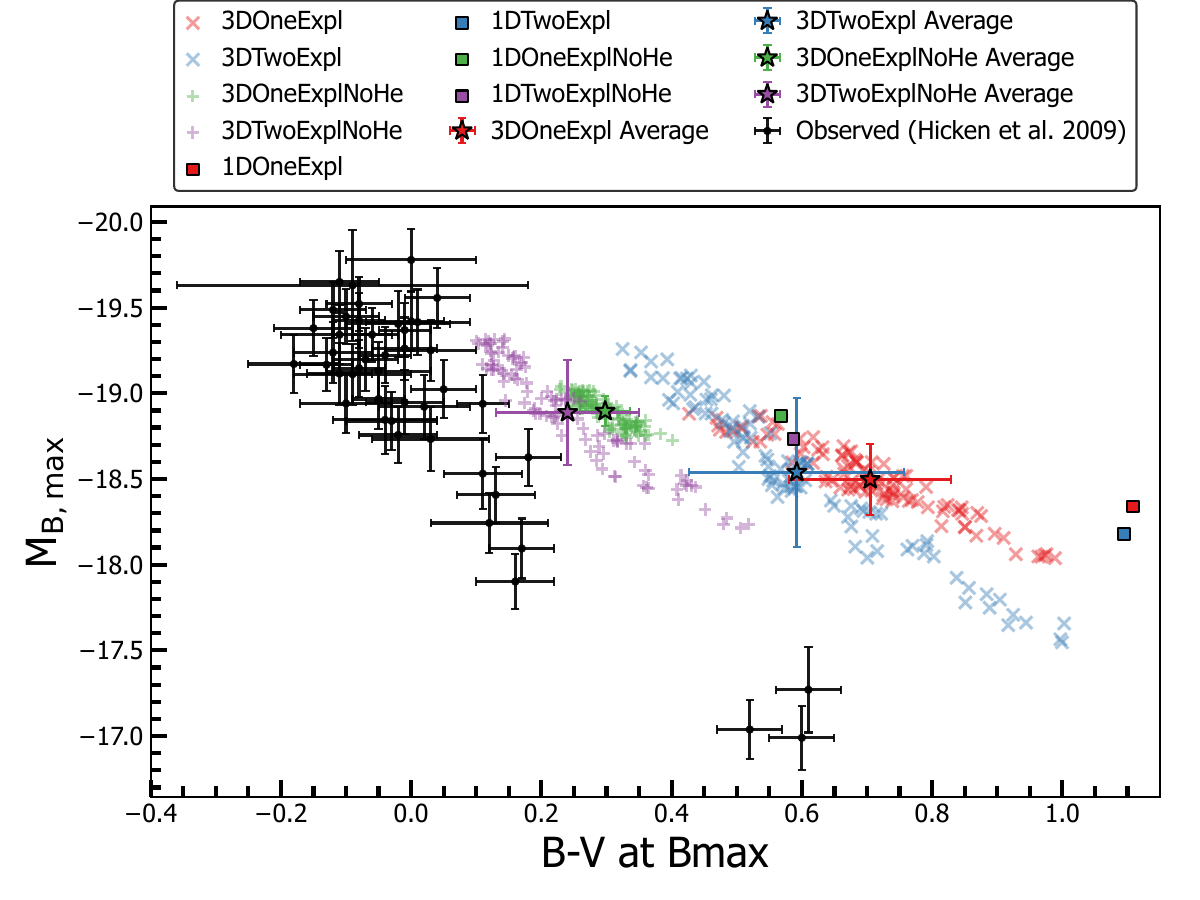}
        \caption{Same as Figure \ref{Fig:delta_m15_B} but for peak B-band magnitude and B-V at B max.}
    \label{Fig: B-V at B max}
\end{figure}

Figure \ref{Fig: B-V at B max} shows the B-V colour at B-band maximum for our simulations. Again, we compare the synthetic observables to the homogeneous sample of nearby SNe~Ia from \cite{Hickensample}. We find that all models have a distribution that is too red to match the observational sample; however the models possess a spread in synthetic observables which is comparable to the spread in the observational sample.  We note that the models encounter difficulty reproducing the bluest portion of the observational sample, which is similar to previous investigations of other models which use a double-detonation mechanism \citep{collins_double_detonation, kromer2010,Boos2024}. Figure \ref{Fig: B-V at B max} also highlights that the colour evolution from 1D calculations is significantly redder than the 3D calculations. Notably, the 1D calculations produce a B-V colour at maximum which cannot match any line-of-sight in the corresponding full 3D calculation.

\subsubsection{Rise time}
\label{sec: Rise Time}

\begin{figure}
    \centering
        \includegraphics[width=1\linewidth]{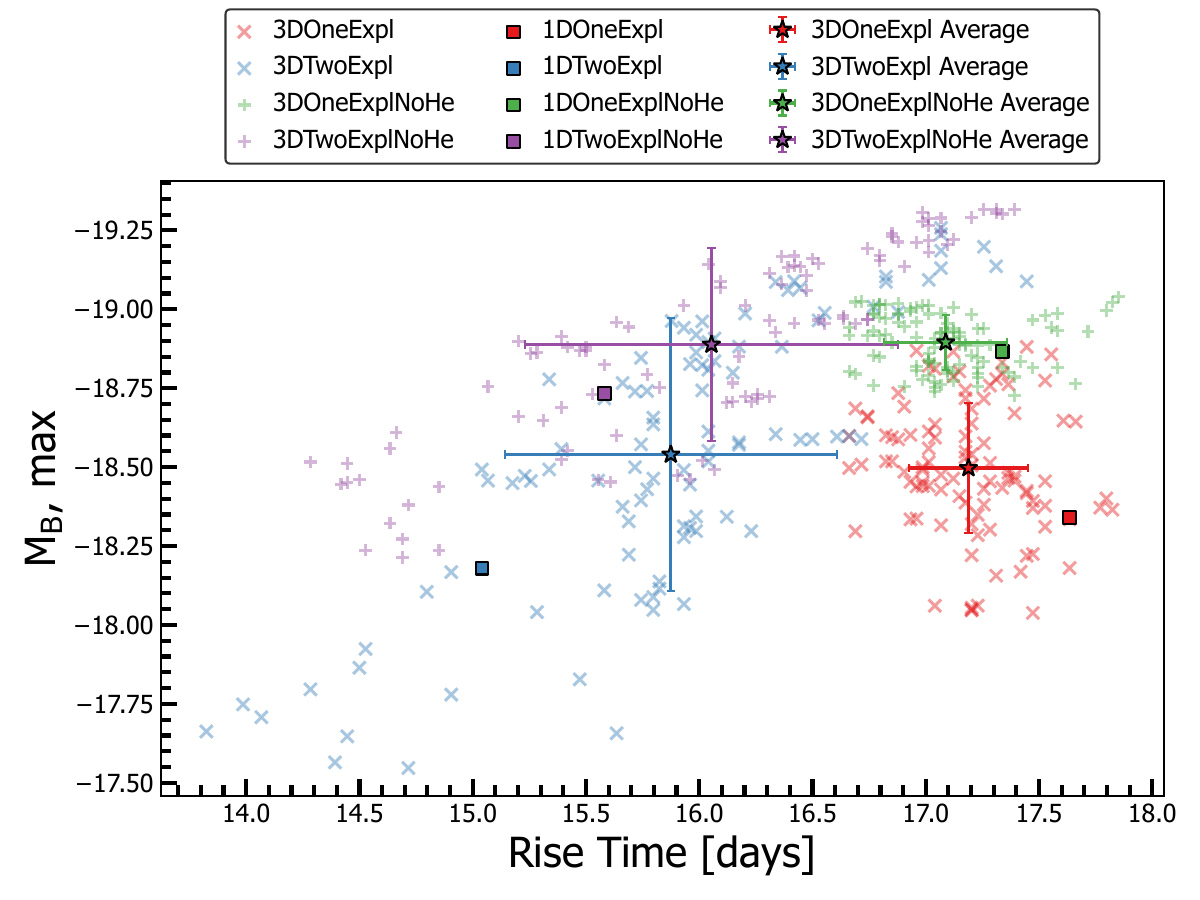}
        \caption{Peak B-band magnitude plotted against the rise time for the 100 different viewing-angles and their standard deviation ($\star$); the full 3D models are represented by $\times$ and the 3D NoHe models are represented by $+$. We also compare the 3D simulations to the 1D simulations ($\square$).}
    \label{Fig: Rise Time}
\end{figure}

Figure \ref{Fig: Rise Time} shows the correlation between peak B-band brightness and the rise time for the 1D and 3D simulations. We find that the angle-averaged rise times for the NoHe models fall within each other's viewing-angle spread (within one standard deviation). Furthermore we find that the NoHe models possess a similar degree of spread to their full model equivalent. Contrasting the 3D modelling to the 1D modelling, we find that the 1D full models produce rise times outside of their 3D equivalent. However, we find that the 1D NoHe models are within that of their 3D equivalent. 

\cite{Firth_2015_rise_time} determined from an observational sample of SNe~Ia that the rise time ranges from 15.98 to 24.7 days; we find that when the secondary survives, all lines-of-sight produce rise times within observational constraints. However, when the secondary also detonates, we obtain a significant spread in rise times for which many lines-of-sight produce rise times which are faster than observational constraints. These faster rise times can be attributed to the compressed distribution of the radioactive $^{56}$Ni produced when the secondary detonates. The spread in the rise time represents one of the largest differences between the \one and \two model however the rise time is particularly sensitive to multidimensional effects and the detailed micro-physics. Band rise times, decline rates and peak magnitudes for each model can be found in Table \ref{Table: Key light curve data}.

\begin{table*}
    \begin{tabular}{|l|l|l|l|l|l|l|l|l|}
    \hline
        ~ &3DOneExpl&3DOneExpl&3DTwoExpl&3DTwoExpl &1DOneExpl&1DOneExpl&1DTwoExpl&1DTwoExpl\\
        ~ &&NoHe&&NoHe &&NoHe&&NoHe\\\hline
        \(M_{bol}\) & $-18.65^{-18.91}_{-18.41}$ & $-18.73^{-18.93}_{-18.58}$ & $-18.74^{-19.17}_{-18.26}$ & $-18.81^{-19.13}_{-18.37}$ & -18.61 & -18.69 & -18.66 & -18.75 \\[0.2cm]
        \(M_{U}\) & $-18.47^{-19.18}_{-17.76}$ & $-18.97^{-19.49}_{-18.63}$ & $-18.61^{-19.49}_{-17.20}$ & $-19.12^{-19.69}_{-18.08}$ & -17.48 & -18.61 & -17.38 & -18.67 \\ [0.2cm]
        \(M_{B}\) & $-18.50^{-18.88}_{-18.04}$ & $-18.90^{-19.04}_{-18.73}$ & $-18.54^{-19.26}_{-17.55}$ & $-18.89^{-19.32}_{-18.21}$ & -18.34 & -18.87 & -18.18 & -18.73 \\ [0.2cm]
        \(M_{V}\) & $-19.28^{-19.49}_{-19.09}$ & $-19.29^{-19.40}_{-19.17}$ & $-19.20^{-19.64}_{-18.63}$ & $-19.20^{-19.53}_{-18.80}$ & -19.50 & -19.32 & -19.35 & -19.21 \\ [0.2cm]
        \(M_{R}\) & $-19.33^{-19.46}_{-19.21}$ & $-19.19^{-19.26}_{-19.12}$ & $-19.37^{-19.65}_{-19.05}$ & $-19.28^{-19.48}_{-19.02}$ & -19.39 & -19.27 & -19.53 & -19.41 \\ \hline
        \(R_{bol}\) & $\phantom{-}17.78^{18.99}_{16.91}$ & $\phantom{-}17.18^{17.91}_{16.45}$ & $\phantom{-}16.84^{19.64}_{15.64}$ & $\phantom{-}16.32^{17.47}_{15.07}$ & \phantom{-}18.39 & \phantom{-}17.82 & \phantom{-}17.47 & \phantom{-}17.18 \\ [0.2cm]
        \(R_{U}\) & $\phantom{-}15.67^{16.61}_{14.93}$ & $\phantom{-}15.22^{15.77}_{14.58}$ & $\phantom{-}14.27^{15.42}_{12.36}$ & $\phantom{-}13.83^{14.82}_{11.99}$ & \phantom{-}17.01 & \phantom{-}15.45 & \phantom{-}14.36 & \phantom{-}13.04 \\ [0.2cm]
        \(R_{B}\) & $\phantom{-}17.19^{17.82}_{16.66}$ & $\phantom{-}17.09^{17.85}_{16.66}$ & $\phantom{-}15.87^{17.45}_{13.82}$ & $\phantom{-}16.05^{17.39}_{14.28}$ & \phantom{-}17.64 & \phantom{-}17.34 & \phantom{-}15.04 & \phantom{-}15.58 \\ [0.2cm]
        \(R_{V}\) & $\phantom{-}20.21^{21.36}_{18.91}$ & $\phantom{-}20.36^{20.91}_{19.53}$ & $\phantom{-}18.65^{20.26}_{16.42}$ & $\phantom{-}18.99^{20.26}_{16.74}$ & \phantom{-}19.82 & \phantom{-}20.34 & \phantom{-}18.34 & \phantom{-}19.53 \\ [0.2cm]
        \(R_{R}\) & $\phantom{-}19.68^{21.23}_{18.42}$ & $\phantom{-}19.22^{20.09}_{18.34}$ & $\phantom{-}19.17^{21.61}_{17.01}$ & $\phantom{-}18.87^{19.82}_{17.18}$ & \phantom{-}19.61 & \phantom{-}19.31 & \phantom{-}20.85 & \phantom{-}19.91 \\ \hline
        $\Delta M_{15}(B)$ & $\phantom{-}1.59^{1.86}_{1.40}$ & $\phantom{-}1.73^{1.88}_{1.53}$ & $\phantom{-}1.72^{2.00}_{1.15}$ & $\phantom{-}1.87^{2.17}_{1.32}$ & \phantom{-}1.88 & \phantom{-}1.73 & \phantom{-}1.79 & \phantom{-}1.78 \\ [0.2cm]
        $\Delta M_{15}(V)$ & $\phantom{-}1.00^{1.23}_{0.84}$ & $\phantom{-}1.07^{1.21}_{0.91}$ & $\phantom{-}1.02^{1.24}_{0.65}$ & $\phantom{-}1.10^{1.31}_{0.78}$ & \phantom{-}1.10 & \phantom{-}1.05 & \phantom{-}1.11 & \phantom{-}1.14 \\ [0.2cm]
        B-V at Bmax & $\phantom{-}0.70^{0.99}_{0.43}$ & $\phantom{-}0.30^{0.40}_{0.23}$ & $\phantom{-}0.59^{1.00}_{0.32}$ & $\phantom{-}0.24^{0.52}_{0.10}$ & \phantom{-}1.11 & \phantom{-}0.57 & \phantom{-}1.09 & \phantom{-}0.59 \\ \hline
    \end{tabular}
    \caption{Model parameters are shown for the \one, \oneno, \two, \twono, \oneO, \onenoO, \twoO and \twonoO models (see \protect\citealt{pakmor_2021} for detailed nucleosynthesis). We show the average absolute magnitude and rise time (days) for several bands. We also show the average decline rate for both the B and V bands as well as B-V at B-band maximum. For the 3D models we also display the maximum and minimum values of each parameter.}
    \label{Table: Key light curve data}
\end{table*}

\subsection{Spectra}
\label{sec:Spectra}

In Section~\ref{sec:Angle-averaged spectra}, we present the angle-averaged spectra from our 3D models and those from our 1D models. Following this, we show the line-of-sight variation for our 3D models in Section \ref{sec:Line of sight dependent spectra}.

\subsubsection{Angle-averaged spectra}
\label{sec:Angle-averaged spectra}

\begin{figure}
\centering
\begin{subfigure}[b]{\textwidth}
   \includegraphics[scale=0.4,width=0.49\linewidth,trim={0cm 1.6cm 0cm 0cm},clip]{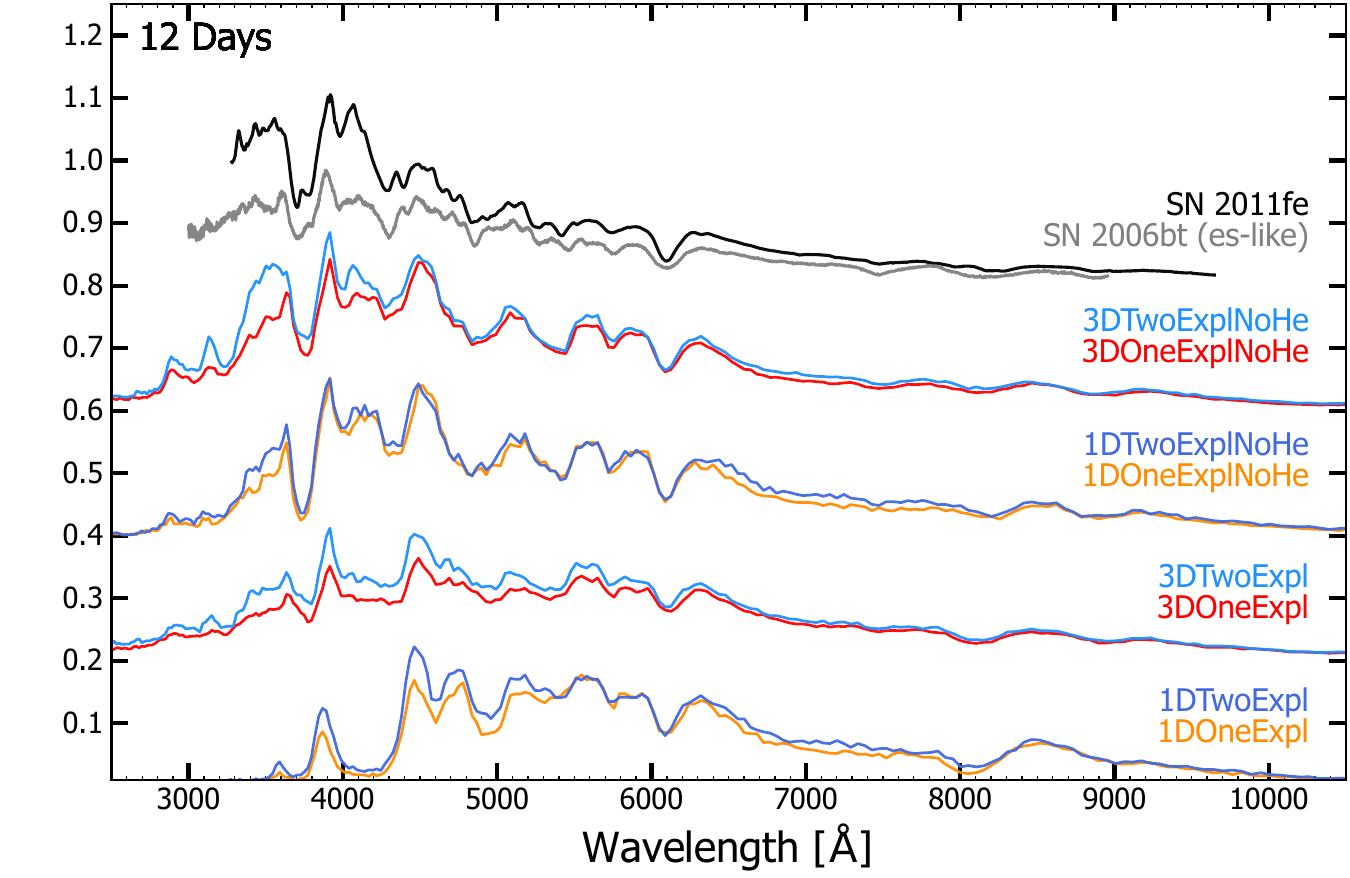 }
   \label{fig:Angle-averaged spectra 12.5 days} 
\end{subfigure}
\begin{subfigure}[b]{\textwidth}
   \includegraphics[scale=0.4,width=0.49\linewidth,trim={0cm 1.6cm 0cm 0cm},clip]{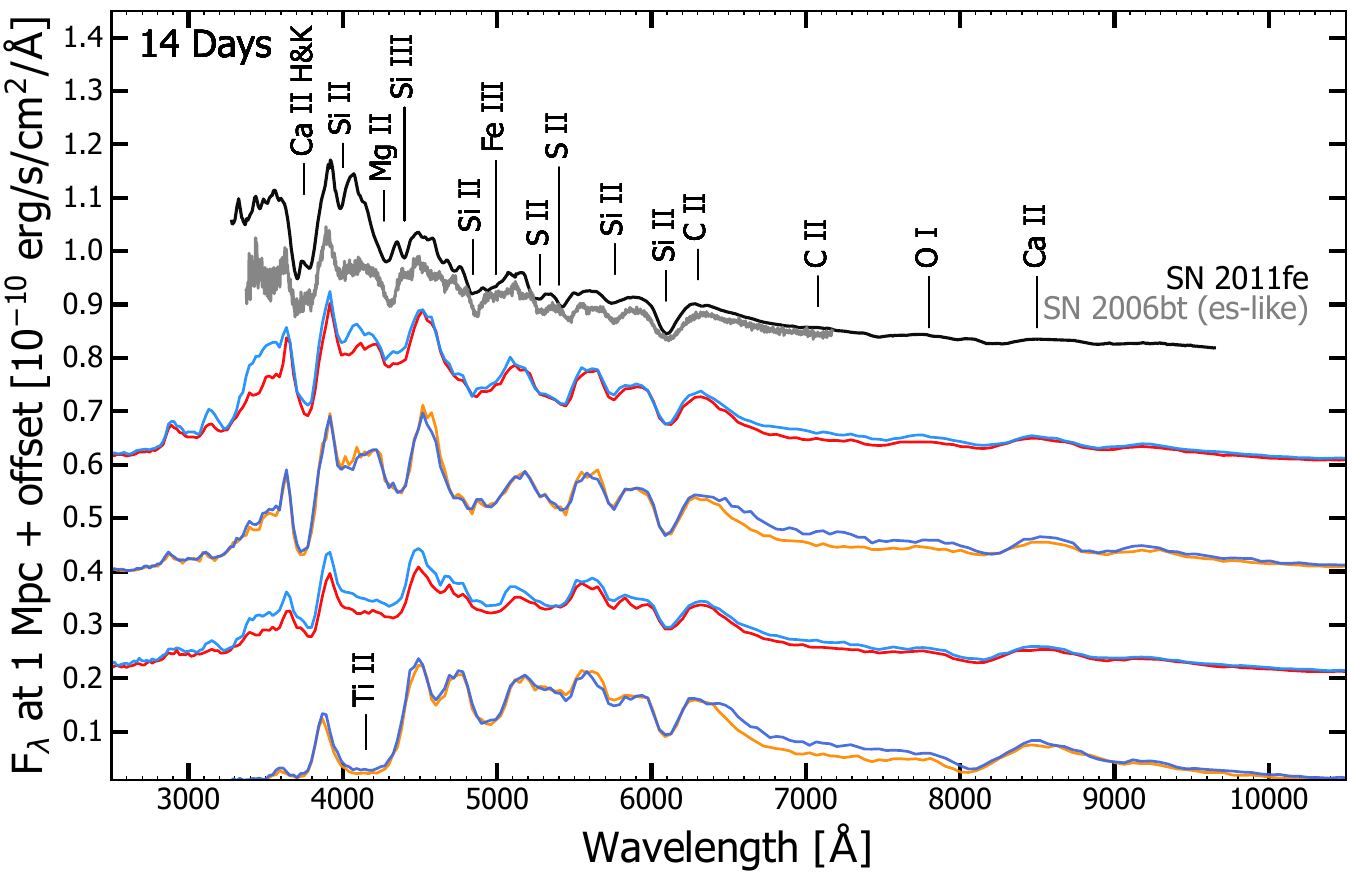}
   \label{fig:Angle-averaged spectra 14.5 days} 
\end{subfigure}
\begin{subfigure}[b]{\textwidth}
   \includegraphics[scale=0.4,width=0.49\linewidth,trim={0cm 0cm 0cm 0cm},clip]{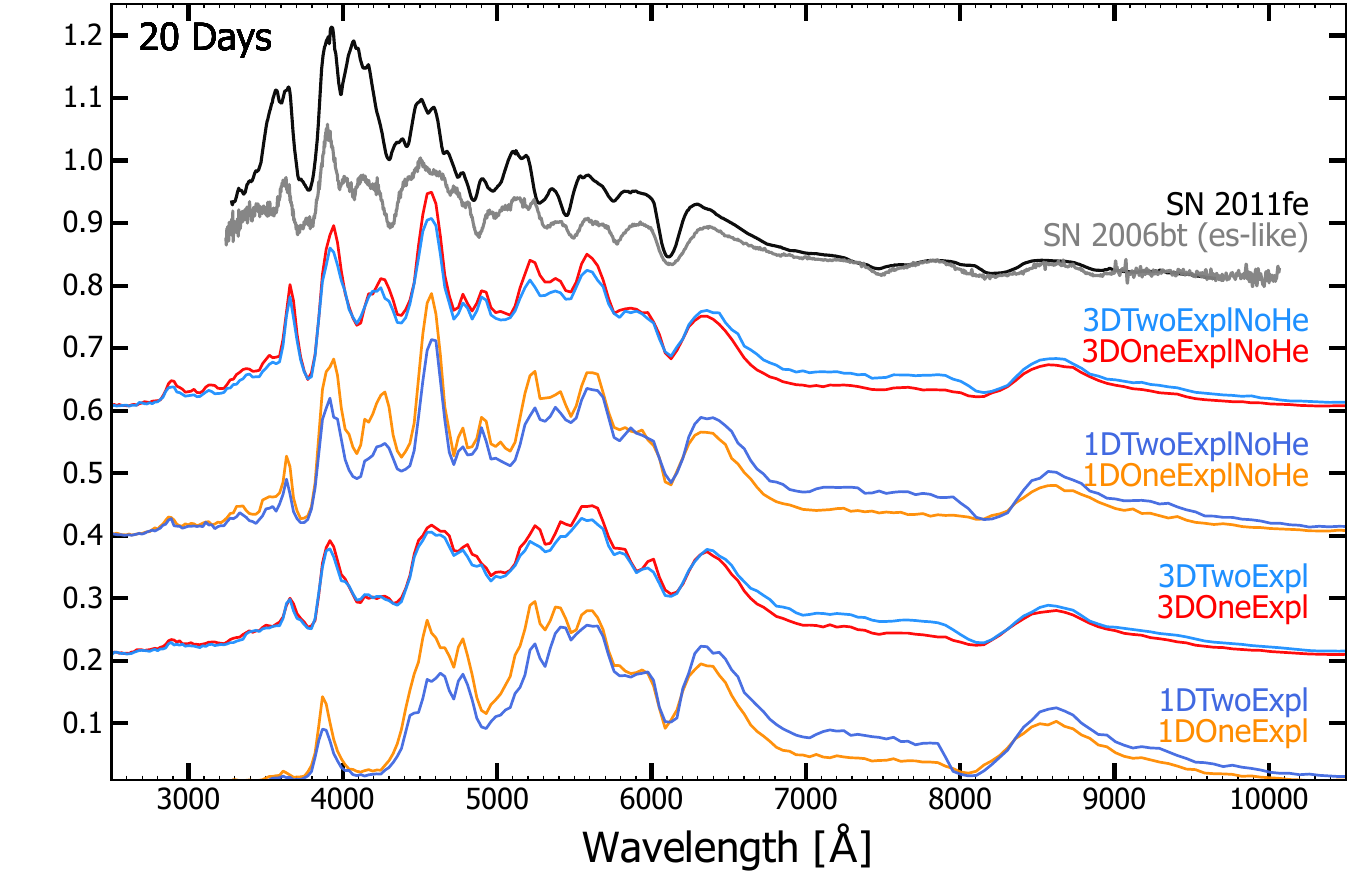}
   \label{fig:Angle-averaged spectra 20 days} 
\end{subfigure}
\caption{3D angle-averaged and 1D spectra for each model shown at 12, 14 and 20 days after explosion compared to \fe and \bt at similar epochs. Important ions which are typically responsible for emission and absorption in SNe~Ia around maximum light are marked on the spectra of \fe. Each set of spectra are offset from one another by the same arbitrary constant. We note the spectrum of SN 2006bt at 14 days is plotted in scaled flux.}
\label{fig: angle averaged spectra}
\end{figure}

\begin{figure*}
    \centering
    
    \begin{subfigure}[b]{0.49\textwidth}
        \includegraphics[height=5.3cm,width=\linewidth,trim={0cm 0cm 0.0cm 0.05cm},clip]{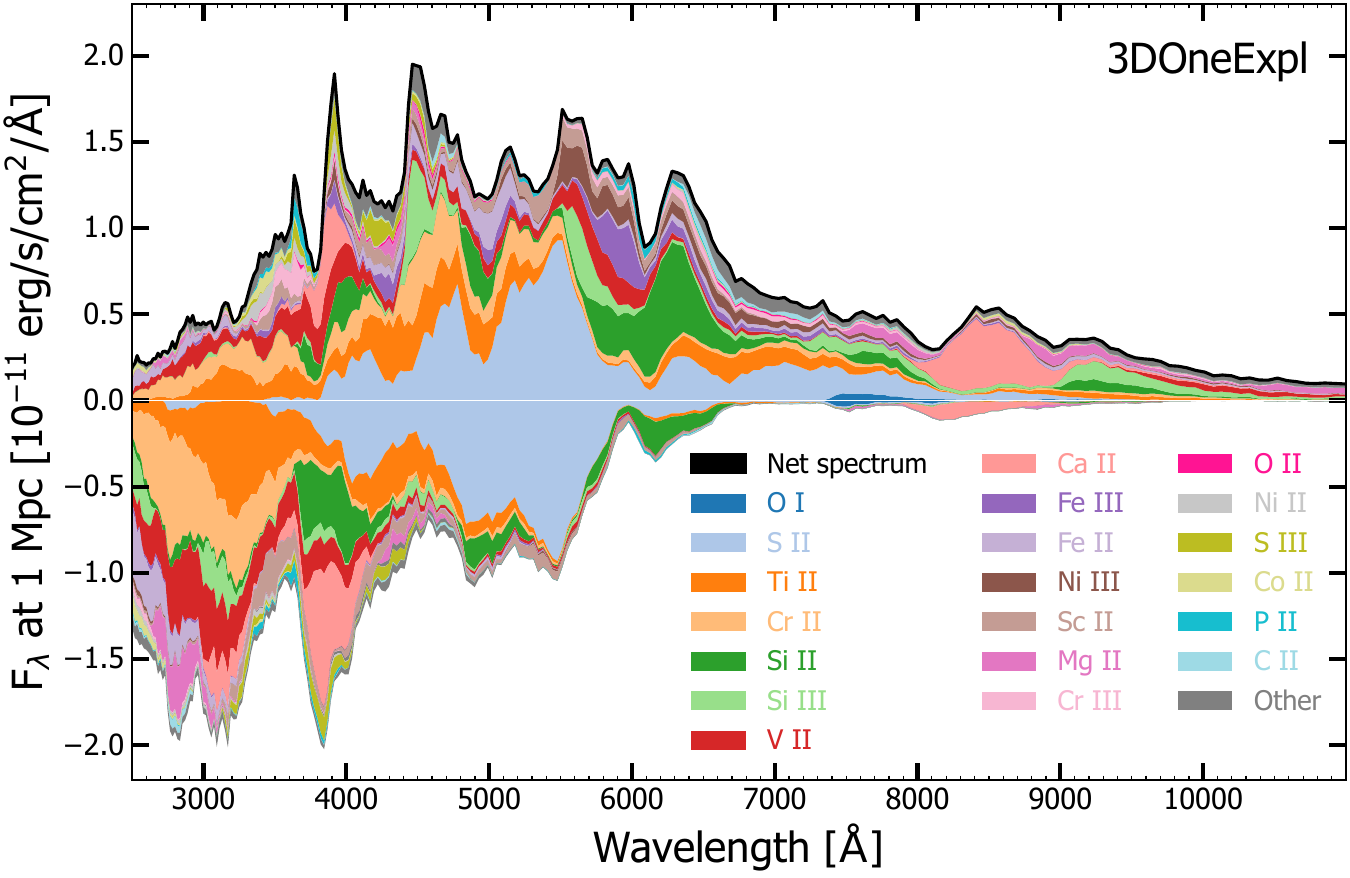}
        \label{fig: 3D One Explosion spectra 2 weeks}
    \end{subfigure}\hfill
    \begin{subfigure}[b]{0.49\textwidth}
        \includegraphics[height=5.3cm,width=\linewidth,trim={0cm 0cm 0.0cm 0.05cm},clip]{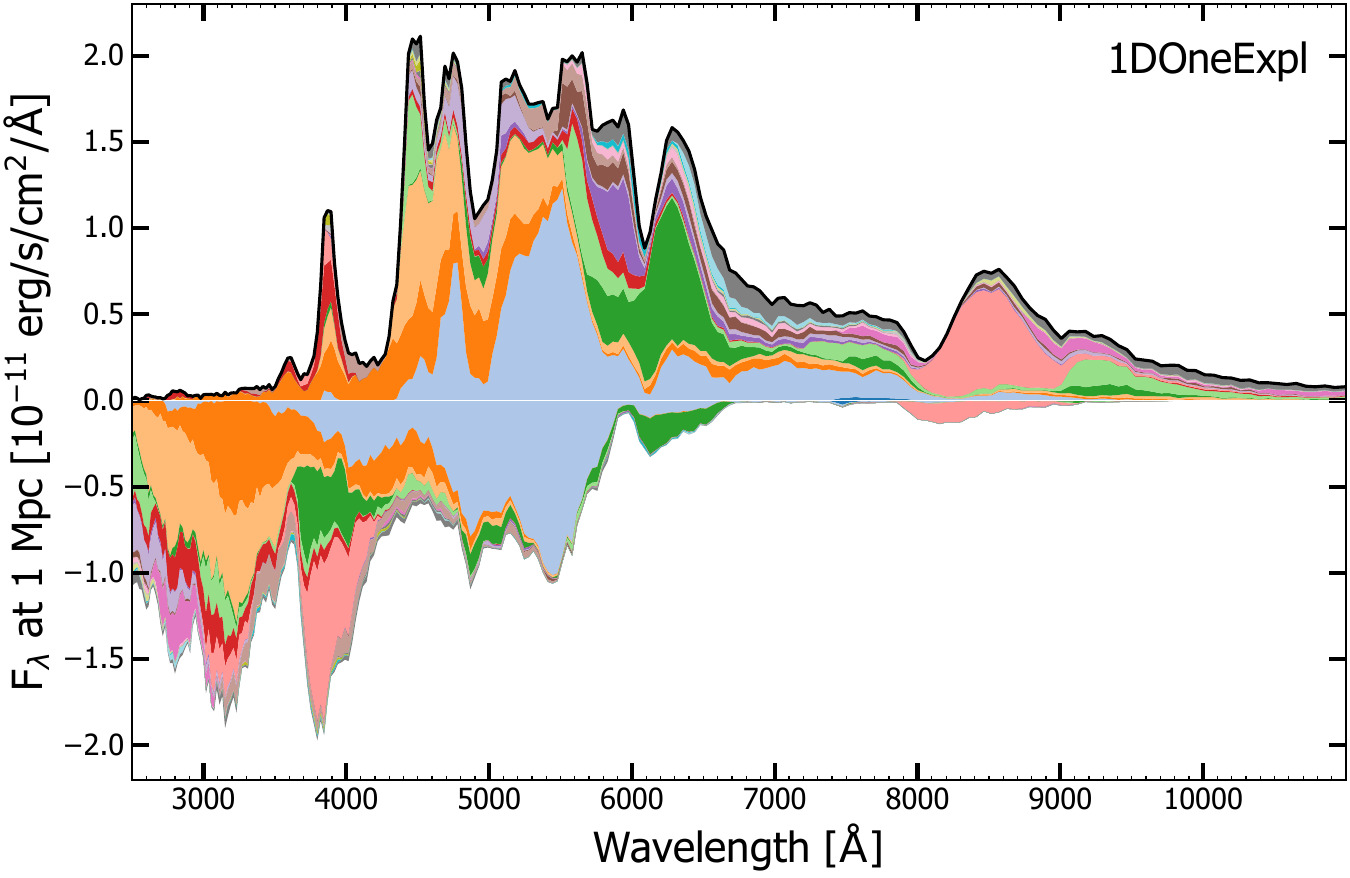}
        \label{fig: 1D One Explosion spectra 2 weeks}
    \end{subfigure}
    
    \begin{subfigure}[b]{0.49\textwidth}
        \includegraphics[height=5.3cm,width=\linewidth,trim={0cm 0cm 0.0cm 0.05cm},clip]{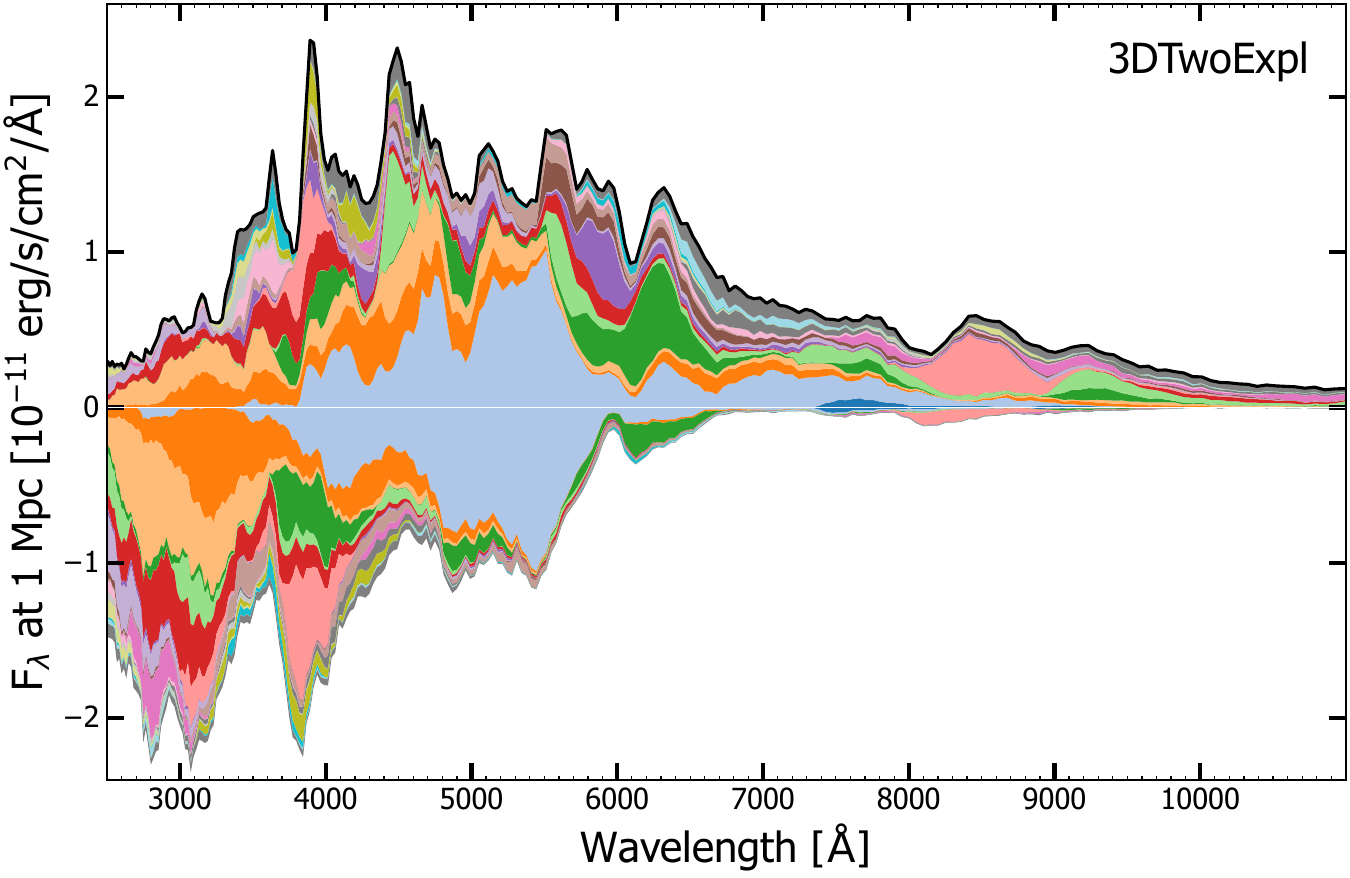}
        \label{fig:Two Explosion spectra  2 weeks}
    \end{subfigure}\hfill
    \begin{subfigure}[b]{0.49\textwidth}
        \includegraphics[height=5.3cm,width=\linewidth,trim={0cm 0cm 0.0cm 0.05cm},clip]{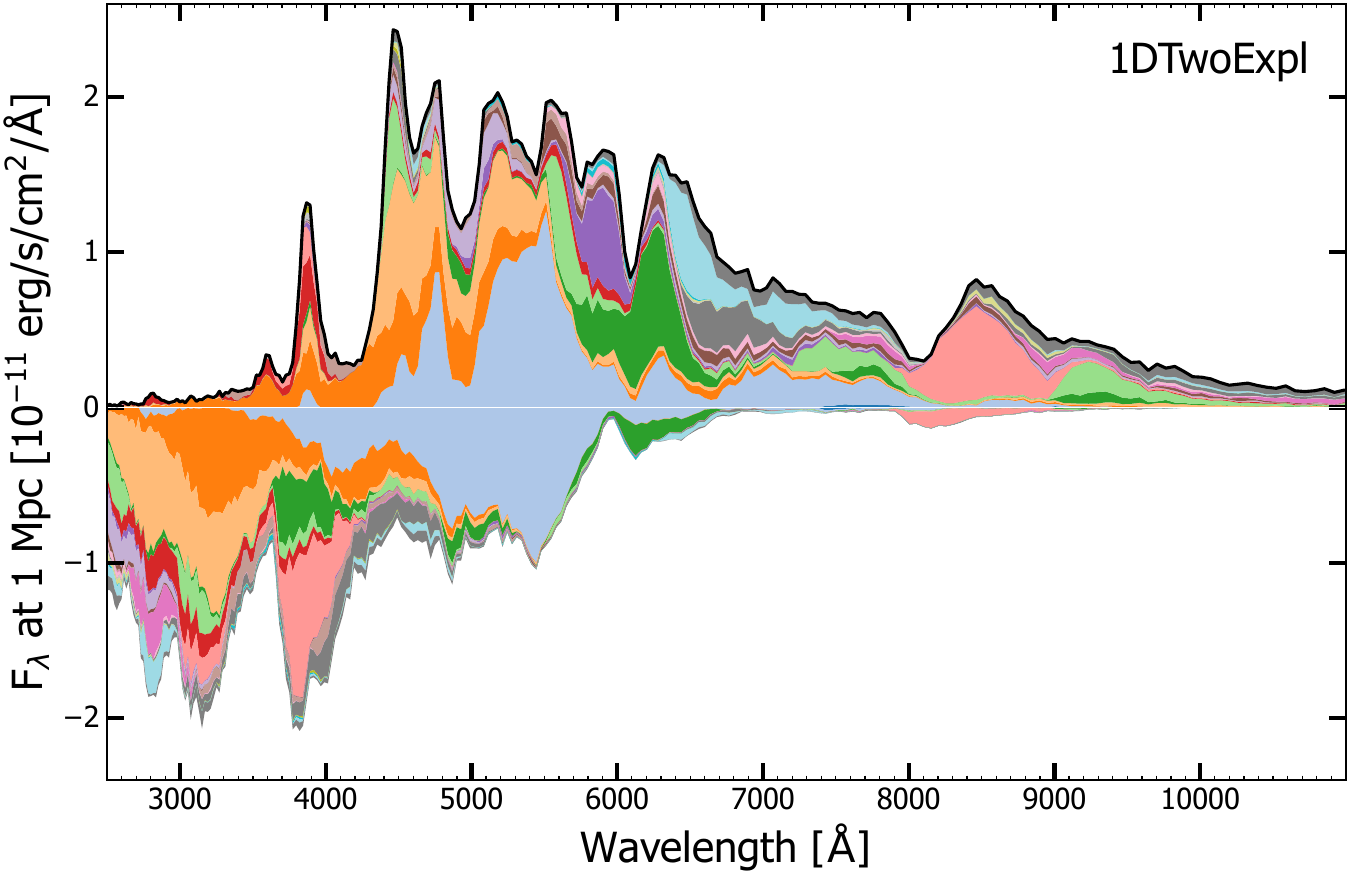}
        \label{fig: 1D Two Explosion spectra 2 weeks}
    \end{subfigure}

        \begin{subfigure}[b]{0.49\textwidth}
        \includegraphics[height=5.3cm,width=\linewidth,trim={0cm 0cm 0.0cm 0.05cm},clip]{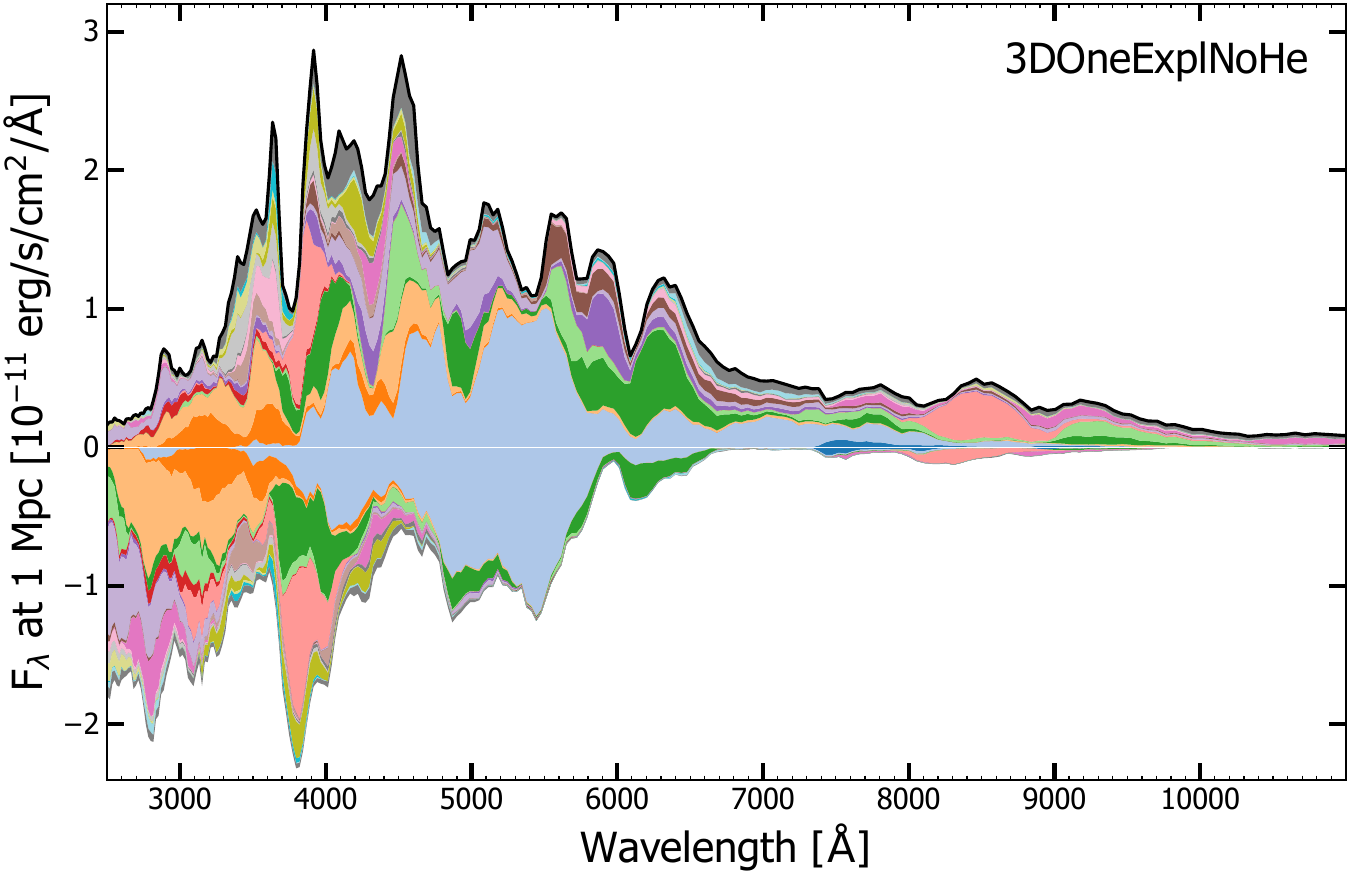}
        \label{fig:One Explosion spectra 2 weeks}
    \end{subfigure}\hfill
    \begin{subfigure}[b]{0.49\textwidth}
        \includegraphics[height=5.3cm,width=\linewidth,trim={0cm 0cm 0.0cm 0.05cm},clip]{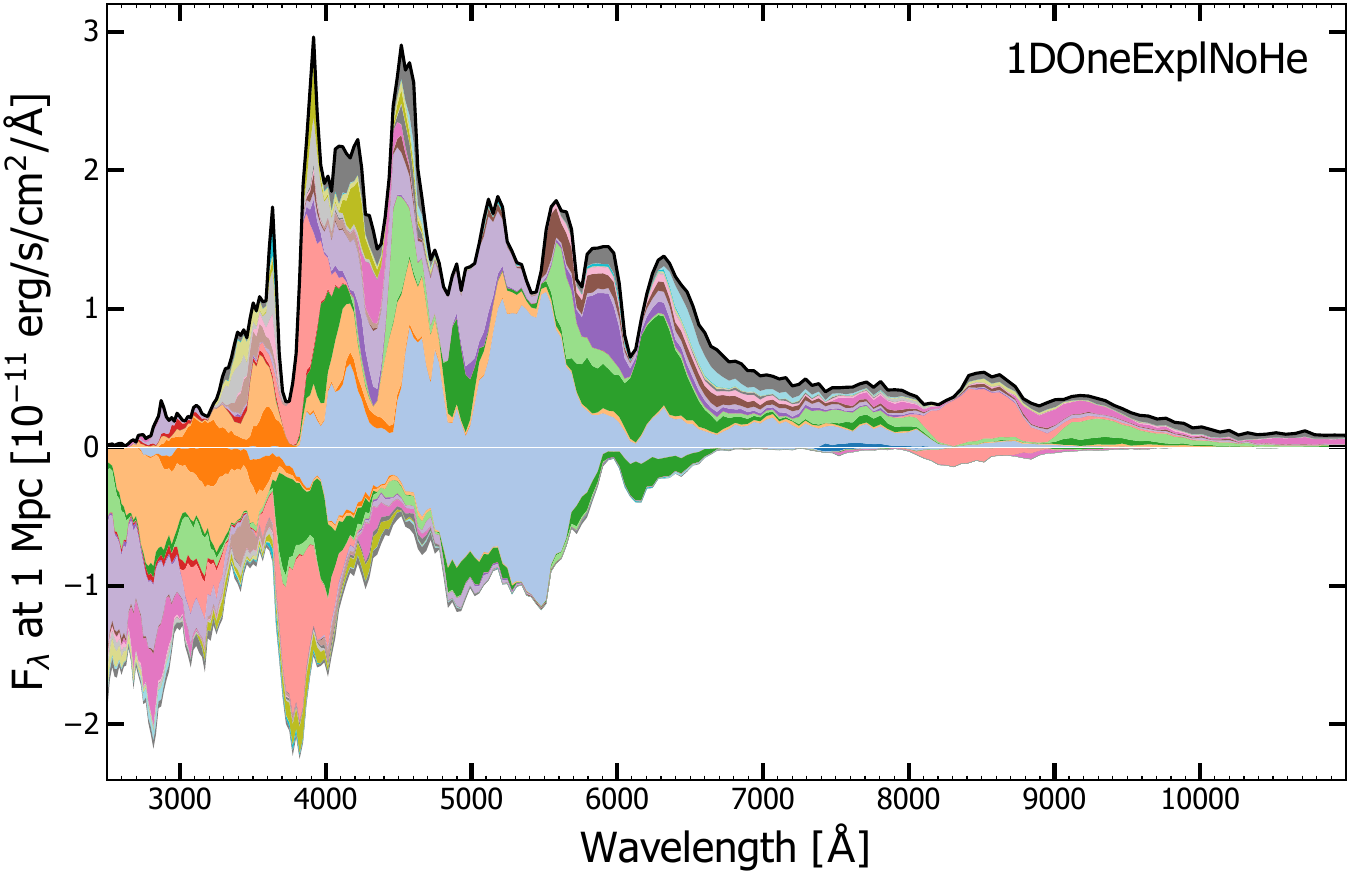}
        \label{fig: 1D One Explosion No He spectra  2 weeks}
    \end{subfigure}

        \begin{subfigure}[b]{0.49\textwidth}
        \includegraphics[height=5.3cm,width=\linewidth,trim={0cm 0cm 0.0cm 0.05cm},clip]{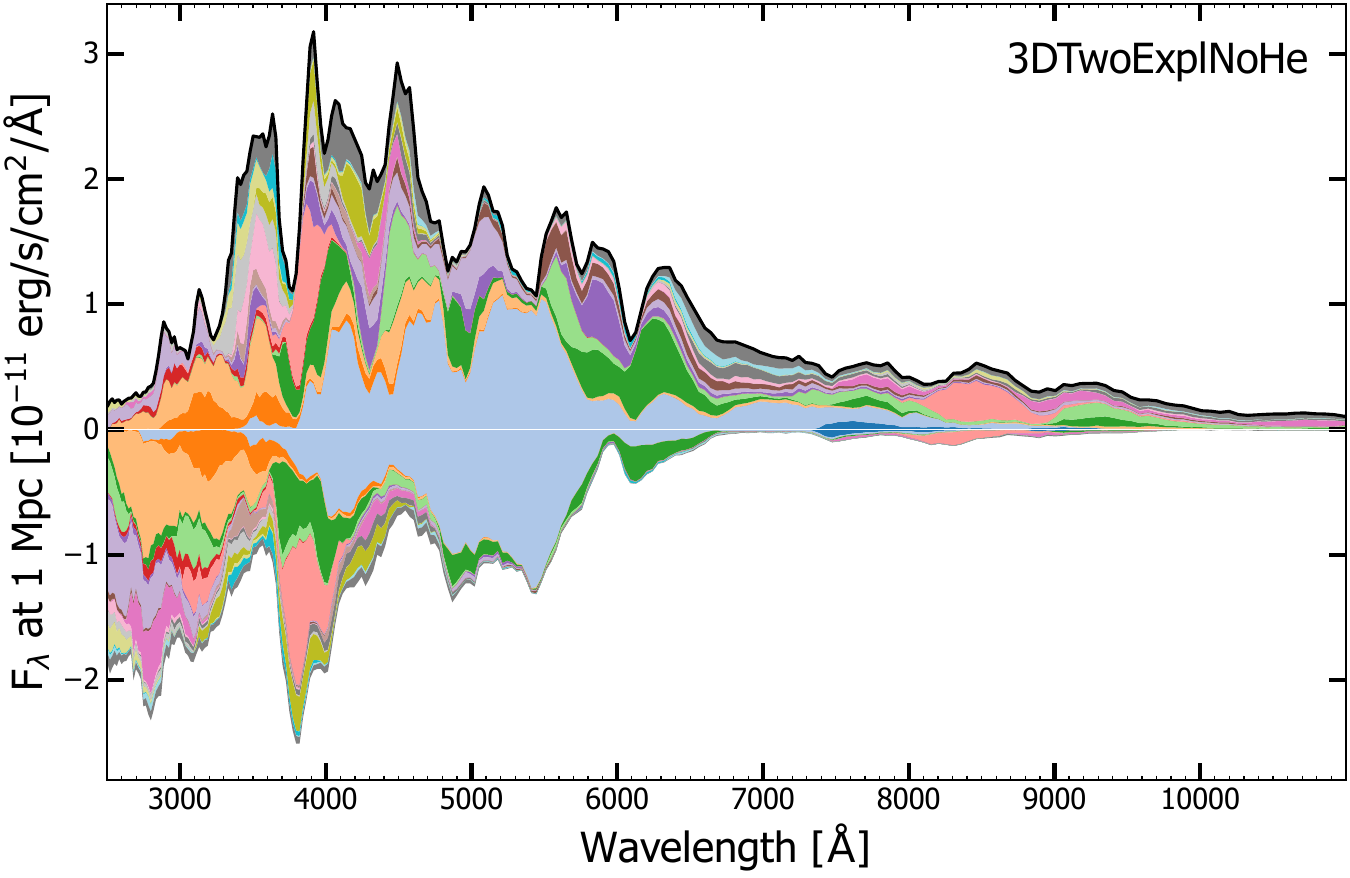}
        \label{fig: 3D Two Explosion spectra 2 weeks}
    \end{subfigure}\hfill
    \begin{subfigure}[b]{0.49\textwidth}
        \includegraphics[height=5.3cm,width=\linewidth,trim={0cm 0cm 0.0cm 0.05cm},clip]{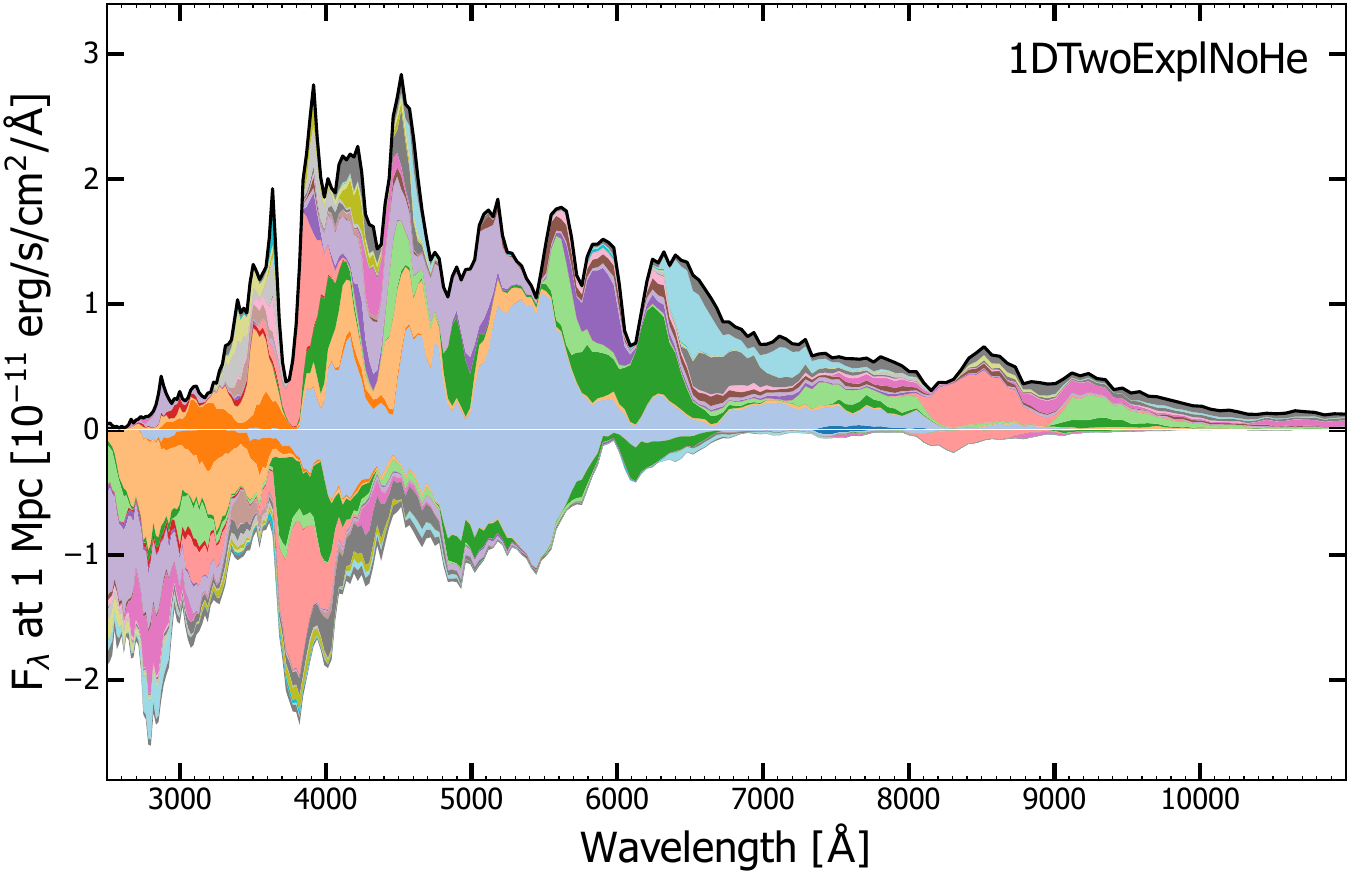}
        \label{fig:1D Two Explosion No He spectra 2 weeks}
    \end{subfigure}
    
    \caption{Spectra for all models at 14 days after explosion for each of the models, where the contributions of individual ions are indicated. Beneath the axis we also show the ions responsible for absorption.}
    \label{fig: angle averaged spectra emission absorption}
\end{figure*}

We compare \one and \two model spectra (1D and 3D, full and NoHE) in Figure~\ref{fig: angle averaged spectra} across three different epochs with each set of spectra being offset by an arbitrary constant. We show each model at 12, 14, and 20 days after explosion and compare to both peculiar and normal SNe~Ia around similar epochs. The key ions that are responsible for shaping the spectrum for each model at 14 days after explosion are indicated in Figure \ref{fig: angle averaged spectra emission absorption}. In this Figure we show equivalent 3D and 1D cases next to each other to facilitate their comparison. We find that the models do produce the expected IME spectral features, including the \ion{Si}{II} 6355\AA$\space$ and \ion{Ca}{II} triplet at 8498\AA, 8542\AA$\space$ and 8662\AA. The 1D simulations of the full models produce a clear \ion{Ti}{II} absorption feature at $\sim$4000\AA-4500\AA. This feature is still present in the full 3D simulations, however, it is significantly diminished. This \ion{Ti}{II} feature is characteristic of double-detonation models and agrees with the previous findings \citep{kromer2010,Townsley2019,Gronow2020,collins_double_detonation}.

The spectra for the 1D and 3D full models display noticeable differences from one another for both the OneExpl and TwoExpl scenarios (Figure~\ref{fig: angle averaged spectra} and upper panels of Figure \ref{fig: angle averaged spectra emission absorption}). Particularly relevant to this are 
\ion{Cr}{II} and \ion{Ti}{II}, which are present in the helium detonation ash and make a significant contribution to absorption and fluorescence. This is more significant in the 1D models, for which very little flux is able to emerge blueward of $\sim$4500\AA\xspace (see upper right panels of Figure \ref{fig: angle averaged spectra emission absorption}). These ions remain important in the 3D cases, but the blue flux suppression (and also re-emission through fluorescence in the 4500--6000\AA\xspace region) is significantly reduced and, as noted above, the \ion{Ti}{II} absorption trough at $\sim$4000\AA-4500\AA\xspace is weaker. When the He ash is not included (NoHe models, see lower panels of Figure \ref{fig: angle averaged spectra emission absorption}) the differences between 1D and 3D models are less pronounced. However, there are still noticeable spectral differences (for example around the \ion{Ca}{II} H\&K feature), and it is still the case that the 1D scenario leads to more flux suppression in the ultraviolet.
These comparisons highlight that multi-dimensional effects need to be considered, and are particularly relevant when investigating the impact of the helium shell ash in our scenario.

The temporal evolution of the average spectra shown in Figure~\ref{fig: angle averaged spectra} reveals that the impact of the detonation of the secondary is also different in 3D compared to 1D. At 12 days, the secondary detonation is more influential in 3D, underscored by the different fluxes in both the \two and \twono models between 3300\AA\xspace and 4500\AA\xspace when compared to that of the \one and \oneno models respectively. Conversely, in 1D the impact of the secondary detonation only grows over time as the differences at 20 days between the NoHe and full models in 1D are the most substantial.
 
\subsubsection{Line-of-sight spectra}
\label{sec:Line of sight dependent spectra}

\begin{figure*}
    \centering
    
    \begin{subfigure}[b]{0.49\textwidth}
        \includegraphics[width=\linewidth,trim={0cm 0cm 0.0cm 0.05cm},clip]{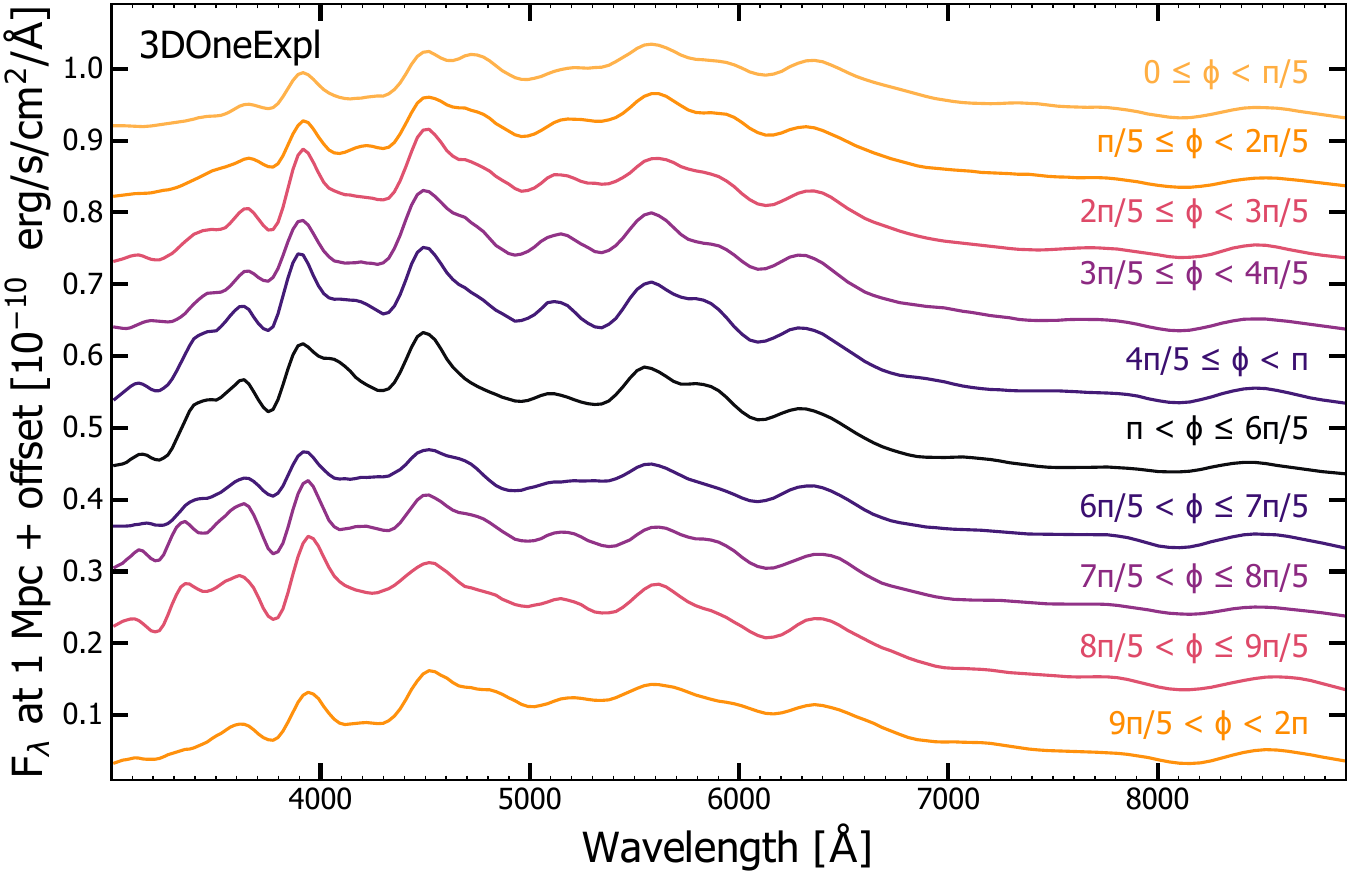}
        \label{fig:subfig1}
    \end{subfigure}\hfill
    \begin{subfigure}[b]{0.49\textwidth}
        \includegraphics[width=\linewidth,trim={0cm 0cm 0.0cm 0.05cm},clip]{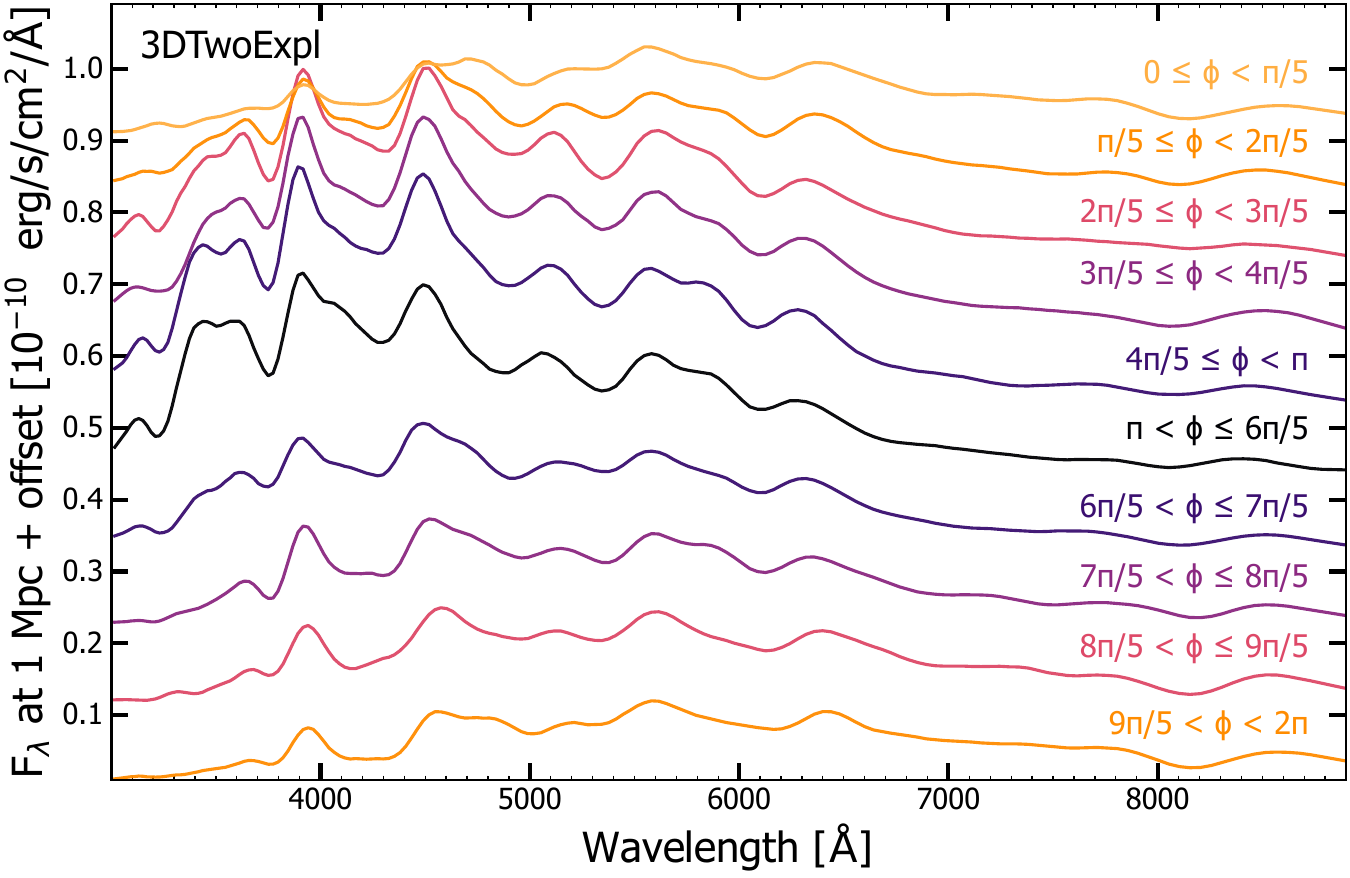}
        \label{fig:subfig2}
    \end{subfigure}
    
    \vspace{0pt} 
    
    \begin{subfigure}[b]{0.49\textwidth}
        \includegraphics[width=\linewidth,trim={0cm 0cm 0.0cm 0.05cm},clip]{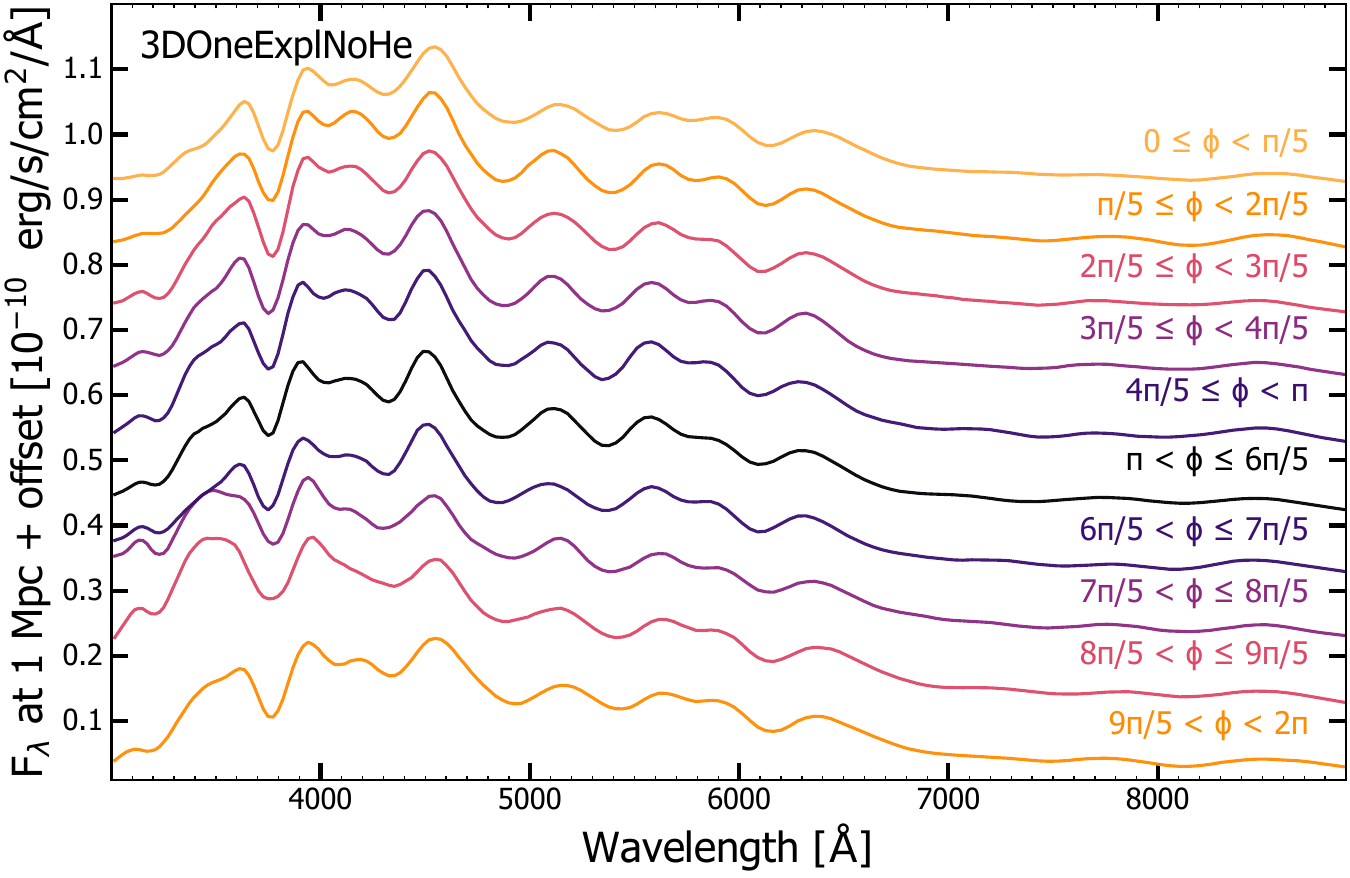}
        \label{fig:subfig3}
    \end{subfigure}\hfill
    \begin{subfigure}[b]{0.49\textwidth}
        \includegraphics[width=\linewidth,trim={0cm 0cm 0.0cm 0.05cm},clip]{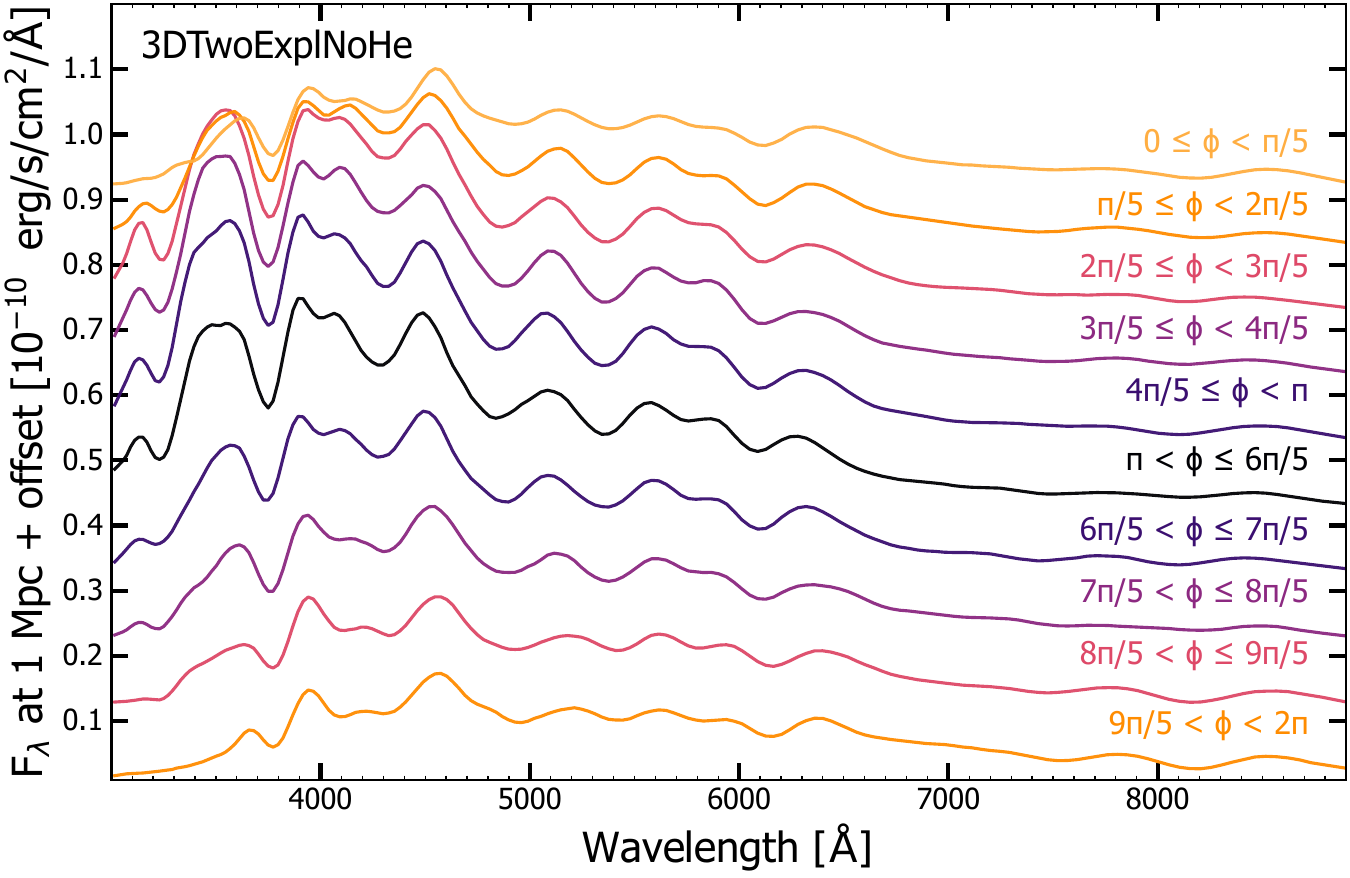}
        \label{fig:subfig4}
    \end{subfigure}
    
    \caption{Spectra of each 3D explosion model for different lines-of-sight at two weeks after explosion. The lines-of-sight shown occur in the $0.0 \le \cos{\theta} < 0.2$ plane. These 10 lines-of-sight represent the most significant spread in the synthetic observables and are offset from one another by the same arbitrary constant.}
    \label{fig:los_spectra}
\end{figure*}

Figure \ref{fig:los_spectra} shows the line-of-sight spectra corresponding to the light curves displayed in Section~\ref{Sec:Line of sight light curves}. For each model, substantial differences in the spectra arise for different observer orientations when compared to the respective model's average spectra (see Figure~\ref{fig: angle averaged spectra emission absorption}). In certain directions, the models exhibit features more comparable to normal SNe~Ia, while in others, they display a closer resemblance to peculiar SNe~Ia. Additionally, the \one and \two models have more viewing-angle variation in their spectra than the \oneno and \twono models. Hence, the increased variation in the full models can be attributed to the helium shell detonation. Across all models, the line-of-sight variation is minor at redder wavelengths ($\gtrapprox$6000\AA) and differences are due to small changes in the flux. In contrast to the 3D and 1D full models, the 3D NoHe simulations produce lines-of-sight that can be well approximated by the 1D NoHe simulations. Again this suggests that the 3D effects are most important when considering the impact of the helium detonation ash -- but we stress that the angle variations are still significant even in the NoHe models.

\subsection{Observational comparisons} 
\label{Sec: Observational Comparisons}
\subsubsection{A comparison to the normal SN 2011fe}
\label{Sec: A comparison to the normal SN 2011fe}

In this section, we compare our models to observations of the normal SN 2011fe \citep{Nugent2011}. For merger-plane orientations, light curve and colour comparisons with SN 2011fe are shown in Figures~\ref{fig:Line of sight Viewing Angles}, \ref{fig:Line of sight Viewing Angles no he} and \ref{fig:Line of sight colour evolution viewing angles}. These show that both the \one and \two models typically produce light curves that are too red. They tend to be too faint and decline too fast in the U and B bands while being too bright in the redder bands. The models also display a  broad I-band peak rather than a distinct secondary maximum, however this may change for a more detailed radiative transfer calculation (see Section~\ref{sec:Radiative Transfer}). 

The effect of the viewing angle variation on peak B-band magnitude and colour can be seen in Figure~\ref{Fig: B-V at B max}. From this, it is apparent that several observer orientations within both models produce significantly brighter and bluer photometric observables. This variation is most striking in the \two model, meaning that, for particular directions, this model comes considerably closer to matching normal SNe~Ia with the bluest lines-of-sight having a B-V at B-band peak of 0.32 compared to 0.42 in the \one model. Hence, the \one and \two models can produce synthetic photometry that aligns more closely with normal Type Ia supernovae if specific lines-of-sight are considered.

\begin{figure}
\centering
\begin{subfigure}[b]{\textwidth}
    \includegraphics[width=0.48\linewidth,trim={0cm 1.6cm 0.05cm 0.03cm},clip]{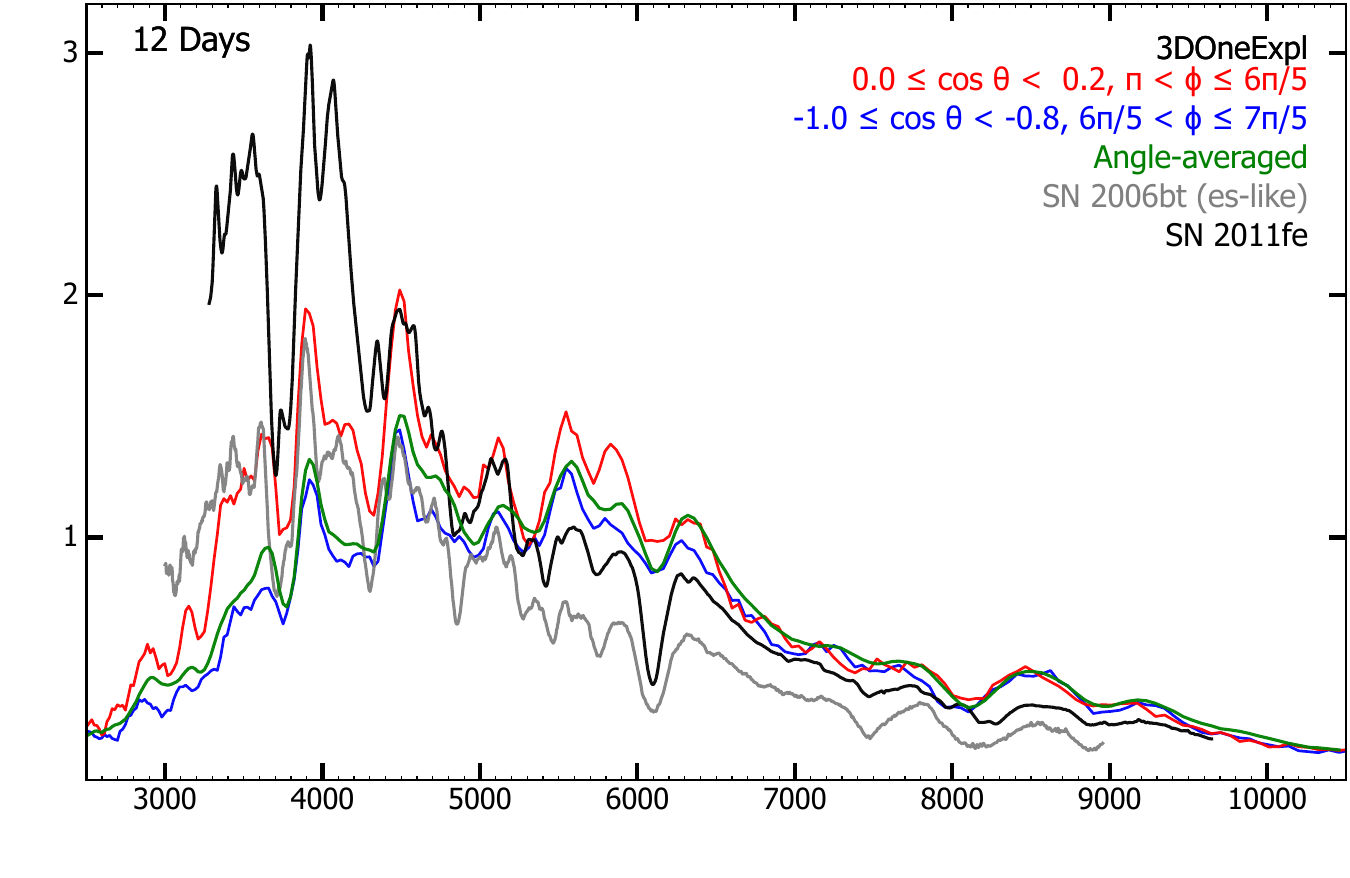}
   \label{fig:Line of sight One Explosion spectra Viewing Angle 12 days} 
   
\end{subfigure}
\begin{subfigure}[b]{\textwidth}
    \includegraphics[width=0.48\linewidth,trim={0cm 1.6cm 0.05cm 0.03cm},clip]{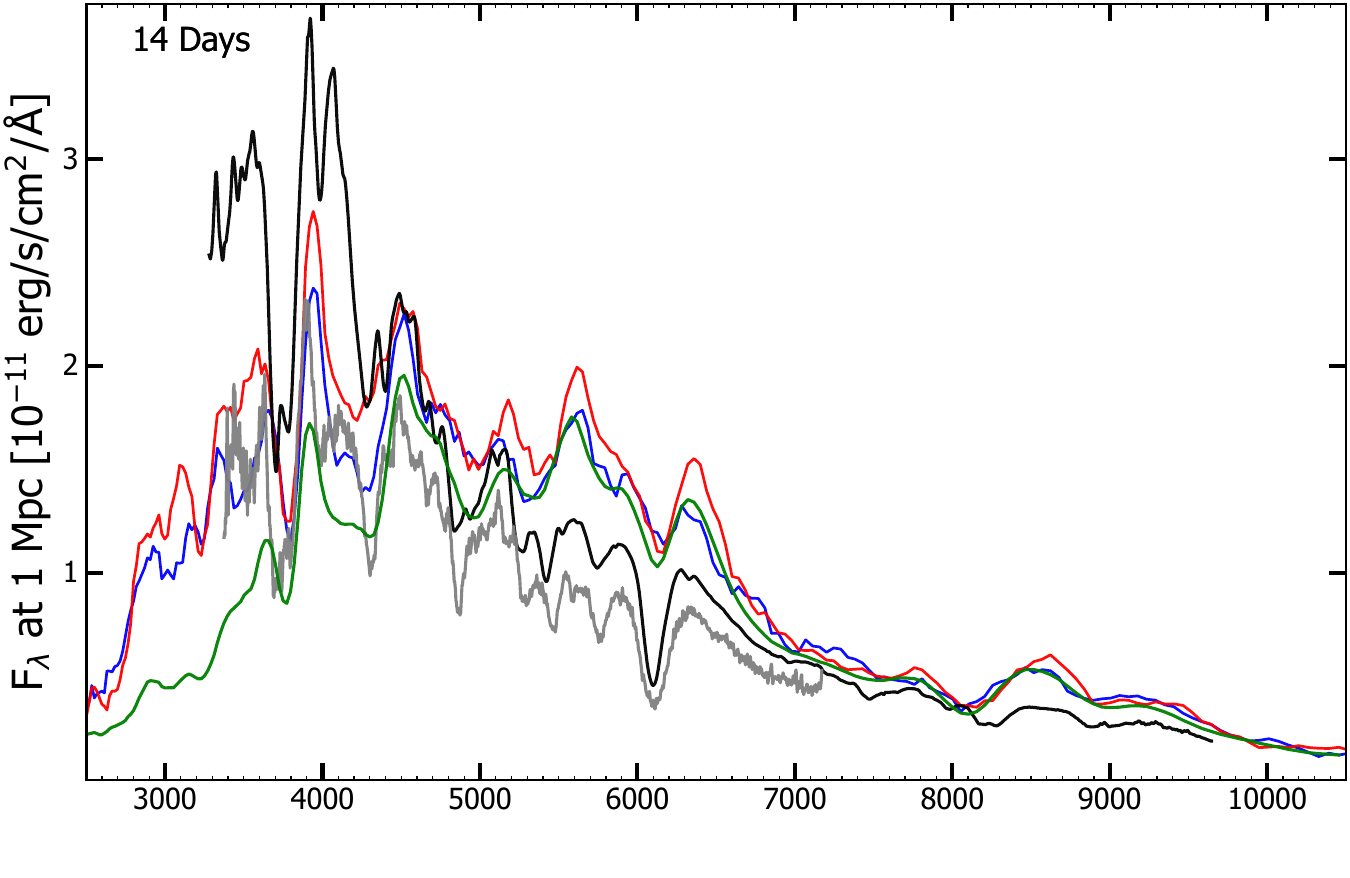}
   \label{fig:Line of sight One Explosion spectra Viewing Angle 14 days} 
\end{subfigure}
\begin{subfigure}[b]{\textwidth}
    \includegraphics[width=0.48\linewidth,trim={0cm 0cm 0.05cm 0.03cm},clip]{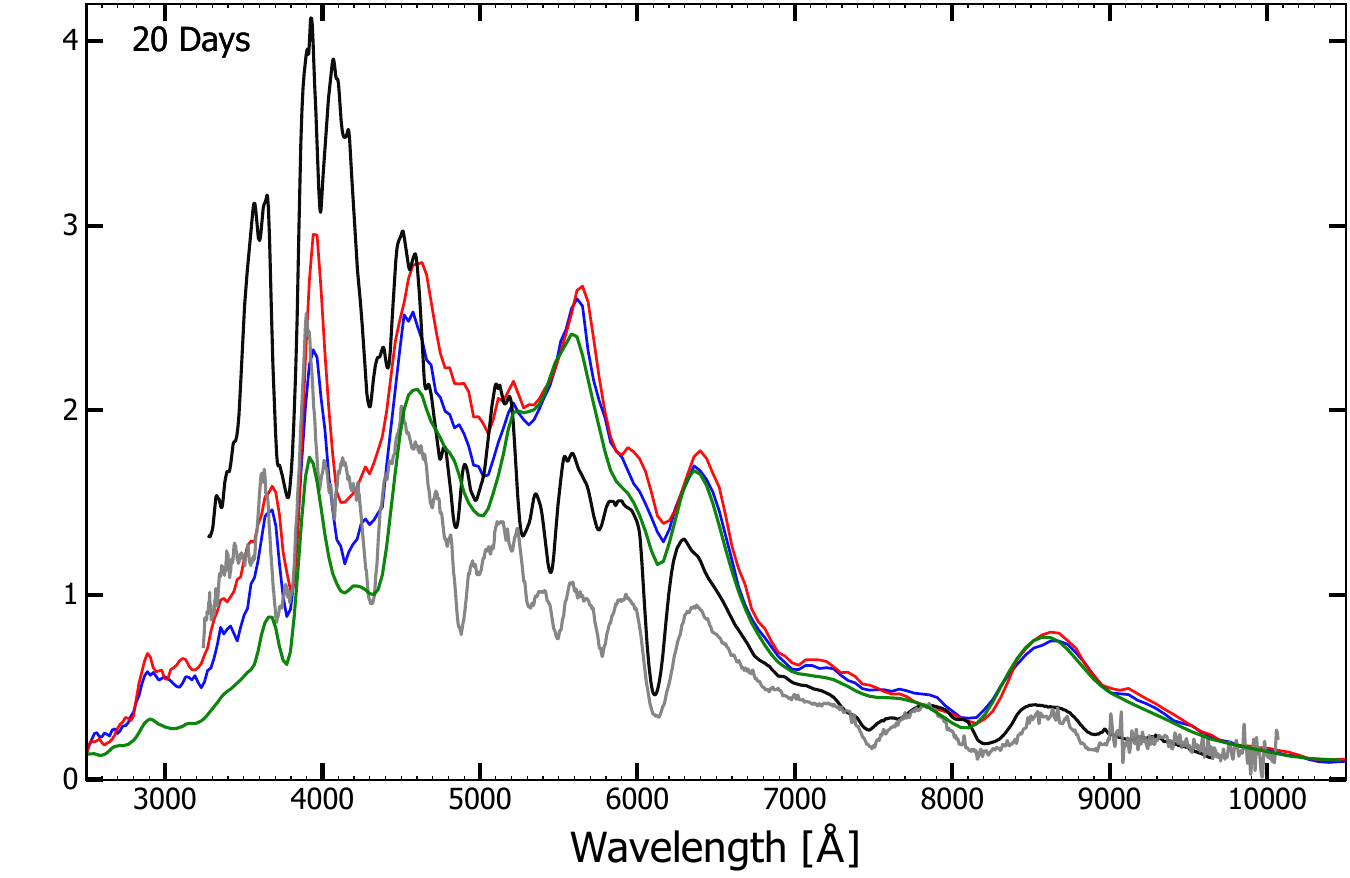}
   \label{fig:Line of sight One Explosion spectra Viewing Angle 21 days} 
   
\end{subfigure}
\caption{3D angle-averaged and 2 unique line-of-sight spectra for the \one model at 12, 14, and 20 days after explosion, compared to SN 2011fe and SN 2006bt at similar epochs. The observer directions $0.0 \le \cos{\theta} < 0.2, 8\pi/5 \le \phi < 9\pi/5$ (red) and $-0.2 \le \cos{\theta} < 0.0, 9\pi/5 \le \phi < 2\pi$ (blue) were selected as lines-of-sight that show the best overall agreement to observations of SN 2011fe and SN 2006bt, respectively. We note a Savitzky-Golay filter has been applied to the model spectra.}

\label{fig:Viewing Angles spectra One Explosion}
\end{figure}

\begin{figure}
\centering
\begin{subfigure}[b]{\textwidth}
    \includegraphics[scale=0.4,width=0.48\linewidth,trim={0cm 1.6cm 0.05cm 0.03cm},clip]{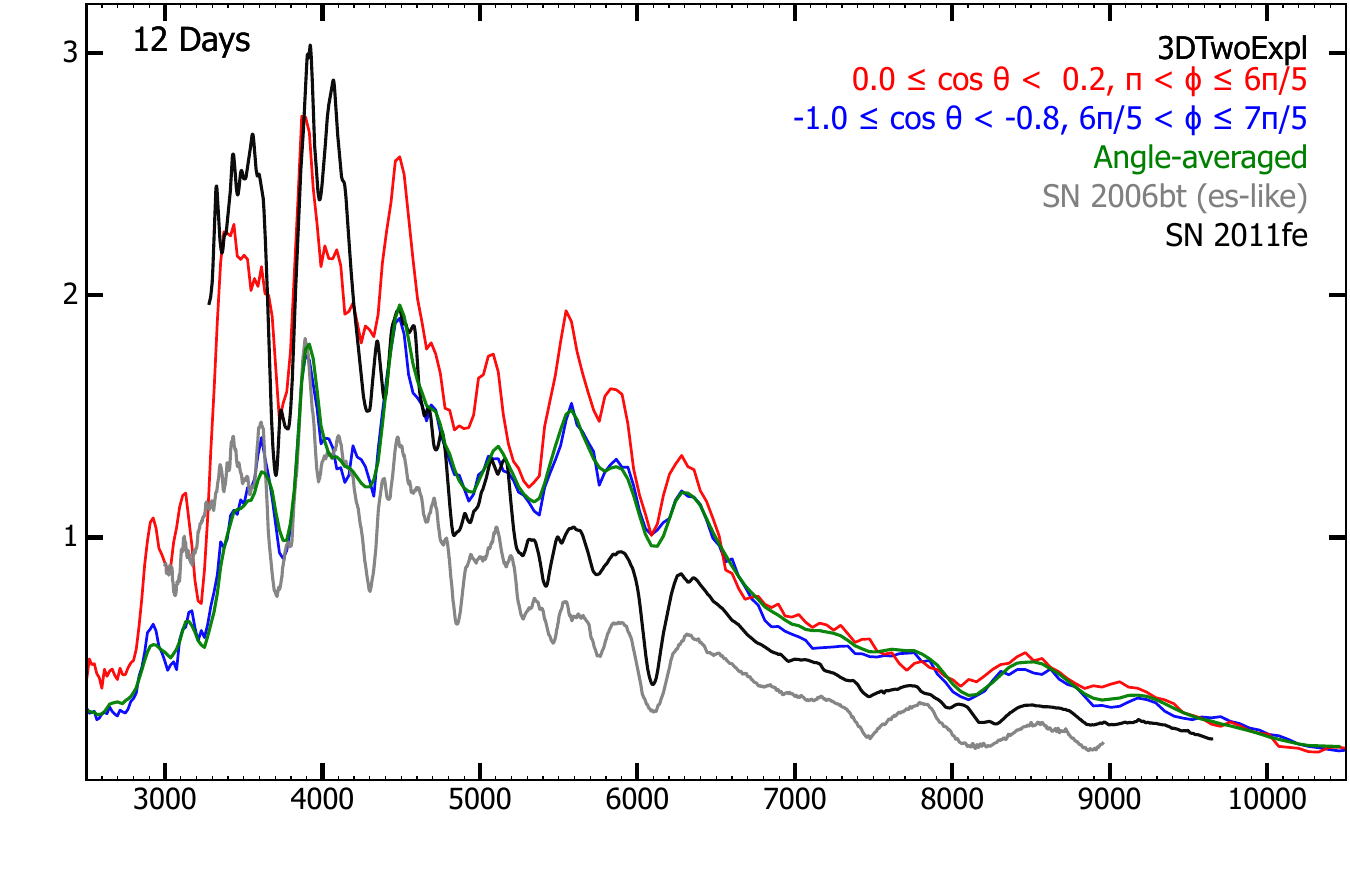}
   \label{fig:Line of sight Two Explosion spectra Viewing Angle 12 days} 
\end{subfigure}
\begin{subfigure}[b]{\textwidth}
    \includegraphics[scale=0.4,width=0.48\linewidth,trim={0cm 1.6cm 0.05cm 0.03cm},clip]{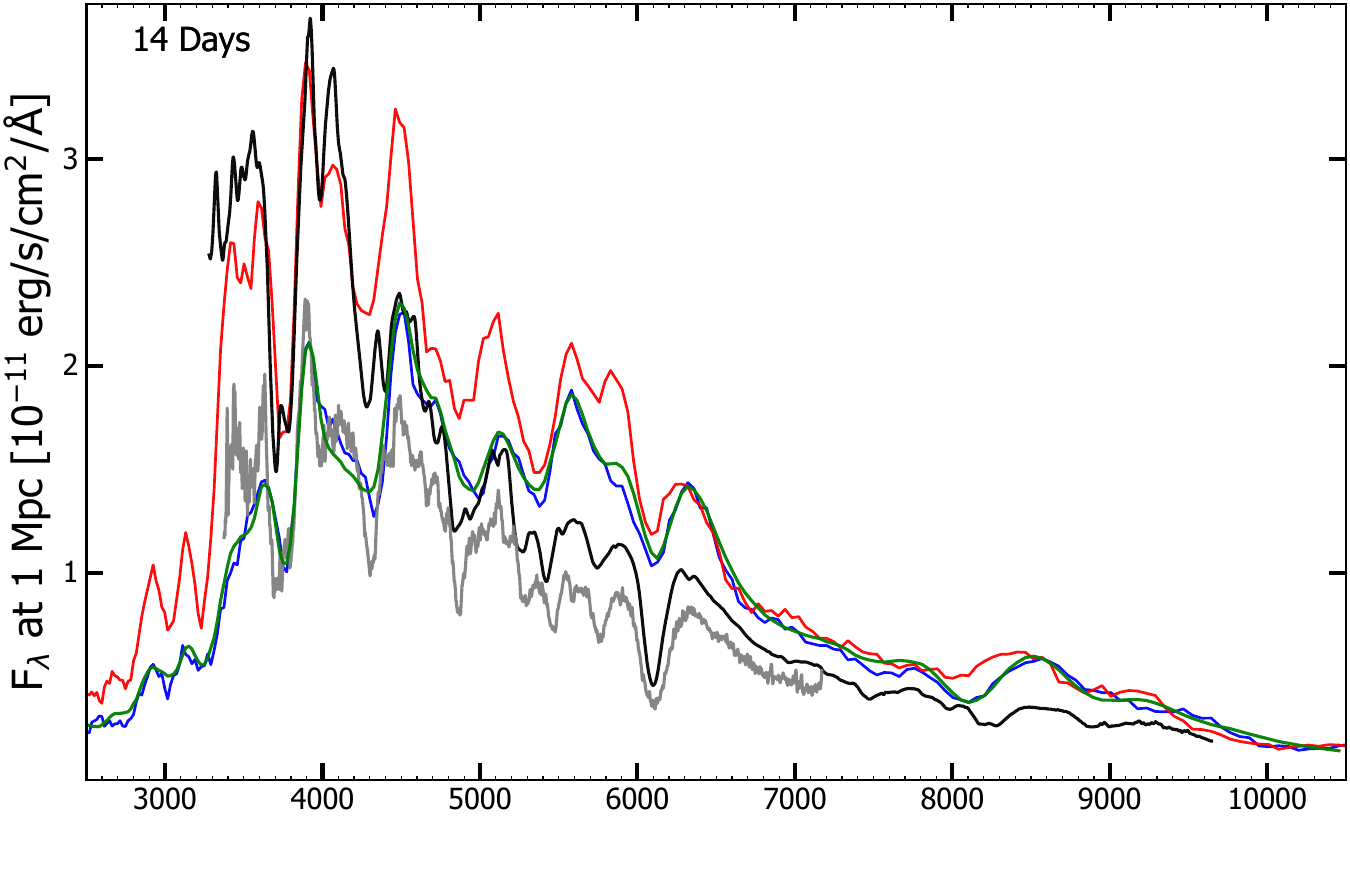}
    \label{fig:Line of sight Two Explosion spectra Viewing Angle 14 days} 
\end{subfigure}
\begin{subfigure}[b]{\textwidth}
    \includegraphics[scale=0.4,width=0.48\linewidth,trim={0cm 0cm 0.05cm 0.03cm},clip]{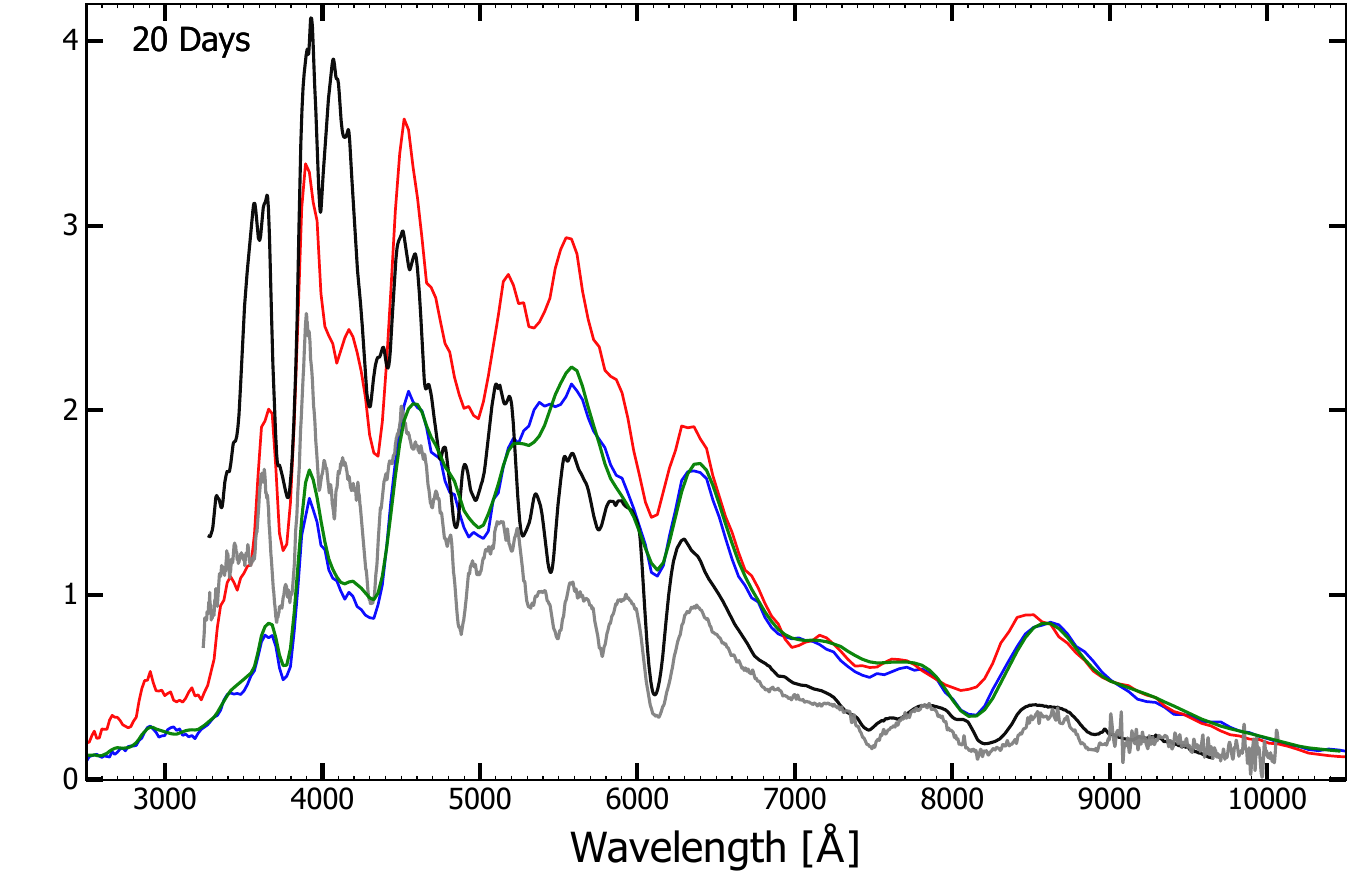}
   \label{fig:Line of sight Two Explosion spectra Viewing Angle 21 days} 
\end{subfigure}
\caption{Same as Figure \ref{fig:Viewing Angles spectra One Explosion} but for the \two model.}
\label{fig:Viewing Angles spectra Two Explosion}
\end{figure}

There is also an improvement in the spectroscopic agreement for some observer orientations relative to the angle-averaged spectra. This is illustrated in Figures~\ref{fig:Viewing Angles spectra One Explosion} and \ref{fig:Viewing Angles spectra Two Explosion} where we display two observer orientations which show most promising agreement with observations over the same three epochs compared in Figure~\ref{fig: angle averaged spectra}. To highlight the degree of viewing angle variation within the models and demonstrate the ability of the model spectra to resemble different classes of SNe Ia these line-of-sight spectra are compared to the angle-averaged spectra. The differences between the \one and \two models compared to observations of SN 2011fe are most prominent between $\sim$3000-4500\AA\xspace, where the \two model yields a much stronger \ion{Ca}{II} H\&K feature and exhibits additional IME features compared to the \one model. At longer wavelengths, both models produce the expected \ion{Si}{II} and \ion{Ca}{II} features, but the features are stronger in the \two model and, as such, closer to matching those of normal SNe~Ia.

Overall, the \two model is a more promising match for SN 2011fe than the \one model as it possesses more lines-of-sight with comparable colour and peak luminosity while having better agreement with spectroscopic features at shorter wavelengths for these lines-of-sight. For example, consider lines-of-sight of the \two model that fall outside 1 standard deviation of the mean value of both B-band peak brightness and B-V at B-band maximum (to the brighter and bluer end of the distribution). These show an average B-V at B-band peak of $\sim$0.35 and peak B-band magnitude of -19.1 mag (see Figure~\ref{fig:Line of sight Viewing Angles no he} ). This makes them significantly bluer in B-V at B-band peak colour than the mean of the distribution (which is B-V of 0.59) and thus in better agreement with normal SNe~Ia. The average peak B-band magnitude for these lines-of-sight is also more in line with the peak brightnesses exhibited by the bulk of SNe Ia \citep{Ashall2016}. However, such lines-of-sight only represent a small fraction of the viewing angles within the model (e.g. only 9\% of the viewing angles gave B-V at peak $\lessapprox$0.4). Thus, although these models demonstrate that the double-degenerate double-detonation scenario has promise for yielding ejecta structures that could be close to what is required for normal Ia's, the particular realisations considered here cannot be viewed as promising standard models for normal SNe~Ia because they predict too large a fraction of the lines-of-sight would show peculiarities and/or very red colours.

\begin{figure}
\centering
\begin{subfigure}[b]{\textwidth}
    \includegraphics[scale=0.4,width=0.48\linewidth,trim={0cm 1.6cm 0.05cm 0.03cm},clip]{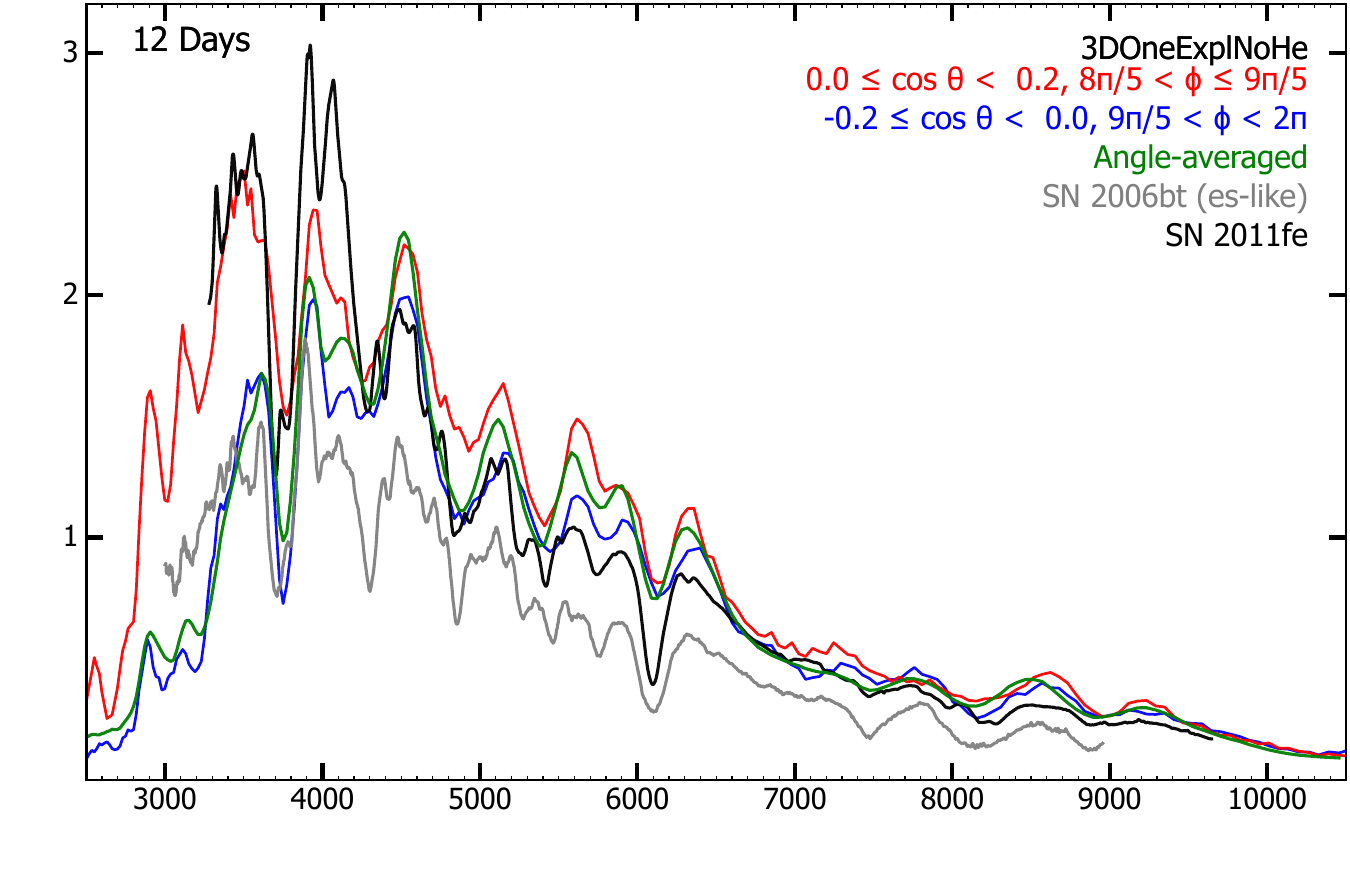}
   \label{fig:Line of sight 3D One Explosion No He spectra Viewing Angle 12 days} 
\end{subfigure}
\begin{subfigure}[b]{\textwidth}
    \includegraphics[scale=0.4,width=0.48\linewidth,trim={0cm 1.6cm 0.05cm 0.03cm},clip]{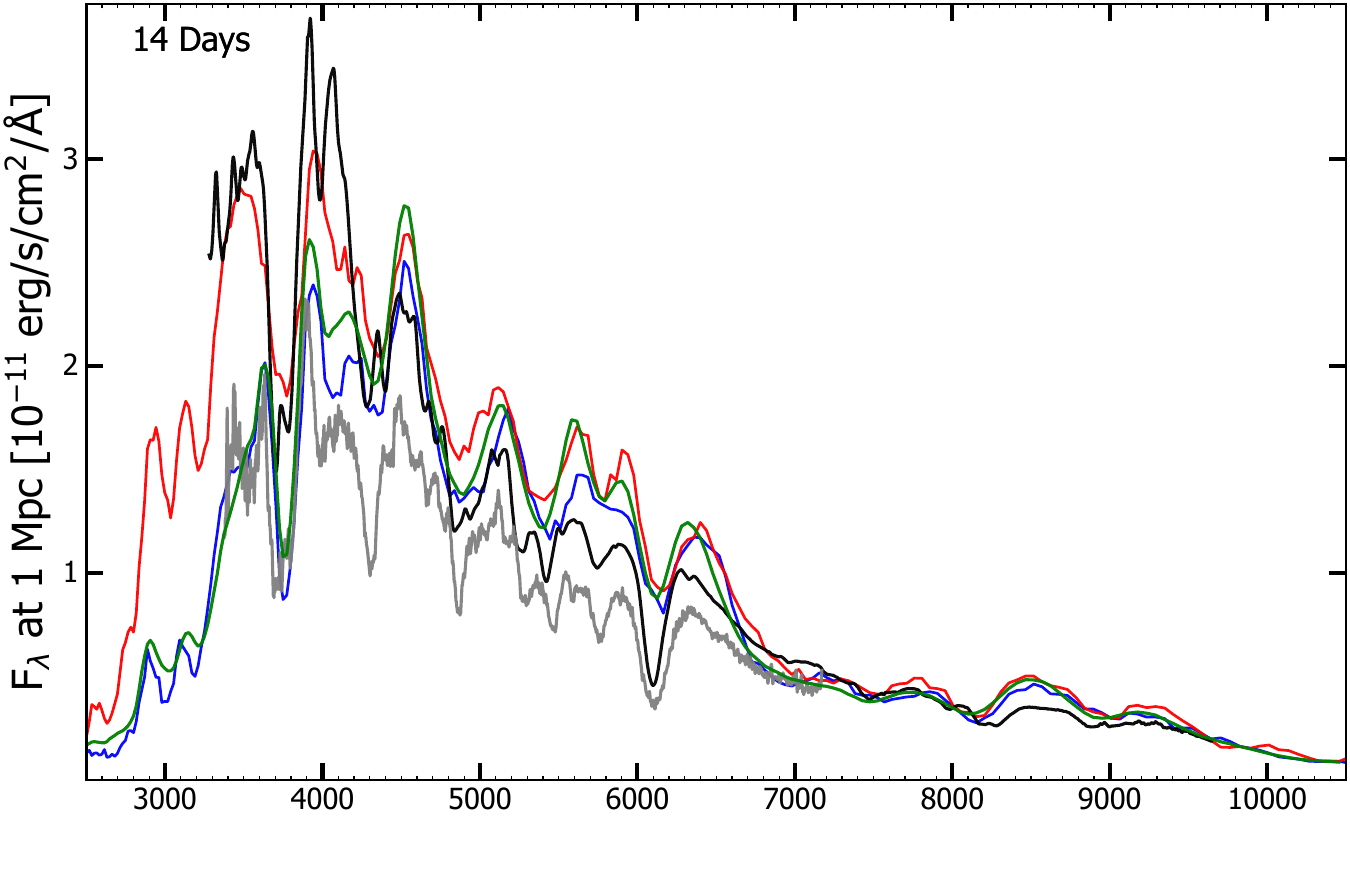}
   \label{fig:Line of sight 3D One Explosion No He spectra Viewing Angle 14 days} 
\end{subfigure}
\begin{subfigure}[b]{\textwidth}
    \includegraphics[scale=0.4,width=0.48\linewidth,trim={0cm 0cm 0.05cm 0.03cm},clip]{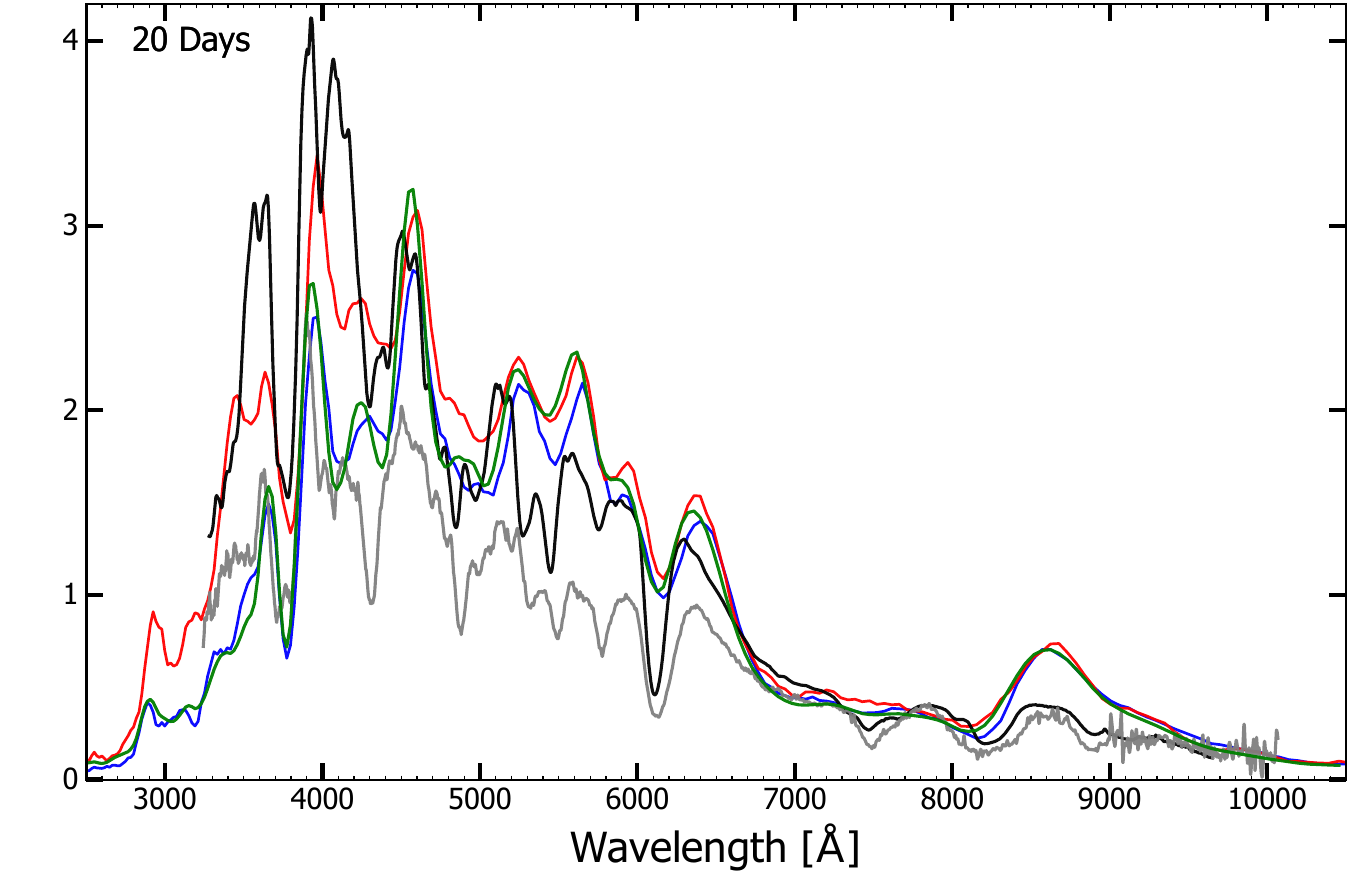}
   \label{fig:Line of sight 3D One Explosion No He spectra Viewing Angle 21 days} 
\end{subfigure}
\caption{Same as Figure \ref{fig:Viewing Angles spectra One Explosion} but for the \oneno model.}
\label{fig:Viewing Angles spectra One No He Explosion}
\end{figure}

\begin{figure}
\centering
\begin{subfigure}[b]{\textwidth}
        \includegraphics[scale=0.4,width=0.48\linewidth,trim={0cm 1.6cm 0.05cm 0.03cm},clip]{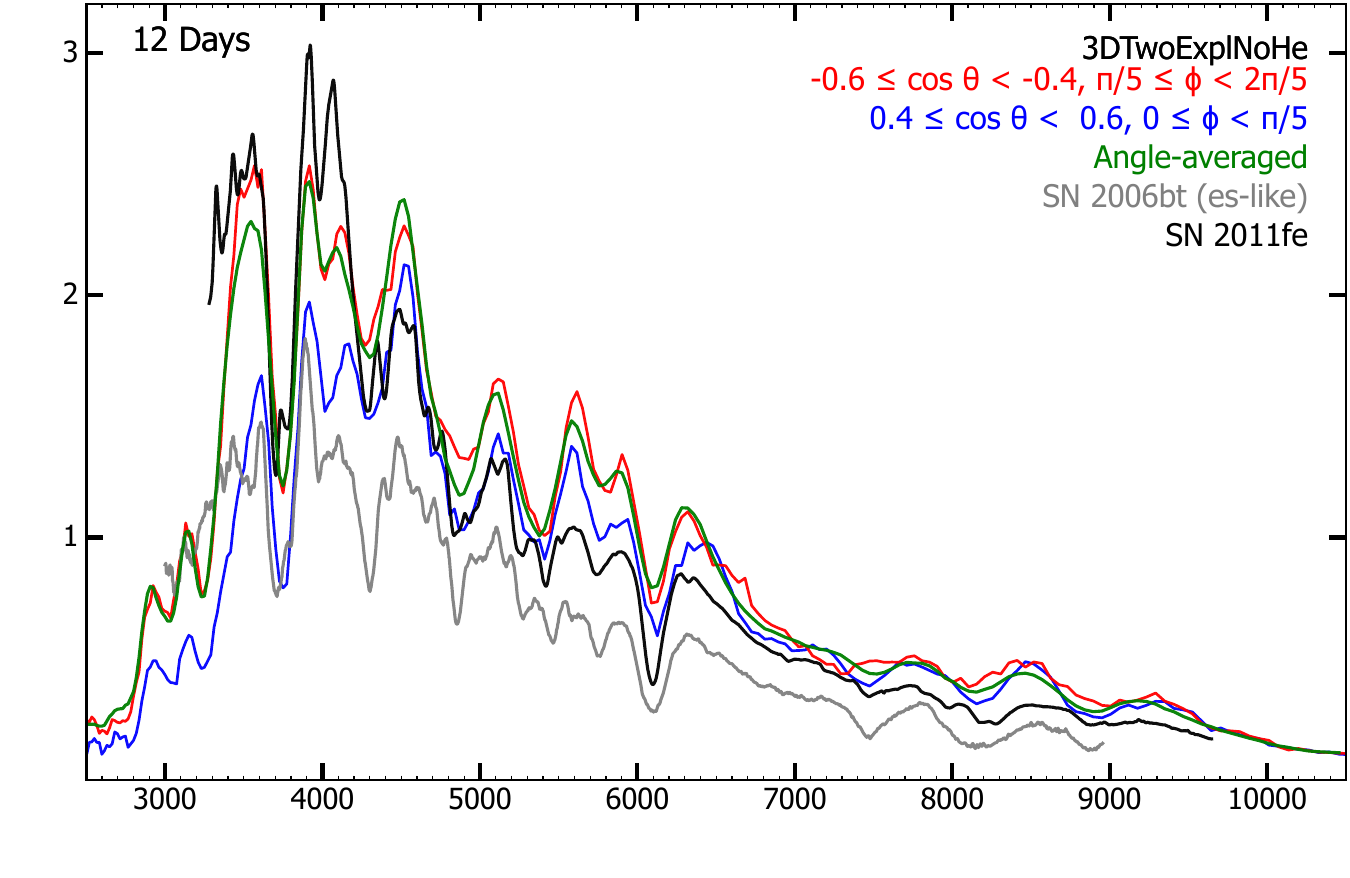}
   \label{fig:Line of sight Two Explosion No He spectra Viewing Angle 12 days} 
\end{subfigure}
\begin{subfigure}[b]{\textwidth}
    \includegraphics[scale=0.4,width=0.48\linewidth,trim={0cm 1.6cm 0.05cm 0.03cm},clip]{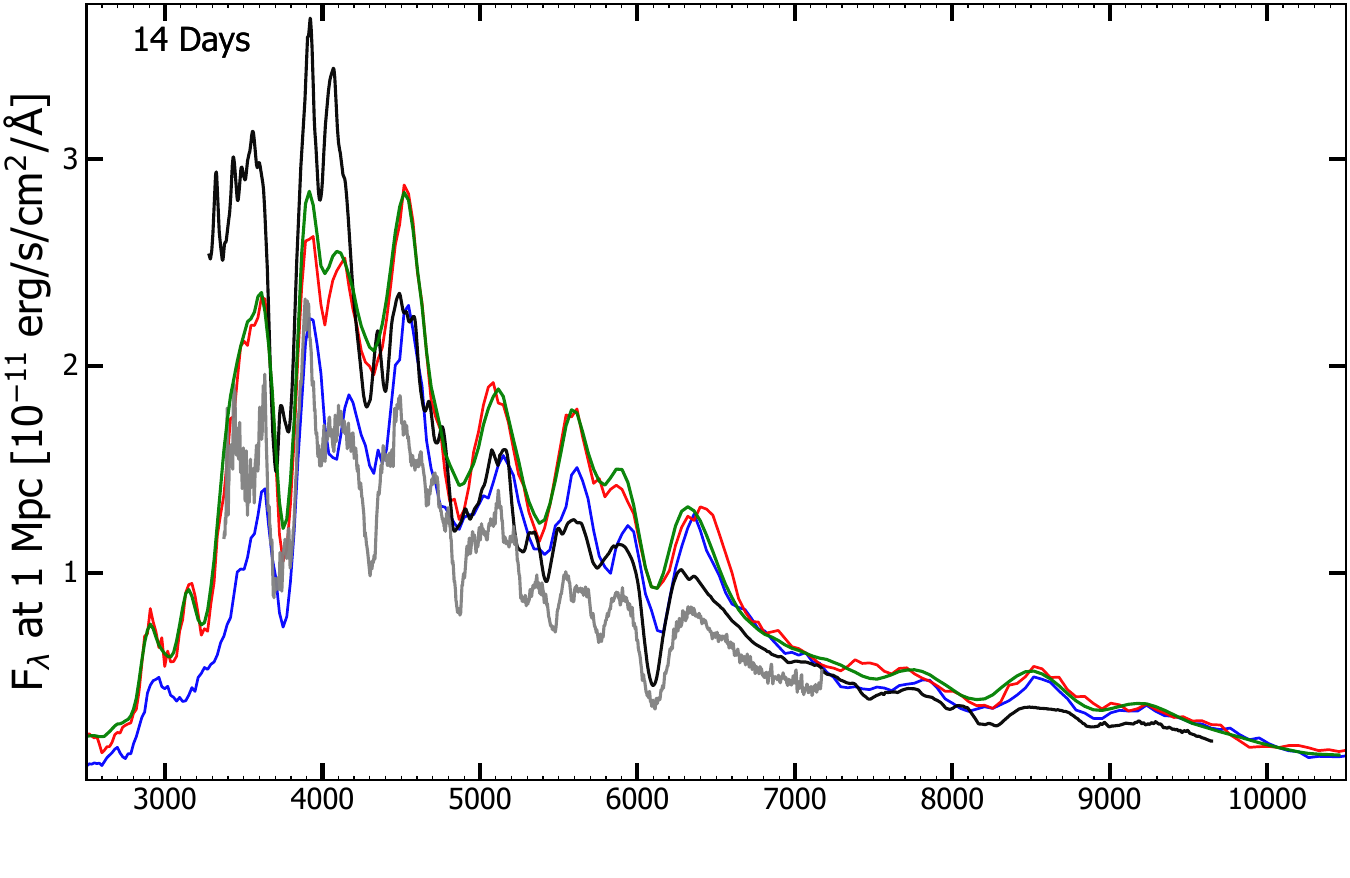}
   \label{fig:Line of sight Two Explosion No He spectra Viewing Angle 14 days} 
\end{subfigure}
\begin{subfigure}[b]{\textwidth}
    \includegraphics[scale=0.4,width=0.48\linewidth,trim={0cm 0cm 0.05cm 0.03cm},clip]{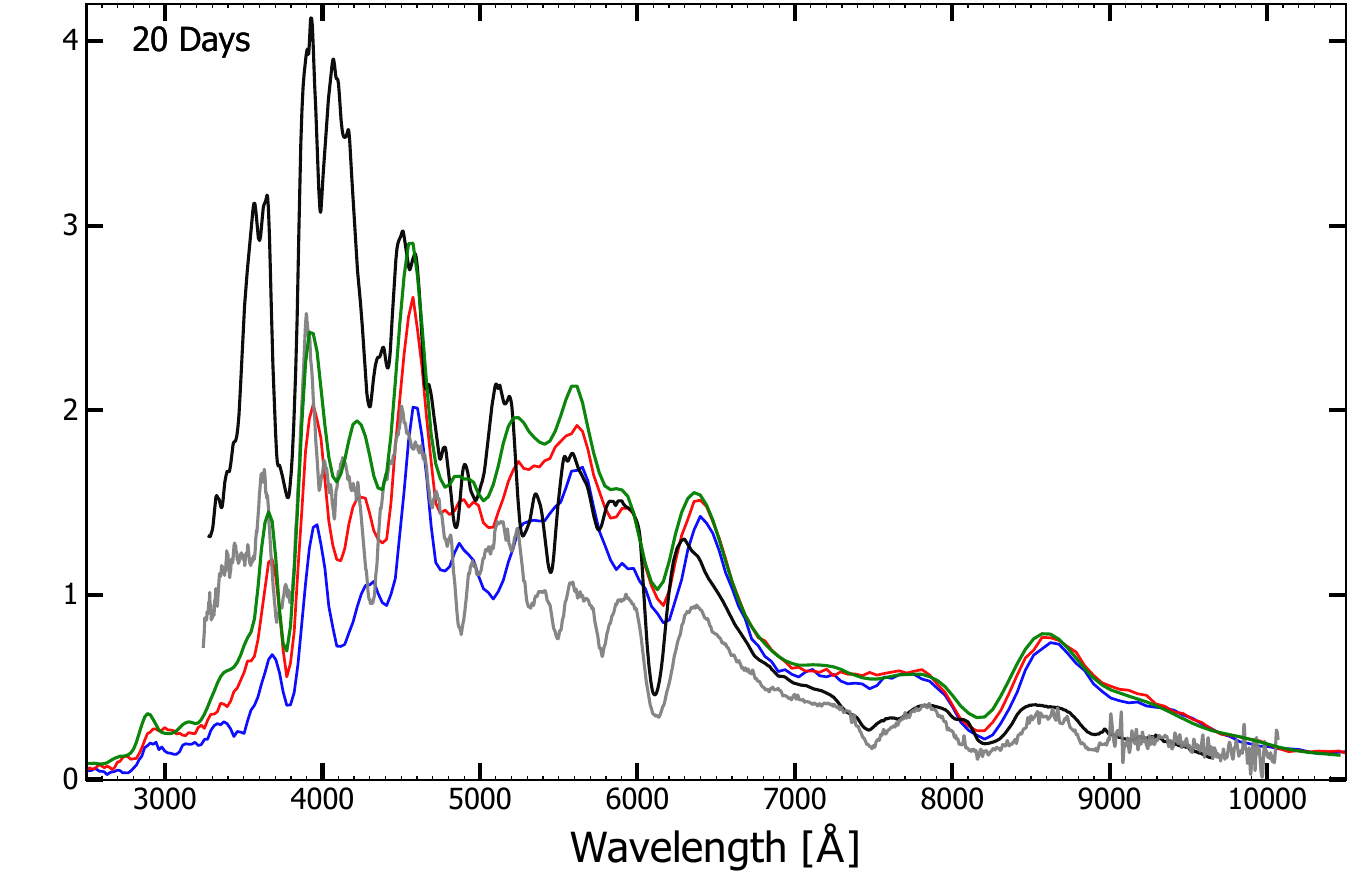}
   \label{fig:Line of sight Two Explosion No He spectra Viewing Angle 21 days} 
\end{subfigure}
\caption{Same as Figure \ref{fig:Viewing Angles spectra One Explosion} but for the \twono model.}
\label{fig:Viewing Angles spectra Two Explosion No He}
\end{figure}

Investigation into the \oneno and \twono models has allowed for better understanding of the impact of the helium shell detonation on the synthetic observables. In particular, our calculations confirm that the removal of the helium shell detonation ash significantly improves the comparison to normal SNe~Ia. Figure~\ref{Fig: B-V at B max} clearly illustrates that once the helium ash is removed, the peak brightness and colour of both the \oneno and \twono models become closer to that of the population of normal SNe~Ia. The investigation of  selected line-of-sight spectra for the 3D NoHe models (see Figures~\ref{fig:Viewing Angles spectra One No He Explosion} and \ref{fig:Viewing Angles spectra Two Explosion No He}) highlights that both models can match observations significantly better than the full models. For some observer orientations the \oneno model has a greatly improved match to observations of SN 2011fe across all three epochs between $\sim$3300\AA\xspace and 4600\AA\xspace compared to that of the \one model, both in terms of luminosity and morphology of spectral features such as the optical \ion{Si}{II} and \ion{Ca}{II} H\&K features. The \twono model demonstrates significantly improved agreement with normal SNe~Ia for most observer orientations. The average spectra also now appear more compatible with normal SNe~Ia. However, the main improvement for viewing angles that already resembled normal SNe~Ia lies in the overall bluer spectral energy distribution (SED). This suggests that further exploration of the double-degenerate double-detonation scenario for models with reduced helium shell masses could be promising in the search for model realisations that may provide good matches to normal SNe~Ia for a larger fraction of observer orientations.

\subsubsection{A comparison to the 02es-like SN 2006bt}
\label{Sec: A comparison to the normal SN 2006bt}

In this Section, we compare the models to members of the 02es-like subclass, specifically SN 2002es itself \citep{Ganeshalingam2012} and SN 2006bt \citep{Foley2010}. 
We note all our models are substantially brighter than SN 2002es itself and, as such, can only quantitatively be compared to the brighter objects within the 02es-like class, such as SN 2006bt. Nevertheless, we will make spectral comparisons with SN 2002es as the prototype for this subclass.

Light curve comparisons with both SN 2006bt and SN 2002es for lines-of-sight around the merger plane are shown in Figures~\ref{fig:Line of sight Viewing Angles} and \ref{fig:Line of sight Viewing Angles no he}. The lines-of-sight in the \one and \two models are more similar to those of the 02es-like subclass than normal SNe~Ia. This similarity arises due to their lower luminosities in the U and B bands and the absence of a secondary maximum in the I-band. Consequently, both models show a redder colour evolution (see Figure~\ref{fig:Line of sight colour evolution viewing angles}), thus aligning them more closely with the 02es-like subclass than normal SNe~Ia. As noted above, orientation effects in both models are substantial. However, a relatively large fraction of lines-of-sight show SEDs that are broadly compatible with SN 2006bt. Consequently, the average line-of-sight within each model shows better agreement with the spectral time series of SN 2006bt rather than SN 2011fe, as can be seen in Figures~\ref{fig:Viewing Angles spectra One Explosion} and \ref{fig:Viewing Angles spectra Two Explosion}. Additionally, Figures~\ref{fig:Viewing Angles spectra One Explosion} and \ref{fig:Viewing Angles spectra Two Explosion} show that specific observer orientations within both models can exhibit better agreement for both models at shorter wavelengths ($\leq$4300\AA) compared to the angle-averaged spectra. However, the models still fail to accurately predict the strength of all features at wavelengths greater than $\sim$4300Å and consequently appear too red after maximum light. Overall, both models possess many observer orientations that can demonstrate good photometric and spectroscopic agreement with the peculiar 02es-like subclass of SNe~Ia. However, since the \one model exhibits a smaller degree of viewing angle variation, it naturally leads to more observer orientations that can replicate the typical observables seen in members of the 02es-like subclass, such as SN 2006bt.

The NoHe models yield bluer light curves, which results in a color evolution that more closely resembles the 02es-like subclass than the full models (see Figure~\ref{fig:Line of sight colour evolution viewing angles}). This is particularly apparent around maximum light and leads to significant changes in the overall SED, as is apparent in our spectra (see Figure~\ref{fig:los_spectra}). Indeed while Figures~\ref{fig:Viewing Angles spectra One Explosion} to \ref{fig:Viewing Angles spectra Two Explosion No He} demonstrate that the \oneno model and, in particular, the \twono model produce spectra most similar to SN 2011fe, all models can yield lines-of-sight that replicate observations of SN 2006bt fairly well. This suggests that, although the current models may contain too much helium ash for good agreement across all observer orientations for each model, fully eliminating (or minimising) the impact of the helium ash is less critical for matching the 02es-like subclass due to their redder colour relative to the normal SNe~Ia.

\subsubsection{A comparison of \ion{Si}{II} velocity at maximum}
\label{sec:SiII_velocity}

\begin{figure}
\centering
\begin{subfigure}[b]{\textwidth}
        \includegraphics[width=0.49\linewidth]{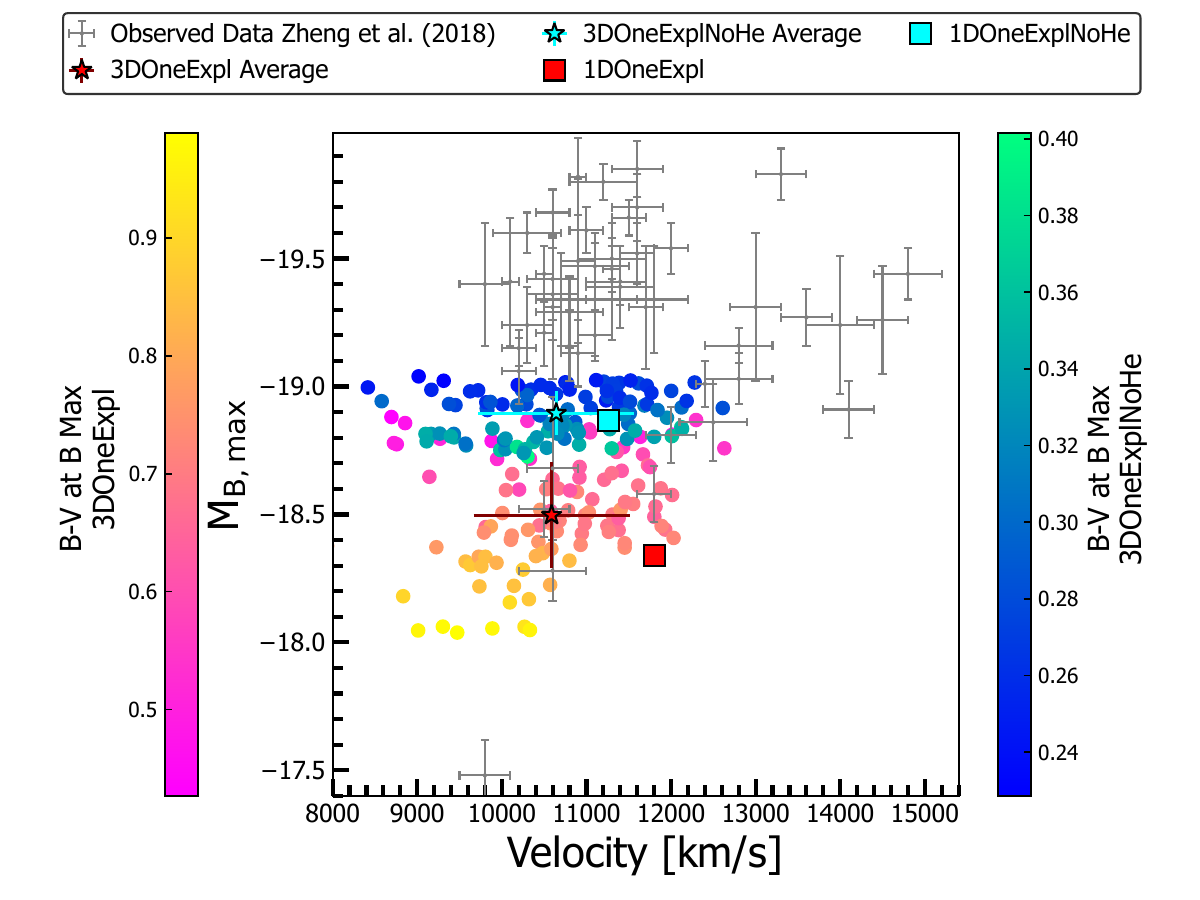} 
\end{subfigure}
\caption{\ion{Si}{II} velocity at at B-band peak compared to an observational sample of SNe~Ia \citep{Zheng2018}, alongside the B-V value at peak B-band maximum as indicated by the color bar.}
\label{fig:Si II Velocity one}
\end{figure}

\begin{figure}
\centering
\begin{subfigure}[b]{\textwidth}
    \includegraphics[,width=0.49\linewidth]{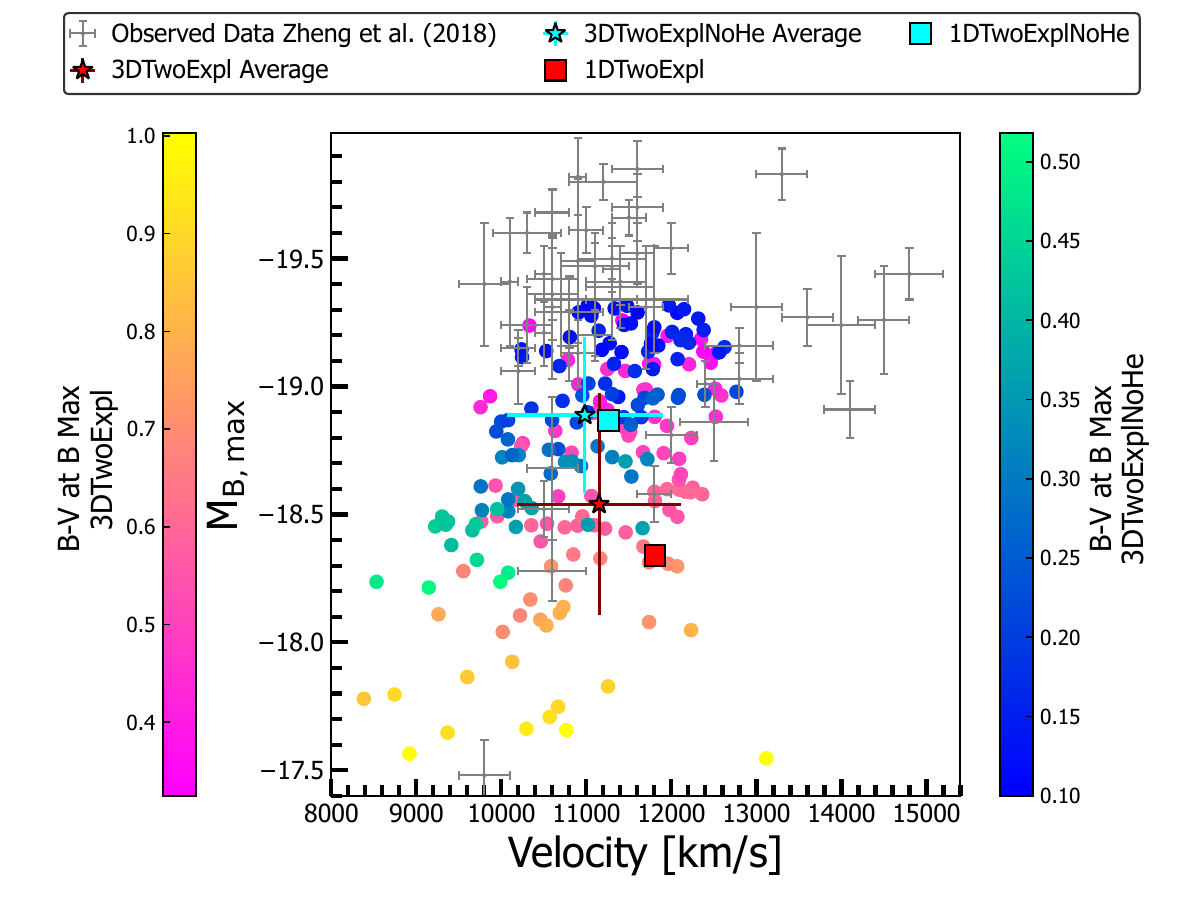}
\end{subfigure}
\caption{Same as Figure \ref{fig:Si II Velocity one} but for the \two and \twono models}
\label{fig:Si II Velocity two}
\end{figure}

Previous analysis of the \ion{Si}{II} 6355\AA~feature has suggested that there are two distinct populations of SNe~Ia, with normal and high-velocity features \citep{Polin2019} where the boundary between them is placed at 11800 km/s at peak brightness \citep{Wang2009}.
Here we investigate the range of \ion{Si}{II} velocities predicted by the models, and examine how observer orientation impacts this quantity.
To this end, we examined all 100 lines-of-sight for each model and fitted a Gaussian to the \ion{Si}{ii} feature at the time of B-band maximum.
The resulting distribution of \ion{Si}{II} 6355\AA\ line velocity versus peak B-band magnitude is shown for our models in Figure \ref{fig:Si II Velocity one}~and~\ref{fig:Si II Velocity two}. For reference, we also indicate the corresponding B-V colour for each line-of-sight and overplot the extinction-corrected observational sample of \cite{Zheng2018}. 

As expected based on the light curve comparisons discussed above, the absolute values of the B-band peak magnitudes are generally lower than in the observed sample.
For all models, there is a significant spread in the \ion{Si}{ii} line velocity, which depends on observer orientation. The range of velocities is roughly similar to the observations, except that the models cannot account for the high velocity population ($\gtrapprox$13000 km/s), and a small number of the model lines-of-sight have velocities lower than any in the observational sample.
Similar to observations, the spread in velocity is significant across the range of luminosity: while there is a weak correlation of velocity with luminosity (and colour) in most of the models, the distribution in velocity spans at least $10^3$~km~s$^{-1}$ at any fixed brightness. 
We also performed a similar analysis of the \ion{Si}{ii} line velocity versus \dmv, which yielded similar findings (i.e. significant scatter introduced by the range of viewing angles in the simulation).
Clearly, our 1D simulations cannot account for the viewing-angle diversity predicted in the line velocities, but 
we note that they do produce velocities that are similar to those typical of the 3D simulations in most cases: for all models except the \one model, the 1D line model velocity is within a standard deviation of the mean from the corresponding 3D simulation (see Figure \ref{fig:Si II Velocity one} and Figure \ref{fig:Si II Velocity two}).

\section{Discussion and Conclusions}
\label{sec:Discussion and Conclusions}

We have carried out a series of 3D radiative transfer calculations for the dynamically driven double-degenerate double-detonation explosion models developed by \cite{pakmor_2021}, where the primary difference between explosion models is the fate of the secondary WD. We have shown that the angle-averaged 3D light curves and spectra are quite different at early times but become similar at later times compared to the previous 1D simulations. The models exhibit significant viewing-angle variation, which is strongest in bluer bands and is reduced when the helium shell detonation ash is removed. The NoHe models investigated show that in many respects, the 3D NoHe simulations are better represented by the equivalent 1D NoHe calculations. This confirms that many orientation-dependent effects originate in the distribution of the helium detonation ash. This is primarily due to the shell detonation producing a large amount of Ti, which generates a considerable amount of line blanketing in 1D. 
Nevertheless, significant observer orientation effects are present in all our simulations affecting light curve morphologies, colours and predicted line velocities for key spectral signatures (such as \ion{Si}{II} 6355\AA).
Our results therefore highlight both the importance of three-dimensional radiative transfer for simulating 3D hydrodynamic models with complex ejecta, and the particular sensitivity of results to properties of the helium shell detonation ash (including its 3D structure).

The \two model can produce a reasonable match to the spectroscopic features of both the normal \fe and the 02es-like \bt for different lines-of-sight. However, in its full form, the model colours are generally too red to match \fe. If the helium ash is removed, the NoHe version of the \two model becomes somewhat closer to matching \fe spectroscopically while also producing more lines-of-sight which can produce peak luminosities closer to that of SN 2011fe. The \one model in its full form produces a poorer match to that of SN 2011fe than the \two model; however, when the helium ash is removed the overall ability for the NoHe version of the \one model to produce lines-of-sight matching \fe improves significantly. This result reinforces the findings of \cite{Townsley2019} that smaller helium shells produce agreement better aligned with normal SNe~Ia. In general, the current models yield colour evolution closer to that of peculiar subclasses such as the 02es-like objects, which is a similar finding to studies of double-detonation models \citep{kromer2010,Polin2019,Gronow2020,collins_double_detonation}. The \one model, in particular, produces more lines-of-sight that resemble the 02es-like subclass. Thus, although the present models are too luminous to account for most members of the 02es-like subclass, similar models should be investigated across a range of merger masses as potential candidates for peculiar SNe~Ia.

Overall, our results suggest that 
the dynamically driven double-degenerate double-detonation models generally show promise and merit further investigation that should include both exploration of lower-mass helium shells (to test their compatibility with normal SNe~Ia) and ranges of progenitor mass (in particular as candidates for subluminous, peculiar SNe~Ia). 
Although our calculations demonstrate the importance of 3D effects, we note that they are limited by the use of an approximate NLTE treatment, the accuracy of which means we have limited this study to relatively early phases.
Due to the highly asymmetric nature of the inner ejecta of both models, it will be of particular interest for future studies to perform radiative transfer calculations at substantially later times using both a multidimensional approach and a complete NLTE treatment in order to assess how nebular phase observations can be used to test the predictions of such merger models further.

\section*{Acknowledgements}

This work was performed using resources provided by the Cambridge Service for Data Driven Discovery (CSD3) operated by the University of Cambridge Research Computing Service (www.csd3.cam.ac.uk), provided by Dell EMC and Intel using Tier-2 funding from the Engineering and Physical Sciences Research Council (capital grant EP/T022159/1), and DiRAC funding from the Science and Technology Facilities Council (www.dirac.ac.uk).
The authors gratefully acknowledge the Gauss Centre for Supercomputing e.V. (www.gauss-centre.eu) for funding this project by providing computing time on the GCS Supercomputer \cite{JUWELS} at Jülich Supercomputing Centre (JSC).
This research has made use of the CfA Supernova Archive, which is funded in part by the National Science Foundation through grant AST 0907903.
SAS and FPC acknowledge funding from STFC grant ST/X00094X/1.
CEC acknowledges funding by the European Union (ERC, HEAVYMETAL, 101071865). Views and opinions expressed are however those of the author(s) only and do not necessarily reflect those of the European Union or the European Research Council. Neither the European Union nor the granting authority can be held responsible for them. The work of FKR is supported by the Klaus-Tschira Foundation and by the European Union (ChETEC-INFRA, project no. 101008324).
LJS acknowledges support by the European Research Council (ERC) under the European Union’s Horizon 2020 research and innovation program (ERC Advanced Grant KILONOVA No. 885281).
LJS acknowledges support by Deutsche Forschungsgemeinschaft (DFG, German Research Foundation) - Project-ID 279384907 - SFB 1245 and MA 4248/3-1.
JMP acknowledges the support of the Department for Economy (DfE). Numpy \citep{harris2020array}, SciPy \citep{2020SciPy-NMeth}, Matpoltlib \citep{harris2020array} and
\href{https://zenodo.org/records/8302355} {\textsc{artistools}}\footnote{\href{https://github.com/artis-mcrt/artistools/}{https://github.com/artis-mcrt/artistools/}} \citep{artistools2023a} was used for data processing and plotting.

\section*{Data Availability}

The light curves and spectra shown here will be shared on reasonable request to the corresponding author.
 
\bibliographystyle{mnras}
\bibliography{bibliography} 
\bsp	
\label{lastpage}
\end{document}